\documentclass[twocolumn, twocolappendix]{aastex63}

\usepackage{CJK}

\usepackage{graphicx}
\usepackage{natbib}
\usepackage{gensymb}
\hypersetup{linkcolor=blue,citecolor=blue,filecolor=blue,urlcolor=blue}
\usepackage{rotating}

\usepackage{longtable}
\usepackage{threeparttablex}
\usepackage{multirow}
\usepackage{footmisc}

\usepackage{amsmath}
\usepackage{verbatim}

\usepackage{relsize} 

\newcommand{\HI}{H{\sc\ i}}

\newcommand{\OI}{O{\sc\ i}}

\newcommand{\SiII}{Si{\sc\ ii}}
\newcommand{\SiIII}{Si{\sc\ iii}}
\newcommand{\SiIV}{Si{\sc\ iv}}

\newcommand{\CII}{C{\sc\ ii}}

\newcommand{\CIII}{C{\sc\ iii}}
\newcommand{\CIV}{C{\sc\ iv}}
\newcommand{\OVI}{O{\sc\ vi}}

\newcommand{\rvir}{R_{\rm 200m}}
\newcommand{\rvirm}{R_{\rm 200m}}
\newcommand{\rvirc}{R_{\rm 200c}}

\newcommand{\kms}{\rm km~s^{-1}}
\newcommand{\mhalo}{M_{\rm h}}
\newcommand{\mhalom}{M_{\rm 200m}}
\newcommand{\mhaloc}{M_{\rm 200c}}

\newcommand{\mstar}{M_*}
\newcommand{\logmstar}{\rm \log M_*}
\newcommand{\mhi}{M_{\rm HI}}
\newcommand{\logmhi}{\rm \log M_{HI}}
\newcommand{\msun}{M_\odot}
\newcommand{\msunyr}{M_\odot~{\rm yr}^{-1}}
\newcommand{\vlsr}{v_{\rm LSR}}
\newcommand{\mdotstar}{\dot{M_*}}
\newcommand{\mfuv}{m_{\rm FUV}}

\newcommand{\Cf}{C_{\rm f}}
\newcommand{\cfHI}{89\% (24/27)}
\newcommand{\cfCII}{20\% (3/15)}
\newcommand{\cfCIV}{22\% (9/41)}
\newcommand{\cfSiII}{5\% (2/40)}
\newcommand{\cfSiIII}{19\% (7/36)}
\newcommand{\cfSiIV}{11\% (4/37)}

\newcommand{\npairtotal}{56}
\newcommand{\npairthiswork}{22}
\newcommand{\npairliterature}{34}
\newcommand{\npairbordoloi}{12}
\newcommand{\npairjohnson}{11}
\newcommand{\npairliang}{5}
\newcommand{\npairqu}{3}
\newcommand{\npairzhengic}{3}
\newcommand{\nunidwarf}{45}
\newcommand{\medrvirval}{136} 
\newcommand{\medrvir}{\langle\rvir\rangle}
\newcommand{\samplemstarrange}{10^{6.5-9.5}}
\newcommand{\medmstar}{10^{8.3}}
\newcommand{\samplemhalorange}{10^{10.0-11.5}}
\newcommand{\medmhalo}{10^{10.9}}
\newcommand{\medlogsfr}{-1.73}
\newcommand{\medmHI}{10^{7.4}}

\newcommand{\impactpara}{b}

\newcommand{\beq}	{\begin{equation}}
\newcommand{\eeq}	{\end{equation}}

\def\fv{\ifmmode {f_{\rm V}} \else $f_{\rm V}$\fi}
\def\f0{\ifmmode {f_{\rm V,0}} \else $f_{\rm V,0}$\fi}
\def\nh{\ifmmode {n_{\rm H}} \else $n_{\rm H}$\fi}
\def\n0{\ifmmode {n_{\rm H,0}} \else $n_{\rm H,0}$\fi}
\def\fhi{\ifmmode {f_{\rm HI}} \else $f_{\rm HI}$\fi}
\def\Teq{\ifmmode {T_{\rm eq}} \else $T_{\rm eq}$\fi}
\def\nhi0{\ifmmode {N_{\rm HI,0}} \else $N_{\rm HI,0}$\fi}

\def\mp{\ifmmode {m_{\rm p}} \else $m_{\rm p}$\fi}
\def\cmv{\ifmmode {\rm cm^{-3}} \else ${\rm cm^{-3}}$\fi}
\def\cmc{\ifmmode {\rm cm^{-2}} \else ${\rm cm^{-2}}$\fi}
\def\rcgm{\ifmmode {r_{\rm CGM}} \else $r_{\rm CGM}$\fi}
\def\mcgm{\ifmmode {M_{\rm CGM}} \else $M_{\rm CGM}$\fi}
\def\fcgm{\ifmmode {f_{\rm CGM}} \else $f_{\rm CGM}$\fi}
\def\mcgmcool{\ifmmode {M_{\rm CGM,cool}} \else $M_{\rm CGM,cool}$\fi}
\def\fcgmcool{\ifmmode {f_{\rm CGM,cool}} \else $f_{\rm CGM,cool}$\fi}

\newcommand{\blue}{\color{blue}}

\shorttitle{Metals in the CGM of Nearby Dwarf Galaxies}
\shortauthors{Zheng et~al.}

\begin{document}
\defcitealias{behroozi13}{B13}
\defcitealias{munshi21}{M21}
\defcitealias{zheng20b}{Z20}
\defcitealias{zheng19b}{Z19}
\defcitealias{bordoloi14}{B14}
\defcitealias{bordoloi18}{B18}
\defcitealias{liang14}{LC14}
\defcitealias{johnson17}{J17}
\defcitealias{qu22}{QB22}
\defcitealias{tchernyshyov22}{T22}
\defcitealias{Karachentsev13}{K13}

\begin{CJK*}{UTF8}{gbsn} 
\title{A Comprehensive Investigation of Metals in the Circumgalactic Medium of Nearby Dwarf Galaxies}

\author[0000-0003-4158-5116]{Yong Zheng (郑永)}
\affiliation{Department of Physics, Applied Physics and Astronomy, Rensselaer Polytechnic Institute, Troy, NY 12180, {\blue zhengy14@rpi.edu}}

\author[0000-0003-3520-6503]{Yakov Faerman}
\affiliation{Department of Astronomy, University of Washington, Seattle, WA 98195, USA}

\author[0000-0002-3391-2116]{Benjamin D. Oppenheimer}
\affiliation{CASA, Department of Astrophysical and Planetary Sciences, University of Colorado, Boulder, CO 80309, USA}

\author[0000-0002-1129-1873]{Mary E. Putman}
\affiliation{Department of Astronomy, Columbia University, New York, NY 10027, USA}

\author[0000-0001-5538-2614]{Kristen B. W. McQuinn}
\affiliation{Rutgers University, Department of Physics and Astronomy, Piscataway, NJ 08854, USA}

\author[0000-0001-6196-5162]{Evan N. Kirby}
\affiliation{Department of Physics and Astronomy, University of Notre Dame, Notre Dame, IN 46556, USA}

\author[0000-0002-1979-2197]{Joseph N. Burchett}
\affiliation{Department of Astronomy, New Mexico State University, Las Cruces, NM 88001, USA}

\author[0000-0003-4122-7749]{O. Grace Telford}
\altaffiliation{Carnegie-Princeton Fellow}
\affiliation{Rutgers University, Department of Physics and Astronomy, Piscataway, NJ 08854, USA}
\affiliation{Department of Astrophysical Sciences, Princeton University, 4 Ivy Lane, Princeton, NJ 08544, USA}
\affiliation{The Observatories of the Carnegie Institution for Science, 813 Santa Barbara Street, Pasadena, CA 91101, USA}

\author[0000-0002-0355-0134]{Jessica K. Werk}
\affiliation{Department of Astronomy, University of Washington, Seattle, WA 98195, USA}

\author[0000-0001-9654-5889]{Doyeon A. Kim}
\affiliation{Department of Astronomy, Columbia University, New York, NY 10027, USA}

\begin{abstract}

Dwarf galaxies are found to have lost most of their metals via feedback processes; however, there still lacks consistent assessment on the retention rate of metals in their circumgalactic medium (CGM). Here we investigate the metal content in the CGM of 45 isolated dwarf galaxies with $M_*=10^{6.5-9.5}~M_\odot$ ($M_{\rm 200m}=10^{10.0-11.5}~M_\odot$) using {\it HST}/COS. While H{\sc\ i} (Ly$\alpha$) is ubiquitously detected ($89\%$) within the CGM, we find low detection rates ($\approx5\%-22\%$) in C{\sc\ ii}, C{\sc\ iv}, Si{\sc\ ii}, Si{\sc\ iii}, and Si{\sc\ iv}, largely consistent with literature values. Assuming these ions form in the cool ($T\approx10^4$ K) CGM with photoionization equilibrium, the observed H{\sc\ i} and metal column density profiles can be best explained by an empirical model with low gas density and high volume filling factor. For a typical galaxy with $M_{\rm 200m}=10^{10.9}~M_\odot$ (median of the sample), our model predicts a cool gas mass of $M_{\rm CGM,cool}\sim10^{8.4}~M_\odot$, corresponding to $\sim2\%$ of the galaxy's baryonic budget. Assuming a metallicity of $0.3Z_\odot$, we estimate that the dwarf galaxy's cool CGM likely harbors $\sim10\%$ of the metals ever produced, with the rest either in more ionized states in the CGM or transported to the intergalactic medium. We further examine the EAGLE simulation and show that H{\sc\ i} and low ions may arise from a dense cool medium, while C{\sc\ iv} arises from a diffuse warmer medium. Our work provides the community with a uniform dataset on dwarf galaxies' CGM that combines our recent observations, additional archival data and literature compilation, which can be used to test various theoretical models of dwarf galaxies.

\end{abstract}


\section{Introduction}
\label{sec:intro}

The circumgalactic medium (CGM) is a large gaseous envelope surrounding a galaxy. It contains the imprints of outflows from feedback processes within a galaxy's disk, and is an important potential source of new star formation fuel for the galaxy. Numerous observations of the halos of Milky Way (MW) mass galaxies have found that the CGM contains a significant amount of baryons and metals \citep{putman12, tumlinson17, peroux20}. On the other hand, studies on the CGM of dwarf galaxies (with stellar masses of $\mstar\lesssim10^{9.5}~\msun$) have so far been limited to a few works \citep[e.g.,][see below for more details]{bordoloi14, burchett16, johnson17, zheng20b, qu22}. A systematic investigation remains to be conducted.

Dwarf galaxies have relatively shallow potential wells, and indeed the stellar mass--stellar metallicity relation for galaxies shows that low-mass galaxies do not retain metals as well as their higher-mass counterparts \citep[e.g.,][]{gallazzi05, kirby11, kirby13}. The gas phase abundance in relation to stellar mass also shows this trend of decreasing metal retention in the interstellar medium (ISM) with decreasing mass \citep[e.g.,][]{tremonti04, lee06, andrews13, mcquinn15b}. These results suggest that most of the metals produced throughout a dwarf galaxy's star formation history now reside beyond the central regions of the galaxy: either in the galaxy's CGM or in the intergalactic medium (IGM).

Indeed, hydrodynamic simulations of dwarf galaxies have demonstrated the efficiency of stellar feedback in redistributing baryons and metals into the CGM and IGM \citep[e.g.,][]{shen14, muratov17, christensen16, christensen18, wheeler19, agertz20, rey20, mina21, andersson22}. For example, \cite{christensen18} found that only $\sim$20\%--40\% of the metals (by mass) are retained in the ISM of dwarf galaxies with $\mstar\lesssim$10$^9~\msun$, while $\sim$10\%--55\% resides in the CGM. Meanwhile, recent simulations of eight dwarf galaxies by \cite{mina21} suggest that the metals in the simulated CGM are likely to be too diffuse to be easily detected. For a range of ions (\HI, \SiII, \CIV, \OVI), they find low column densities of CGM gas as a function of impact parameter, with values typically lower than the published dwarf galaxy CGM literature (see below).   

\begin{table*}
\small
\centering
\label{tb:literature}
\caption{Compilation of Data Sample and Literature References}
\begin{tabular}{rlllllc}
\hline
References & $z_{\rm gal}$ & $\mstar (\msun)$ & $\impactpara/\rvir$ & Ions & Selection Criteria & Pairs Adopted\\ 
(1) & (2) &(3) & (4)& (5)& (6) & (7) \\
\hline 
\hline 
New  & \multirow{3}{*}{$\sim$0} & \multirow{3}{*}{$\samplemstarrange$} & \multirow{3}{*}{0.08--1.0} & \CII, \CIV, & No galaxies at $|\delta d|<$100 kpc and &  \multirow{3}{*}{\npairthiswork} \\
Observations or& & &  & \SiII, \SiIII &  $|\delta v|<$150 $\kms$, and no $L\gtrsim0.1L_*$ & \\
Archival Data & & &  & \SiIV & galaxies within 300 kpc and 300$\kms$ & \\
\hline 
\citetalias{qu22} (NGC3109, & \multirow{2}{*}{$\sim$0}  & \multirow{2}{*}{$10^{7.4-8.3}$} & \multirow{2}{*}{0.2--0.7} & \CII, \CIV, \OI, & $>$1.3Mpc and $>$300$\kms$ from MW & \multirow{2}{*}{\npairqu\ (6)}\\ 
Sextans A \& B) & & & & \SiII, \SiIII, \SiIV & $\sim$2Mpc and $>$600$\kms$ from M31 & \\
\hline 
\multirow{2}{*}{\citetalias{zheng20b} (IC1613)} &  \multirow{2}{*}{$\sim$0}  & \multirow{2}{*}{$10^{8}$} & \multirow{2}{*}{0.05--0.5} &\CII, \CIV & On the outskirts of LG, no known & \multirow{2}{*}{\npairzhengic\ (6)}\\  
& & & & \SiII, \SiIII, \SiIV & galaxies within 400 kpc & \\
\hline 
\multirow{2}{*}{\citetalias{johnson17}} &  \multirow{2}{*}{0.09--0.3}  & \multirow{2}{*}{$10^{7.7-9.2}$} & \multirow{2}{*}{0.1--1.7} & \HI, \CIV, \OVI,  & No $L\geqslant0.1L_*$ galaxies within & \multirow{2}{*}{\npairjohnson\ (18)}\\  
& & & & \SiII, \SiIII, \SiIV  & $|\delta d|$=500 kpc and $|\delta v|$=300 $\kms$ & \\
\hline 
\citetalias{bordoloi14}+\citetalias{bordoloi18} &  $\leqslant$0.1 & $10^{8-9.9}$ & 0.06--1.1 & \HI, \CIV & No known galaxies within 300 kpc & {\npairbordoloi\ (43)}\\
\hline 
\multirow{2}{*}{\citetalias{liang14}} & \multirow{2}{*}{$\leqslant$0.176}  & \multirow{2}{*}{$10^{5.2-11.1}$} & \multirow{2}{*}{0.2--6.0} & \HI, \CII, \CIV,  & No known galaxies within & \multirow{2}{*}{\npairliang\ (195)}\\  
& & & & \SiII, \SiIII, \SiIV & $|\delta d|$=500 kpc and $|\delta v|$=500 $\kms$ & \\
\hline 
\hline 
\textbf{Full Sample} & \multirow{2}{*}{0.0--0.3}& \multirow{2}{*}{$10^{6.5-9.5}$}& \multirow{2}{*}{0.05--1.0}& \HI, \CII, \CIV & All the above combined, with & \multirow{2}{*}{\npairtotal} \\
 (this work) & & & & \SiII, \SiIII, \SiIV & $\mstar \leqslant 10^{9.5} \msun$, S/N$\geqslant$8, and $\impactpara\leqslant\rvir$ & \\
\hline
\end{tabular}
\tablecomments{
\small
Col. (1): References: \citetalias{qu22} for \cite{qu22}, \citetalias{zheng20b} for \cite{zheng20b}, \citetalias{johnson17} for \cite{johnson17},  \citetalias{bordoloi14} for \cite{bordoloi14} (\CIV), 
\citetalias{bordoloi18} for \cite{bordoloi18} (\HI), 
and \citetalias{liang14} for \cite{liang14}. The last row summarizes the Full Sample we use in this work, which consists of pairs from recent observations, archival data, and those adopted from the literature.  
Col. (2): Galaxy redshift. 
Col. (3): Galaxy stellar mass range; when applicable, we have corrected the corresponding values to \cite{kroupa01} IMF (see \S\ref{sec:data_literature_pairs}, \S\ref{sec:galaxy_property}). 
Col. (4): Impact parameters probed by QSO sight lines. 
Col. (5): List of ions included in each reference. 
Col. (6): Selection criteria of nearly isolated dwarf galaxies.  
Col. (7): Dwarf-QSO paris adopted in this work that meet the following criteria: (i) $\mstar \leqslant 10^{9.5} ~\msun$, (ii) S/N$\geqslant$ 8, and (iii) $\impactpara\leqslant\rvir$. Numbers in parentheses indicate the total dwarf-QSO pairs included in each reference. See Figure \ref{fig:logM_rho_all} and \S\ref{sec:data} for more details.}
\end{table*}

Recent years have seen emerging efforts to observationally search for metals in the CGM of dwarf galaxies at $z\lesssim0.3$ (see references in Table \ref{tb:literature}). The COS-Dwarfs program studies \CIV\ and \HI\ absorption in the CGM of 43 galaxies with $\mstar=$10$^{8-9.9}~\msun$ at $z\leqslant0.1$ \citep[][hereafter \citetalias{bordoloi14}+\citetalias{bordoloi18}]{bordoloi14, bordoloi18}. They detect \CIV\ out to $\sim0.5$ virial radius at a sensitivity limit of 50--100 m\AA\ as measured in \CIV\ 1548\AA\ equivalent width (EW). A power-law fit to the observed data shows that \CIV's EW drops quickly as a function of impact parameter ($\impactpara$). 

In a sample of 195 galaxy-QSO pairs at $z<0.176$, \citet[][hereafter \citetalias{liang14}]{liang14} found low detection rates in \CIV\ as well as in other ions (\SiII, \SiIII, \CII, and \CIV) while reporting ubiquitous \HI\ detections in Ly$\alpha$ 1215\AA\ (see also \citealt{wilde21} for an extensive study on CGM \HI\ absorbers over $\mstar\sim10^{7-11}~\msun$). However, note that the majority of \citetalias{liang14}'s galaxy-QSO pairs are not focused on dwarf galaxies, and most pairs are probed with spectra at low signal-to-noise ratio (S/N).

\citet[][hereafter \citetalias{johnson17}]{johnson17} examined 18 star-forming field dwarf galaxies with $\mstar\approx$10$^{7.9-9.2}~\msun$ and studied absorption in \HI, \SiII, \SiIII, \SiIV, \CIV, and \OVI. Their work echoes \citetalias{bordoloi14}+\citetalias{bordoloi18}'s and \citetalias{liang14}'s results that the detection rates of \SiII, \SiIII, \SiIV\ and \CIV\ are very low and drop with increasing $\impactpara$. However, they report a 50\% detection of \OVI\ in these field dwarf galaxies within the virial radii, suggesting that the dwarf galaxies' CGM may be dominated by gas with high-ionization states. Similarly, \citet{tchernyshyov22} also found a high detection rate of \OVI\ in their sample of over $100$ dwarf galaxies and the \OVI\ column densities increase with host galaxies' stellar masses.

In the Local Group, multiple attempts to find metals in dwarf galaxies' CGM have yielded mixed results \citep{richter17, zheng19b, zheng20b, qu22}. \cite{richter17} found no detections of metals in 19 nearby dwarfs, most likely due to the fact that the galaxies probed are mainly spheroidal type and thus contain little gas, and the sight lines are at large impact parameters ($\impactpara\gtrsim0.5\rvir$). \cite{zheng20b} (hereafter \citetalias{zheng20b}) observed six QSOs at 0.05--0.5 virial radii from the dwarf galaxy IC1613 and found significant detections toward most sight lines (see also a tentative detection in WLM in \citealt{zheng19b}). Recently, \citet[][hereafter \citetalias{qu22}]{qu22} examined the CGM of three dwarf galaxies (Sextans A, Sextans B, and NGC 3109) in loose associations, but only detect one \CIV\ absorber toward Sextans A at $\impactpara=21$ kpc (0.2 virial radius). \citetalias{qu22} explored analytical CGM models as established in \cite{Qu18a, Qu18b}, and found that a multitemperature CGM model with photoionization, cooling, and feedback can best explain the nondetections of \CIV. 

\begin{figure*}
	\centering
	\includegraphics[width=0.95\textwidth]{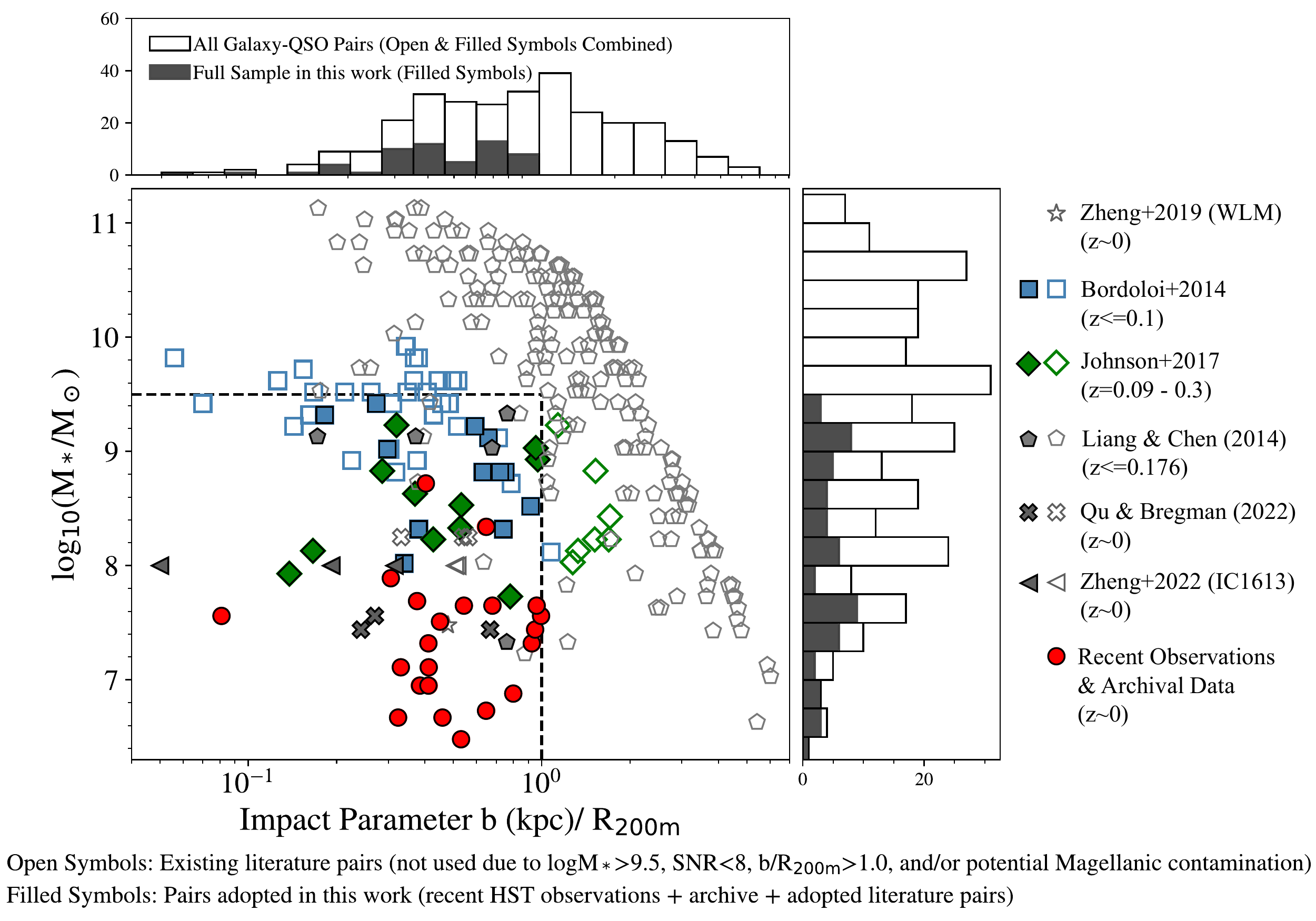}
	\caption{Parameter space ($\log\mstar$ vs. $\impactpara/\rvir$) probed by this work and existing literature. The space enclosed within the vertical and horizontal dash lines ($\impactpara/\rvir=0.05-1.0$ and $\logmstar\leq9.5$) indicates the unique parameter space explored in this work. The histograms on the top and the right show the distribution of $\impactpara/\rvir$ and $\log\mstar$, respectively. Throughout this work, the same symbols are used to consistently represent data from different references: square for \citetalias{bordoloi14}, diamond for \citetalias{johnson17}, pentagon for \citetalias{liang14}, thick X for \citetalias{qu22}, star for \citet{zheng19b}, left triangle for \citetalias{zheng20b}, and circle for new pairs added from recent observations or archival data. For illustration purposes, symbol colors may vary from figure to figure. }
	\label{fig:logM_rho_all}
\end{figure*}

The mixed results discussed above present an ambiguous picture of whether dwarf galaxies retain a significant amount of metals in their CGM. Furthermore, as shown in Figure \ref{fig:logM_rho_all} and Table \ref{tb:literature}, it is not straightforward to directly compare various studies due to the different mass ranges, impact parameters, and QSO spectral quality used in the existing literature, let alone different methods to compute galaxy and absorber properties (e.g., stellar mass, ion column density, virial radius).

To mitigate these issues, in this work we conduct a comprehensive analysis of the cool metal content in the CGM of dwarf galaxies with $\mstar=\samplemstarrange~\msun$ using data from our recent {\it HST}/COS observations, additional archival data, and a thorough compilation of relevant literature values from \citetalias{bordoloi14}+\citetalias{bordoloi18}, \citetalias{liang14}, \citetalias{johnson17}, \citetalias{zheng20b}, and \citetalias{qu22} (see Table \ref{tb:literature}) with consistent quality control. The choice of imposing a mass threshold at $\mstar=10^{9.5}~\msun$ is to include massive dwarfs similar to the Large Magellanic Cloud ($\mstar=$10$^{9.2}~\msun$; \citealt{mcconnachie12}) while excluding higher-mass galaxies such as the dwarf spiral M33 ($\mstar=$10$^{9.5-9.8}~\msun$; \citealt{corbelli03}). We note that our final sample does not include either the LMC or M33 because they are not sufficiently isolated (see \S\ref{sec:data_new_pairs}). 

In this work, we define a galaxy's virial radius, $\rvir$, as the radius within which the average density is 200 times the mean matter density of the Universe at $z=0$, following the definition used by the COS-Dwarfs survey (\citetalias{bordoloi14}). 
We adopt $\rm \Omega_m=0.308$, $\rm \Omega_b=0.0487$, and $H_0=67.8$ $\kms~{\rm Mpc}^{-1}$ \citep{planck15}, and assume a \cite{kroupa01} initial mass function (IMF) for relevant quantities, unless otherwise specified. This paper is organized as follows. \S\ref{sec:data} describes the {\it HST}/COS data, relevant spectral analyses, and dwarf sample selection. \S\ref{sec:result} shows the results of ion radial column density profiles. Then, in \S\ref{sec:theory} and \S\ref{sec:sims_eagle}, we examine the CGM properties of dwarf galaxies from theoretical perspectives. Lastly, in \S\ref{sec:discuss_mass}, we compare our CGM ion mass estimates with previous observational values reported in the literature. We conclude in \S\ref{sec:summary}.

\section{Data Sample \& Measurements}
\label{sec:data}

In the following, we refer to our data sample as the Full Sample (see Table \ref{tb:literature}), which comprises \npairtotal\ dwarf-QSO pairs that include: (i) \npairthiswork\ new pairs either from our recent {\it HST} programs (\#\href{https://mast.stsci.edu/search/ui/#/hst/results?proposal_id=15156}{HST-GO-15156}, PI Zheng; \#\href{https://mast.stsci.edu/search/ui/#/hst/results?proposal_id=15227}{HST-GO-15227}, PI Burchett; \#\href{https://mast.stsci.edu/search/ui/#/hst/results?proposal_id=16301}{HST-GO-16301}, PI Putman) or a thorough search of the Barbara A. Mikulski Archive for Space Telescopes (MAST) for available QSOs (\S\ref{sec:data_new_pairs}), and (ii) \npairliterature\ additional pairs compiled from existing literature (\S\ref{sec:data_literature_pairs}). The Full Sample is highlighted as filled symbols in Figure \ref{fig:logM_rho_all} and tabulated in Table \ref{tb:pair_info}. Overall in this work we probe a unique parameter space of low stellar mass ($\mstar=\samplemstarrange$) and small impact parameter ($\impactpara/\rvir$=0.05--1.0).

\subsection{\npairthiswork\ Pairs from New Observations or HST Archive}
\label{sec:data_new_pairs}

In our recent programs (\#\href{https://mast.stsci.edu/search/ui/#/hst/results?proposal_id=15156}{15156}, \#\href{https://mast.stsci.edu/search/ui/#/hst/results?proposal_id=15227}{15227}, \#\href{https://mast.stsci.edu/search/ui/#/hst/results?proposal_id=16301}{16301}), 
we observed a total of 20 QSOs in the vicinity of 11 nearby dwarf galaxies using {\it HST}/COS G130M and G160M gratings. To supplement the sample, we conducted a thorough archival search in MAST for additional QSO sight lines that were publicly available as of 2022 March 31. We looked for QSO sight lines around isolated dwarf galaxies (see definition below) within 8 Mpc from the Sun as cataloged in \citet[][hereafter \citetalias{Karachentsev13}]{Karachentsev13}.

A dwarf galaxy is deemed ``isolated" if it does not have neighboring galaxies within a distance of $\delta d_{\rm neigh}$=100 kpc and its systemic velocity is more than $|\delta v|=150~\kms$ from other galaxies. Setting $\delta d_{\rm neigh}$=100 kpc is to ensure that the inner CGM of two dwarf galaxies do not overlap, given that the median $\rvir$ for dwarf galaxies in our sample is $\medrvirval$ kpc (see \S\ref{sec:galaxy_property}). We set a velocity threshold of $|\delta v|=150~\kms$ because it is nearly twice the escape velocity allowed by an $\mstar\sim$10$^8~\msun$ galaxy at 0.5$\rvir$ and mitigates contamination from other dwarf galaxy halos in velocity space. Note that a dwarf galaxy meeting these criteria may still reside in a loose association, as is the case for Sextans A and B (see \citetalias{qu22}). 
Additionally, we exclude dwarf galaxies that are in the halos of more-massive ($L\gtrsim0.1L_*$) hosts within $\pm300~\kms$ and 300 kpc.
We also do not consider dwarf galaxies that are likely to be satellites of either the MW or M31, meaning that the galaxies should be farther from the MW or M31 than the virial radius of the corresponding galaxy. Lastly, we only consider dwarf galaxies with \HI\ 21cm detection to ensure that the galaxies may still have an intact CGM. Overall, the above selection criteria ensure minimal ambiguity of absorber origins in both position and velocity space when there is detection near a dwarf galaxy.


Our initial search following the criteria above results in 244 potential UV sight lines near 52 low-mass isolated galaxies. We further limit the data sample to QSOs with S/N$\geqslant8$ per resolution element (see below for definition of S/N). The choice of S/N$\geqslant8$ is to ensure sufficient data remained after the cut, while allowing for a consistent sensitivity floor among all data points included in this work. We also implement the same S/N cut in our literature compilation (see \S\ref{sec:data_literature_pairs}). We emphasize that the consistency in sensitivity allows us to use censored data (e.g., nondetections with upper limits) to assess the metal contents in the dwarf galaxy halos with minimal bias (see \S\ref{sec:result_pymc}).

\begin{figure*}
	\centering
	\includegraphics[width=\linewidth]{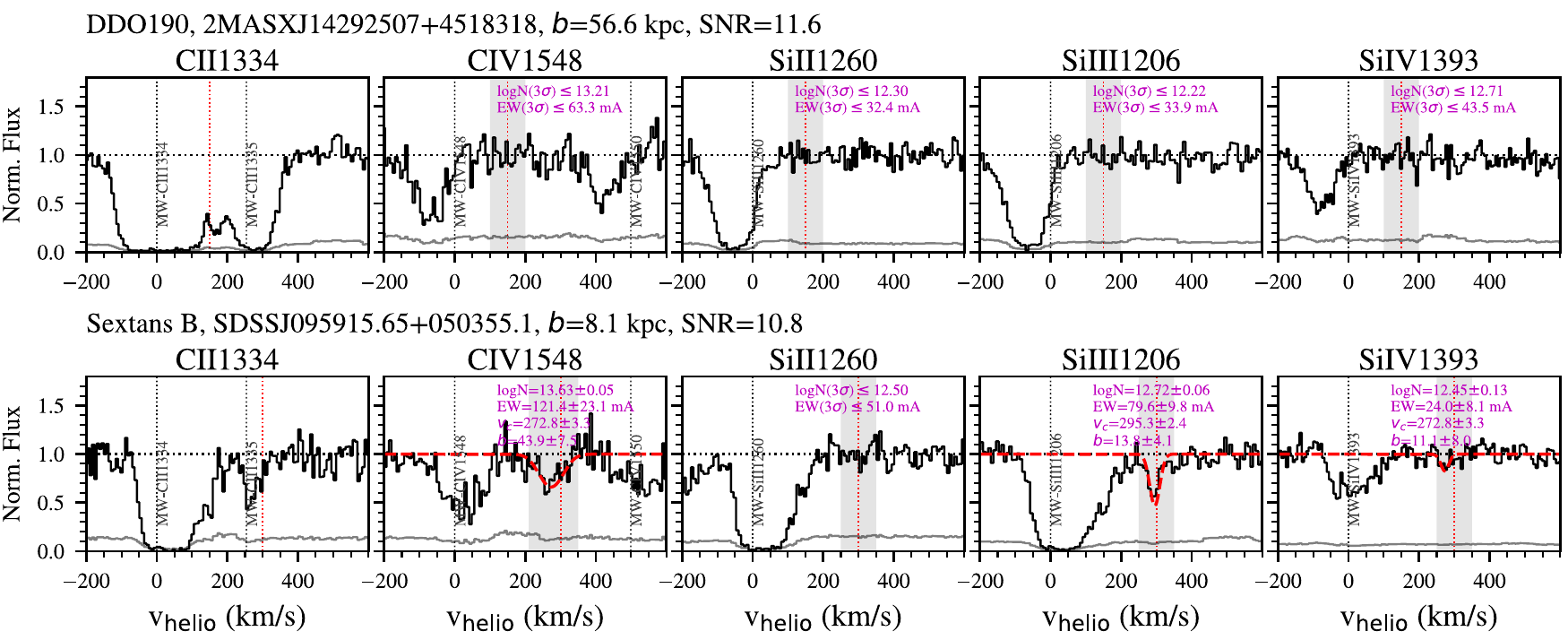}
	\caption{Top: Example ion spectra with nondetections for DDO190. nondetections represent the majority of the sight lines in our data sample. We estimate the equivalent width (EW) and AOD column density over $\pm 50~\kms$ (gray shades) from the systemic velocity of the host dwarf (vertical red lines) and indicate the $3\sigma$ upper limits. We do not measure the absorption in \CII\ 1334 because it is most likely a contamination given that there is no corresponding detection in \SiII\ which is at similar ionization state. Bottom: Detections of metal absorbers in Sextans B. Voigt profile fits are included where there are significant ($\geqslant3\sigma$) detections. We do not use \CII\ 1334 because it is blended with a CII* 1335 line from the MW. }
	\label{fig:dw_det_spec}
\end{figure*}

We follow the same procedures as used in \cite{zheng19b} and \citetalias{zheng20b} to process the {\it HST}/COS QSO spectra, including co-addition, continuum normalization, and upper-limit estimates of ion $\log N$ and EW values based on the apparent optical depth (AOD) method \citep{savage96} for nondetections and Voigt profile fitting for detections. Below we briefly summarize the key aspects of the data reduction, but refer the reader to \cite{zheng19b} and \citetalias{zheng20b} for more details. 

\textsl{Co-addition and S/N}: For archival sight lines included in the Hubble Spectroscopic Legacy Archive (HSLA; \citealt{Peeples17}), we use HSLA's coadded spectra. For new observations or archival spectra with additional epochs of observations not included in the HSLA, we download and coadd the spectra from all epochs using an IDL package \texttt{coadd\_x1d.pro} \citep{danforth10}. \citetalias{zheng20b} conduct a detailed comparison between HSLA and \texttt{coadd\_x1d.pro} and show that these two methods yield consistent coadded spectra. After co-addition, the spectra are binned by 3 pixels to increase the S/N while remaining Nyquist-sampled with 2 pixels per resolution element \citep[COS Data Handbook;][]{cos_data}. We adopt HSLA's definition of S/N, which is estimated by first calculating local S/Ns averaged over 10 \AA\ windows in absorption-line-free regions every 100 \AA\ from 1150 to 1750 \AA. Then the median S/N of the local S/Ns is adopted as the S/N of the whole spectrum.

\textsl{Continuum Normalization, AOD Measurements and Voigt profile fits}: We conduct continuum normalization for the coadded spectra using an open-source Python package \texttt{Linetools} \citep{prochaska16}, and focus on a set of lines typically observed in galaxies' CGM, including \SiII\ 1190/1193/1260/1526 \AA, \SiIII\ 1206 \AA, \SiIV\ 1393/1402 \AA, \CII\ 1334 \AA, and \CIV\ 1548/1550 \AA. Unlike dwarf galaxies at slightly higher redshift (e.g., $z\sim0.1-0.3$; see \S\ref{sec:data_literature_pairs}), \HI\ Ly$\alpha$ absorption is unavailable at $z\sim0$ due to strong contamination from the MW ISM. So in this work the \HI\ data points are from literature references with \HI\ Ly$\alpha$ measurements (see Table \ref{tb:literature}). For metal ions, we look for absorption features in QSO spectra within $\pm50~\kms$ of the systemic velocity of the host galaxy. For nondetections, such as QSO 2MASXJ14292507+4518318 near the dwarf galaxy DDO 190 (top panel in Figure \ref{fig:dw_det_spec}), we calculate the $3\sigma$ upper limit in column density $\log N$ and EW for the corresponding ions using the AOD method.

\textsl{Potential Foreground Contamination}: For each dwarf-QSO pair, whether with absorber detections or upper limits, we examine the pair's surrounding environment to look for potential contamination from interloping absorbers at higher $z$, from nearby galaxies that may pass the isolation criteria outlined above, and from foreground \HI\ clouds \citep{putman12}. For galaxies with systemic velocities at $|\vlsr|<470~\kms$ that are within the coverage of the HI4PI survey \citep{hi4pi16}, we generate velocity channel maps spanning $\pm50~\kms$ from the galaxies' systemic velocities in steps of 10 $\kms$ and extract \HI\ spectra at the locations of the QSOs to look for extended \HI\ foreground emission. We do not find any significant foreground contamination at 3$\sigma$ ($\sigma=53$mK; \citealt{hi4pi16}). For the three dwarf galaxies with $\vlsr>470~\kms$ (i.e. PGC039646, UGC07485, NGC5408; see Table \ref{tb:dwarf_info}), we do not find significant \HI\ halo cloud contamination within $\pm100~\kms$ of the galaxies' systemic velocities based on the HIPASS survey \citep{hipass_data}. 


In total, we find \npairthiswork\ new dwarf-QSO pairs with robust detections or upper limits from our recent {\it HST/COS} observations and archival data. These pairs are shown in red filled circles in Figure \ref{fig:logM_rho_all} and listed in Table \ref{tb:pair_info} with Reference of ``New/Arx.". In Table \ref{tb:logN}, we tabulate the results of $\log N$ measurements, where nondetections are quoted in $3\sigma$ upper limits. All but one QSO sight line are nondetections; the only absorber detections are found toward QSO SDSSJ095915.65+050355.1 that are confirmed to reside in the CGM of Sextans B at $\impactpara=8.1$ kpc, the spectra of which are shown in the bottom panel of Figure \ref{fig:dw_det_spec}. In this case, we perform Voigt profile fitting using the \texttt{ALIS} package\footnote{https://github.com/rcooke-ast/ALIS}. Note that we run Voigt profile fits for all transition lines available, while in Figure \ref{fig:dw_det_spec} only the strongest lines are shown. All of the {\it HST}/COS spectra used in this section (i.e., our new observations or additional archival search) can be found in MAST: \dataset[10.17909/ve0k-ps78]{http://dx.doi.org/10.17909/ve0k-ps78}. We do not include the EW values in Table \ref{tb:logN} as they are not used in relevant analyses; but the data can be found on our github repository (\href{https://github.com/yzhenggit/zheng_dwarfcgm_survey.git}{yzhenggit/zheng\_dwarfcgm\_survey}) for interested readers. In addition to the \npairthiswork\ new pairs, we find a few other absorber detections that are unlikely to be associated with the CGM of our dwarf galaxies based on close inspection of their surrounding environments. We briefly discuss these absorbers and their potential associations in Appendix \ref{sec:app_absorber}, but do not include them in our following analysis. 

\startlongtable
\begin{deluxetable*}{cccccccccc}
\tabletypesize{\footnotesize}
\label{tb:pair_info}
\tablecaption{Properties of Dwarf-QSO Pairs}
\tablehead{
\colhead{PID} & \colhead{Galaxy} & \colhead{$\log\mstar$} & \colhead{$M_{\rm *,ref}$} & \colhead{$\log\mhalom$} & \colhead{QSO}& \colhead{S/N} & \colhead{$b$} & \colhead{$\rvir$} & Reference \\ 
& & ($\log\msun$) & & ($\log\msun$) & & & (kpc) & (kpc) & \\ 
\colhead{(1)} & \colhead{(2)} & \colhead{(3)} & \colhead{(4)} & \colhead{(5)}  & \colhead{(6)} & \colhead{(7)} & \colhead{(8)} & \colhead{(9)} & \colhead{(10)} 
 }
\startdata 
01 &               KKH086 & 6.48 & Dale09 & 10.01 & SDSS-J135726.27+043541.4 & 18.9 & 36.0 & 67.8 & New/Arx. \\
02 &                  GR8 & 6.67 & Dale09 & 10.08 &          PGC-1440438 & 10.9 & 23.2 & 71.6 & New/Arx. \\
03 &                  GR8 & 6.67 & Dale09 & 10.08 & SDSSJ130223.12+140609.0 & 9.4 & 32.9 & 71.6 & New/Arx. \\
04 &               DDO187 & 6.73 & Dale09 & 10.10 & SDSS-J141038.39+230447.1 & 10.3 & 47.0 & 72.7 & New/Arx. \\
05 &              UGCA292 & 6.88 & Dale09 & 10.16 & 2MASS-J12421031+3214268 & 8.2 & 60.9 & 76.1 & New/Arx. \\
06 &             UGC08833 & 6.95 & Dale09 & 10.20 & SDSSJ135341.03+361948.0 & 18.2 & 30.2 & 78.4 & New/Arx. \\
07 &             UGC08833 & 6.95 & Dale09 & 10.20 &              CSO1022 & 11.0 & 32.3 & 78.5 & New/Arx. \\
08 &            PGC039646 & 7.11 & K13 & 10.28 &        MS1217.0+0700 & 11.3 & 34.4 & 83.4 & New/Arx. \\
09 &            PGC039646 & 7.11 & K13 & 10.28 &           PG1216+069 & 22.5 & 27.6 & 83.4 & New/Arx. \\
10 &               DDO181 & 7.32 & Dale09 & 10.39 &          PG-1338+416 & 16.6 & 37.3 & 90.8 & New/Arx. \\
11 &               DDO099 & 7.32 & Dale09 & 10.39 & SDSSJ114646.00+371511.0 & 9.3 & 84.0 & 90.8 & New/Arx. \\
12 &            Sextans A & 7.44 & Dale09 & 10.46 &       RXSJ09565-0452 & 17.4 & 90.9 & 95.7 & New/Arx. \\
13 &             UGC06541 & 7.51 & Dale09 & 10.49 &              MRK1447 & 11.7 & 44.0 & 97.8 & New/Arx. \\
14 &            Sextans B & 7.56 & Dale09 & 10.52 & SDSS-J100535.24+013445.7 & 10.5 & 99.8 & 100.3 & New/Arx. \\
15 &            Sextans B & 7.56 & Dale09 & 10.52 & SDSSJ095915.65+050355.1 & 10.8 & 8.1 & 100.2 & New/Arx. \\
16 &               DDO190 & 7.65 & Dale09 & 10.57 & 2MASXJ14292507+4518318 & 11.6 & 56.6 & 104.2 & New/Arx. \\
17 &               DDO190 & 7.65 & Dale09 & 10.57 &       QSO-B1411+4414 & 30.0 & 100.2 & 104.3 & New/Arx. \\
18 &               DDO190 & 7.65 & Dale09 & 10.57 &           PG1415+451 & 12.9 & 70.8 & 104.2 & New/Arx. \\
19 &             UGC08638 & 7.69 & Dale09 & 10.59 & SDSS-J133833.06+251640.6 & 10.8 & 39.8 & 105.9 & New/Arx. \\
20 &             UGC07485 & 7.89 & K13 & 10.69 &           PG1222+216 & 19.1 & 35.0 & 114.3 & New/Arx. \\
21 &              NGC5408 & 8.34 & K13 & 10.93 &           PKS1355-41 & 17.5 & 89.0 & 137.4 & New/Arx. \\
22 &              NGC4144 & 8.72 & Dale09 & 11.13 &          PG-1206+459 & 20.0 & 64.5 & 160.2 & New/Arx. \\
23 &            Sextans A & 7.44 & Dale09 & 10.45 &            MARK 1253 & 8.6 & 63.4 & 95.0 & QB22 \\
24 &            Sextans A & 7.44 & Dale09 & 10.45 &          PG 1011-040 & 29.7 & 23.0 & 95.0 & QB22 \\
25 &            Sextans B & 7.56 & Dale09 & 10.52 &          PG 1001+054 & 15.6 & 27.1 & 100.3 & QB22 \\
26 &               IC1613 & 8.00 & M12 & 10.75 & 2MASX J01022632-0039045 & 8.0 & 37.7 & 119.5 & Z20 \\
27 &               IC1613 & 8.00 & M12 & 10.75 &       LBQS-0101+0009 & 7.5 & 22.9 & 119.5 & Z20 \\
28 &               IC1613 & 8.00 & M12 & 10.75 &       LBQS-0100+0205 & 7.6 & 6.0 & 119.5 & Z20 \\
29 &                   D9 & 7.73 & J17 & 10.61 &          PG-1522+101 & 12.8 & 84.0 & 107.5 & J17 \\
30 &                   D1 & 7.93 & J17 & 10.71 &          PKS0637-752 & 24.8 & 16.0 & 116.1 & J17 \\
31 &                   D2 & 8.13 & J17 & 10.82 &          PKS0637-752 & 24.8 & 21.0 & 126.3 & J17 \\
32 &                   D4 & 8.23 & J17 & 10.87 &           PG1001+291 & 20.3 & 56.0 & 131.2 & J17 \\
33 &                   D7 & 8.33 & J17 & 10.92 &          PKS0405-123 & 45.3 & 72.0 & 136.4 & J17 \\
34 &                   D8 & 8.53 & J17 & 11.03 &            Q1545+210 & 10.2 & 79.0 & 148.4 & J17 \\
35 &                   D5 & 8.63 & J17 & 11.08 &          PKS0637-752 & 24.8 & 57.0 & 154.2 & J17 \\
36 &                   D3 & 8.83 & J17 & 11.19 &        HB89-0232-042 & 11.8 & 48.0 & 167.8 & J17 \\
37 &                  D12 & 8.93 & J17 & 11.24 &          PG-1522+101 & 12.8 & 169.0 & 174.3 & J17 \\
38 &                  D13 & 9.03 & J17 & 11.29 &       LBQS-1435-0134 & 12.0 & 173.0 & 181.1 & J17 \\
39 &                   D6 & 9.23 & J17 & 11.40 &       LBQS-1435-0134 & 12.0 & 63.0 & 197.1 & J17 \\
40 &             316\_200 & 8.02 & B14 & 10.76 & SDSSJ105945.23+144142.9 & 9.1 & 41.0 & 120.6 & B14+B18 \\
41 &             124\_197 & 8.32 & B14 & 10.92 & SDSSJ031027.82-004950.7 & 9.2 & 101.0 & 136.4 & B14+B18 \\
42 &             172\_157 & 8.32 & B14 & 10.92 & SDSSJ092909.79+464424.0 & 16.0 & 52.0 & 136.4 & B14+B18 \\
43 &              87\_608 & 8.52 & B14 & 11.02 & SDSSJ100102.55+594414.3 & 10.8 & 135.0 & 147.2 & B14+B18 \\
44 &             257\_269 & 8.82 & B14 & 11.18 & SDSSJ080908.13+461925.6 & 11.6 & 125.0 & 166.5 & B14+B18 \\
45 &             135\_580 & 8.82 & B14 & 11.18 & SDSSJ094733.21+100508.7 & 11.0 & 120.0 & 166.5 & B14+B18 \\
46 &             329\_403 & 8.82 & B14 & 11.18 & SDSSJ015530.02-085704.0 & 8.1 & 105.0 & 166.5 & B14+B18 \\
47 &             322\_238 & 9.02 & B14 & 11.29 & SDSSJ134231.22+382903.4 & 9.9 & 54.0 & 181.1 & B14+B18 \\
48 &              93\_248 & 9.12 & B14 & 11.34 & SDSSJ135712.61+170444.1 & 14.8 & 124.0 & 188.2 & B14+B18 \\
49 &             210\_241 & 9.22 & B14 & 11.39 & SDSSJ134206.56+050523.8 & 8.5 & 116.0 & 195.6 & B14+B18 \\
50 &               70\_57 & 9.32 & B14 & 11.44 & SDSSJ133053.27+311930.5 & 9.0 & 37.0 & 203.2 & B14+B18 \\
51 &              316\_78 & 9.42 & B14 & 11.50 &           PG1049-005 & 10.5 & 58.0 & 212.8 & B14+B18 \\
52 & SDSSJ122815.96+014944.1 & 7.33 & LC14 & 10.40 &                3C273 & 105.6 & 69.5 & 91.5 & LC14 \\
53 & SDSSJ112418.74+420323.1 & 9.03 & LC14 & 11.29 &           PG1121+422 & 18.2 & 122.6 & 181.1 & LC14 \\
54 & SDSSJ112644.33+590926.0 & 9.13 & LC14 & 11.34 &          SBS1122+594 & 10.5 & 32.4 & 188.2 & LC14 \\
55 & SDSSJ121413.94+140330.4 & 9.13 & LC14 & 11.34 &           PG1211+143 & 21.5 & 70.1 & 188.2 & LC14 \\
56 & SDSSJ000545.07+160853.3 & 9.33 & LC14 & 11.45 &           PG0003+158 & 24.2 & 156.2 & 204.8 & LC14 \\
\enddata
\tablenotetext{}{Note: Col (1): PID = ID for each dwarf-QSO pair. Col (2): Dwarf galaxy name. 
Cols (3) and (4): Dwarf galaxy stellar mass and the corresponding reference, with
Dale09=3.6$\micron$ flux from \cite{dale09} and converted to $\mstar$ with the adopted distance listed in Table \ref{tb:dwarf_info}; 
K13=Ks mag from \cite{Karachentsev13} and converted to $\mstar$ with adopted distances; M12=$\mstar$ mass from \cite{mcconnachie12}, rescaled with adopted distance.
Col (5): Dwarf galaxy halo mass based on the stellar mass-halo mass (SMHM) relation from \cite{munshi21}; see \S\ref{sec:galaxy_property}. 
Col (6): QSO Name. 
Col (7): QSO signal-to-noise ratio per resolution element. 
Col (8): QSO impact parameter. 
Col (9): Virial radius, defined as the radius within which the mean density is 200 times the matter matter density of the Universe at $z=0$. 
Col (10): Reference from which we adopt the corresponding dwarf-QSO pair that meets our selection criteria (see \S\ref{sec:data}).} 
\end{deluxetable*}

\startlongtable
\begin{deluxetable*}{ccccccccc}
\tabletypesize{\footnotesize}
\label{tb:logN}
\tablecaption{Column Density Measurements}
\tablehead{
\colhead{PID} & QSO & \colhead{logN(\HI)} & \colhead{logN(\CII)} & \colhead{logN(\CIV)} & \colhead{logN(\SiII)} & \colhead{logN(\SiIII)} & \colhead{logN(\SiIV)} & \colhead{Reference}\\
 &  & \colhead{(cm$^{-2}$)} & \colhead{(cm$^{-2}$)} & \colhead{(cm$^{-2}$)} & \colhead{(cm$^{-2}$)} & \colhead{(cm$^{-2}$)} &  \colhead{(cm$^{-2}$)} & \\
 \colhead{(1)}&  \colhead{(2)} & \colhead{(3)} & \colhead{(4)} & \colhead{(5)} & \colhead{(6)} & \colhead{(7)} & \colhead{(8)} & \colhead{(9)}  
 }
\startdata 
01 & SDSS-J135726.27+043541.4  & - & - & $\leq$13.09 & - & $\leq$12.72 & $\leq$12.67 & New/Arx.\\
02 & PGC-1440438  & - & - & $\leq$13.23 & $\leq$12.51 & $\leq$12.37 & $\leq$12.42 & New/Arx.\\
03 & SDSSJ130223.12+140609.0  & - & - & $\leq$13.03 & $\leq$12.50 & $\leq$12.42 & $\leq$12.92 & New/Arx.\\
04 & SDSS-J141038.39+230447.1  & - & - & $\leq$13.05 & $\leq$12.36 & - & $\leq$12.76 & New/Arx.\\
05 & 2MASS-J12421031+3214268  & - & $\leq$13.47 & - & $\leq$12.49 & $\leq$12.45 & $\leq$12.82 & New/Arx.\\
06 & SDSSJ135341.03+361948.0  & - & - & - & $\leq$12.30 & $\leq$12.20 & $\leq$12.24 & New/Arx.\\
07 & CSO1022  & - & - & - & $\leq$12.37 & $\leq$12.26 & $\leq$12.69 & New/Arx.\\
08 & MS1217.0+0700  & - & $\leq$13.22 & - & $\leq$12.30 & $\leq$12.31 & $\leq$12.74 & New/Arx.\\
09 & PG1216+069  & - & $\leq$13.13 & - & $\leq$12.08 & $\leq$12.09 & $\leq$12.43 & New/Arx.\\
10 & PG-1338+416  & - & - & - & $\leq$12.21 & - & - & New/Arx.\\
11 & SDSSJ114646.00+371511.0  & - & - & - & $\leq$12.44 & $\leq$12.38 & $\leq$12.79 & New/Arx.\\
12 & RXSJ09565-0452  & - & - & - & $\leq$12.38 & $\leq$12.17 & $\leq$12.48 & New/Arx.\\
13 & MRK1447  & - & - & $\leq$13.13 & $\leq$12.33 & $\leq$12.23 & $\leq$12.64 & New/Arx.\\
14 & SDSS-J100535.24+013445.7  & - & - & $\leq$13.02 & $\leq$12.42 & $\leq$12.39 & $\leq$12.65 & New/Arx.\\
15 & SDSSJ095915.65+050355.1  & - & - & 13.63$\pm$0.05 & $\leq$12.50 & 12.72$\pm$0.06 & 12.45$\pm$0.13 & New/Arx.\\
16 & 2MASXJ14292507+4518318  & - & - & $\leq$13.21 & $\leq$12.30 & $\leq$12.22 & $\leq$12.71 & New/Arx.\\
17 & QSO-B1411+4414  & - & $\leq$12.69 & $\leq$12.80 & $\leq$11.93 & $\leq$11.72 & $\leq$12.29 & New/Arx.\\
18 & PG1415+451  & - & $\leq$13.13 & $\leq$13.00 & $\leq$12.33 & $\leq$12.22 & $\leq$12.74 & New/Arx.\\
19 & SDSS-J133833.06+251640.6  & - & - & $\leq$13.21 & $\leq$12.25 & $\leq$12.35 & $\leq$12.82 & New/Arx.\\
20 & PG1222+216  & - & $\leq$13.13 & $\leq$12.78 & $\leq$12.20 & $\leq$12.16 & - & New/Arx.\\
21 & PKS1355-41  & - & $\leq$13.17 & - & $\leq$12.14 & $\leq$12.16 & $\leq$12.47 & New/Arx.\\
22 & PG-1206+459  & - & - & - & $\leq$12.10 & $\leq$11.95 & - & New/Arx.\\
23 & MARK 1253  & - & - & $\leq$13.20 & $\leq$12.10 & $\leq$12.40 & $\leq$12.80 & QB22\\
24 & PG 1011-040  & - & - & 13.04$\pm$0.08 & $\leq$11.90 & $\leq$11.90 & $\leq$12.30 & QB22\\
25 & PG 1001+054  & - & - & $\leq$13.10 & $\leq$12.20 & $\leq$12.10 & $\leq$12.60 & QB22\\
26 & 2MASX J01022632-0039045  & - & 14.52$\pm$0.04 & 13.78$\pm$0.05 & 13.53$\pm$0.03 & 13.38$\pm$0.06 & 13.02$\pm$0.07 & Z20\\
27 & LBQS-0101+0009  & - & 14.21$\pm$0.05 & 13.64$\pm$0.07 & 13.19$\pm$0.06 & 13.30$\pm$0.05 & $\leq$12.79 & Z20\\
28 & LBQS-0100+0205  & - & $\leq$13.74 & 13.57$\pm$0.09 & $\leq$13.03 & 12.96$\pm$0.06 & 13.00$\pm$0.07 & Z20\\
29 & PG-1522+101  & 13.04$\pm$0.07 & - & $\leq$12.87 & $\leq$12.26 & $\leq$11.86 & $\leq$12.84 & J17\\
30 & PKS0637-752  & 15.70$\pm$0.40 & - & 13.73$\pm$0.04 & $\leq$12.26 & 13.14$\pm$0.03 & - & J17\\
31 & PKS0637-752  & 15.06$\pm$0.02 & - & - & $\leq$11.96 & 12.48$\pm$0.05 & $\leq$12.53 & J17\\
32 & PG1001+291  & 14.10$\pm$0.01 & - & $\leq$13.18 & $\leq$11.96 & - & $\leq$12.71 & J17\\
33 & PKS0405-123  & 13.94$\pm$0.01 & - & $\leq$12.57 & $\leq$11.96 & - & $\leq$12.53 & J17\\
34 & Q1545+210  & $\leq$12.44 & - & $\leq$13.05 & $\leq$12.26 & $\leq$11.86 & - & J17\\
35 & PKS0637-752  & 14.32$\pm$0.03 & - & - & $\leq$11.96 & - & $\leq$12.53 & J17\\
36 & HB89-0232-042  & 13.88$\pm$0.01 & - & $\leq$13.53 & $\leq$12.26 & $\leq$12.16 & $\leq$13.23 & J17\\
37 & PG-1522+101  & 13.63$\pm$0.02 & - & $\leq$13.42 & $\leq$11.96 & - & $\leq$12.53 & J17\\
38 & LBQS-1435-0134  & 12.74$\pm$0.14 & - & $\leq$12.87 & $\leq$11.96 & $\leq$11.86 & $\leq$12.53 & J17\\
39 & LBQS-1435-0134  & 14.00$\pm$0.01 & - & $\leq$13.42 & $\leq$11.96 & $\leq$11.86 & $\leq$12.23 & J17\\
40 & SDSSJ105945.23+144142.9  & $>$14.23$\pm$0.04 & - & $\leq$13.24 & - & - & - & B14+B18\\
41 & SDSSJ031027.82-004950.7  & 14.10$\pm$0.03 & - & $\leq$13.22 & - & - & - & B14+B18\\
42 & SDSSJ092909.79+464424.0  & $>$14.57$\pm$0.02 & - & 13.73$\pm$0.05 & - & - & - & B14+B18\\
43 & SDSSJ100102.55+594414.3  & 13.92$\pm$0.03 & - & $\leq$13.07 & - & - & - & B14+B18\\
44 & SDSSJ080908.13+461925.6  & 13.78$\pm$0.03 & - & $\leq$13.06 & - & - & - & B14+B18\\
45 & SDSSJ094733.21+100508.7  & 13.63$\pm$0.05 & - & $\leq$13.20 & - & - & - & B14+B18\\
46 & SDSSJ015530.02-085704.0  & $\leq$12.88 & - & $\leq$13.33 & - & - & - & B14+B18\\
47 & SDSSJ134231.22+382903.4  & 14.26$\pm$0.03 & - & $\leq$13.25 & - & - & - & B14+B18\\
48 & SDSSJ135712.61+170444.1  & $\leq$12.74 & - & $\leq$13.11 & - & - & - & B14+B18\\
49 & SDSSJ134206.56+050523.8  & $>$14.65$\pm$0.02 & - & $\leq$13.53 & - & - & - & B14+B18\\
50 & SDSSJ133053.27+311930.5  & $>$14.65$\pm$0.02 & - & 14.27$\pm$0.03 & - & - & - & B14+B18\\
51 & PG1049-005  & $>$14.43$\pm$0.03 & - & $\leq$13.16 & - & - & - & B14+B18\\
52 & 3C273  & 13.86$\pm$0.01 & $\leq$12.57 & $\leq$12.48 & $\leq$11.56 & $\leq$11.70 & $\leq$11.71 & LC14\\
53 & PG1121+422  & 13.97$\pm$0.01 & $\leq$13.05 & $\leq$12.94 & - & - & $\leq$12.44 & LC14\\
54 & SBS1122+594  & 14.26$\pm$0.01 & 13.72$\pm$0.05 & 14.16$\pm$0.01 & - & 13.14$\pm$0.02 & 13.27$\pm$0.03 & LC14\\
55 & PG1211+143  & 14.20$\pm$0.00 & $\leq$12.87 & $\leq$12.75 & $\leq$11.56 & $\leq$11.63 & $\leq$12.23 & LC14\\
56 & PG0003+158  & - & $\leq$12.92 & $\leq$12.95 & $\leq$11.91 & $\leq$11.70 & - & LC14\\
\enddata
\tablenotetext{}{Note: 
Col (1): PID = dwarf-QSO pair ID as used in Table \ref{tb:pair_info}. 
Col (2): QSO name. 
Cols (3)--(8): ion column densities ($\log N$). For nondetections, $3\sigma$ upper limits are indicated. For those nondetection values from the corresponding references where $1\sigma$ or $2\sigma$ are provided, we have converted their values to $3\sigma$ instead to be consistent with the rest of the measurements. Additionally, in cases where only EW values are provided in the corresponding references, we have calculated $\log N$ from EW assuming opitcally thin for the lines of interest based on equation 3 in \cite{savage96}. See \S \ref{sec:data_literature_pairs}.
Col (9): Reference for each set of measurements. }
\end{deluxetable*}

\startlongtable
\begin{deluxetable*}{ccccccccccc}
\tabletypesize{\footnotesize}
\label{tb:dwarf_info}
\tablecaption{Properties of Dwarf Galaxies}
\tablehead{
\colhead{Galaxy} & \colhead{R.A.} & \colhead{Decl.} & $\vlsr$ & \colhead{$d$} & \colhead{$d_{\rm ref}$} & $\logmhi$& $M_{\rm HI, ref}$ & $\log$SFR & SFR$_{\rm ref}$ & Reference \\ 
& (deg) & (deg) & ($\kms$) & (Mpc) & & ($\log \msun$) &  & ($\log (\msunyr)$) & & \\ 
\colhead{(1)} & \colhead{(2)} & \colhead{(3)} & \colhead{(4)} & \colhead{(5)}  & \colhead{(6)} & \colhead{(7)} & \colhead{(8)} & \colhead{(9)} & \colhead{(10)} & \colhead{(11)}
 }
\startdata 
              KKH086 & 208.64 & 4.24 & 295.1 & $2.59\pm0.19$ & D09 & 6.10 & H18 & -4.30 & L11 & New/Arx.\\
                 GR8 & 194.67 & 14.22 & 223.1 & $2.08\pm0.02$ & D09 & 7.03 & H18 & -2.87 & L11 & New/Arx.\\
              DDO187 & 213.99 & 23.06 & 171.0 & $2.21\pm0.07$ & D09 & 7.13 & H18 & -3.22 & L11 & New/Arx.\\
             UGCA292 & 189.67 & 32.77 & 314.8 & $3.85\pm0.55$ & T16 & 7.49 & K13$^{\dagger}$ & -2.73 & L11 & New/Arx.\\
            UGC08833 & 208.70 & 35.84 & 231.7 & $3.26\pm0.09$ & T16 & 7.21 & H18 & -3.07 & L11 & New/Arx.\\
           PGC039646 & 184.81 & 6.29 & 671.7 & 4.51 & K13$^*$ & 6.45 & H18 & -3.39 & K13$^{\dagger\dagger}$ & New/Arx.\\
              DDO099 & 177.72 & 38.88 & 255.9 & $2.65\pm0.10$ & T16 & 7.74 & B08 & -2.49 & L11 & New/Arx.\\
              DDO181 & 204.97 & 40.74 & 224.2 & $3.14\pm0.05$ & D09 & 7.37 & K13$^{\dagger}$ & -2.65 & L11 & New/Arx.\\
           Sextans A & 152.75 & -4.69 & 317.3 & $1.44\pm0.05$ & T16 & 7.95 & H12 & -2.12 & L11 & New/Arx.\\
            UGC06541 & 173.37 & 49.24 & 254.2 & $4.22\pm0.25$ & T16 & 7.04 & K13$^{\dagger}$ & -2.31 & L11 & New/Arx.\\
           Sextans B & 150.00 & 5.33 & 294.0 & $1.43\pm0.02$ & T16 & 7.57 & H18 & -2.58 & L11 & New/Arx.\\
              DDO190 & 216.18 & 44.53 & 162.0 & $2.84\pm0.04$ & J09 & 7.65 & SI02 & -2.43 & L11 & New/Arx.\\
            UGC08638 & 204.83 & 24.78 & 285.4 & $4.30\pm0.06$ & T16 & 7.27 & H18 & -2.46 & L11 & New/Arx.\\
            UGC07485 & 186.09 & 21.16 & 963.9 & 7.94 & K13$^*$ & 7.12 & H18 & -2.72 & K13$^{\dagger\dagger}$ & New/Arx.\\
             NGC5408 & 210.84 & -41.38 & 502.9 & $5.31\pm0.17$ & T16 & 8.48 & K13$^{\dagger}$ & -1.05 & K08 & New/Arx.\\
             NGC4144 & 182.50 & 46.46 & 271.5 & $4.61\pm0.11$ & J09 & 8.36 & K13$^{\dagger}$ & -1.67 & L11 & New/Arx.\\
              IC1613 & 16.20 & 2.13 & -236.4 & $0.76\pm0.02$ & T16 & 7.66 & H18 & -2.23 & L11 & Z20\\
                  D9 & 231.10 & 9.98 & z=0.139 & - & - & - & - & - & - & J17\\
                  D1 & 98.94 & -75.27 & z=0.123 & - & - & - & - & - & - & J17\\
                  D2 & 98.94 & -75.27 & z=0.161 & - & - & - & - & - & - & J17\\
                  D4 & 151.01 & 28.92 & z=0.138 & - & - & - & - & - & - & J17\\
                  D7 & 61.96 & -12.20 & z=0.092 & - & - & - & - & - & - & J17\\
                  D8 & 236.94 & 20.86 & z=0.095 & - & - & - & - & - & - & J17\\
                  D5 & 98.93 & -75.27 & z=0.144 & - & - & - & - & - & - & J17\\
                  D3 & 38.78 & -4.04 & z=0.296 & - & - & - & - & - & - & J17\\
                 D12 & 231.10 & 9.99 & z=0.240 & - & - & - & - & - & - & J17\\
                 D13 & 219.44 & -1.80 & z=0.116 & - & - & - & - & - & - & J17\\
                  D6 & 219.46 & -1.78 & z=0.184 & - & - & - & - & - & - & J17\\
            316\_200 & 164.90 & 14.74 & z=0.010 & - & - & - & - & -0.90 & - & B14+B18\\
            124\_197 & 47.66 & -0.86 & z=0.026 & - & - & - & - & -1.40 & - & B14+B18\\
            172\_157 & 142.30 & 46.70 & z=0.017 & - & - & - & - & -0.20 & - & B14+B18\\
             87\_608 & 150.60 & 59.74 & z=0.011 & - & - & - & - & -1.00 & - & B14+B18\\
            135\_580 & 147.00 & 9.97 & z=0.010 & - & - & - & - & -1.00 & - & B14+B18\\
            257\_269 & 122.18 & 46.31 & z=0.024 & - & - & - & - & -0.60 & - & B14+B18\\
            329\_403 & 28.82 & -8.85 & z=0.013 & - & - & - & - & -0.80 & - & B14+B18\\
            322\_238 & 205.58 & 38.54 & z=0.012 & - & - & - & - & -1.80 & - & B14+B18\\
             93\_248 & 209.37 & 17.07 & z=0.026 & - & - & - & - & -0.40 & - & B14+B18\\
            210\_241 & 205.49 & 5.03 & z=0.025 & - & - & - & - & -0.50 & - & B14+B18\\
              70\_57 & 202.74 & 31.33 & z=0.034 & - & - & - & - & -0.60 & - & B14+B18\\
             316\_78 & 162.95 & -0.84 & z=0.039 & - & - & - & - & -0.30 & - & B14+B18\\
SDSSJ122815.96+014944.1 & 187.07 & 1.83 & z=0.003 & - & - & - & - & - & - & LC14\\
SDSSJ112418.74+420323.1 & 171.08 & 42.06 & z=0.025 & - & - & - & - & -1.17 & - & LC14\\
SDSSJ112644.33+590926.0 & 171.68 & 59.16 & z=0.004 & - & - & - & - & -0.98 & - & LC14\\
SDSSJ121413.94+140330.4 & 183.56 & 14.06 & z=0.064 & - & - & - & - & - & - & LC14\\
SDSSJ000545.07+160853.3 & 1.44 & 16.15 & z=0.037 & - & - & - & - & -0.76 & - & LC14\\
\enddata
\tablenotetext{}{Note: Cols (1)--(4): Galaxy names, R.A., Decl., $\vlsr$ or $z$ of dwarf galaxies. 
Cols (5-6): distances, distance uncertainties, and the corresponding references when available: T16=\cite{tully16}, D09=\cite{dalcanton09}, J09=\cite{jacobs09}, 
K13$^*$=\cite{Karachentsev13}'s distance estimated based on the Tully--Fisher relation. Note that most local galaxies are with distances estimated based on the tip of the red-giant branch (TRGB) method except for PGC039646 and UGC07485. Cols (7)--(8): galaxy \HI\ mass in the ISM and the corresponding reference when available: B08=\HI\ flux from \cite{begum08} and converted to $\logmhi$ with adopted distance; H18=\HI\ flux from \cite{Haynes18} (ALFALFA) and converted to $\logmhi$ with adopted distance; H12=\cite{hunter12} and mass rescaled with adopted distances; K13$^{\dagger}$=\HI\ flux from \cite{Karachentsev13} and converted to $\logmhi$ with adopted distance; SI02=\HI\ flux from \cite{SI02} and converted to $\logmhi$ with adopted distance. 
Cols (9)--(10): star formation rate (SFR) and the corresponding reference when available: L11=FUV mag (Galactic extinction corrected) from \cite{lee11}, converted to SFR with adopted distance and assuming a \cite{kroupa01} IMF; K08=H$\alpha$ luminosity from \cite{kennicutt08} (no FUV mag available), and converted to SFR assuming a \cite{kroupa01} IMF; K13$^{\dagger\dagger}$=FUV mag from \cite{Karachentsev13}, corrected for Galactic extinction using their equation 8, and converted to SFR assuming \cite{kroupa01} IMF. See more details in \S \ref{sec:galaxy_property}.}
\end{deluxetable*}

\subsection{\npairliterature\ Additional Pairs Compiled from Literature}
\label{sec:data_literature_pairs}

\begin{figure*}[t]
	\centering
	\includegraphics[width=\linewidth]{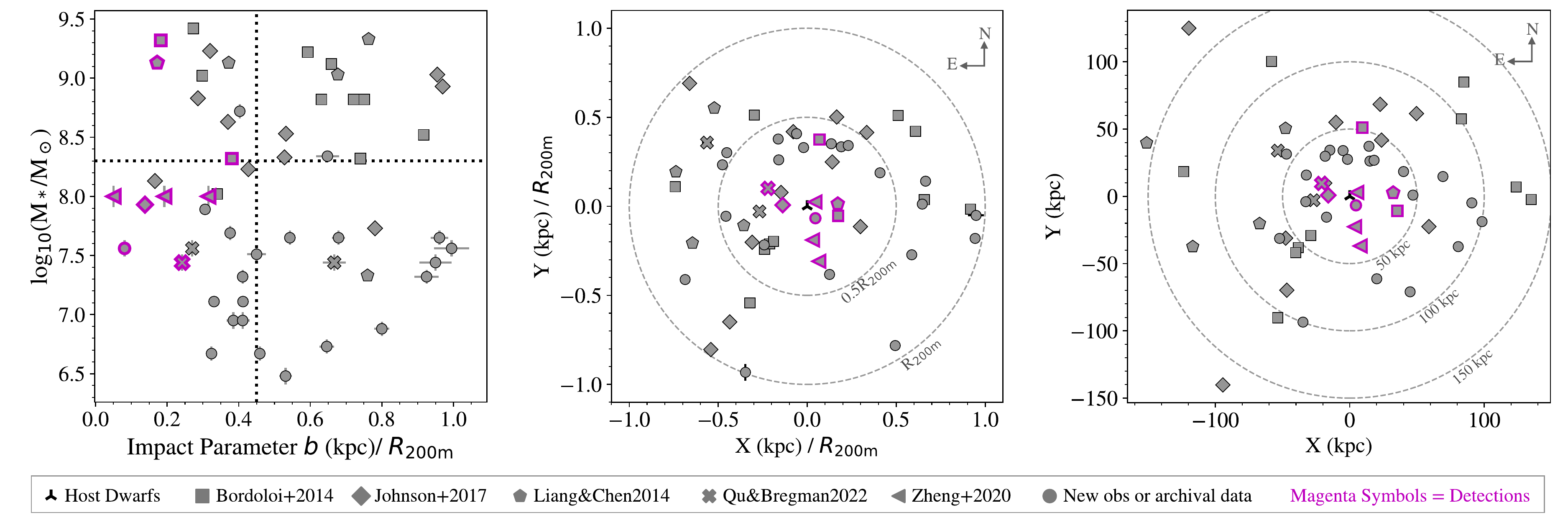}
	\caption{\textbf{Left:} $\mstar$ \textsl{vs.} $\impactpara/\rvir$ for the \npairtotal\ dwarf-QSO pairs in our Full Sample as described in \S\ref{sec:data} and Table \ref{tb:literature}. The data points here correspond to the filled symbols shown in Figure \ref{fig:logM_rho_all}. The dotted lines indicate the median impact parameter ($\langle \impactpara/\rvir\rangle=$0.45) and median stellar mass ($\langle\mstar\rangle=$10$^{8.3}~\msun$) in the Full Sample. \textbf{Middle:} Relative projected positions of QSOs with respect to host dwarf galaxies, normalized by the corresponding virial radii. \textbf{Right:} Same as the middle panel, but in kpc. In each panel, we use magenta-outlined symbols to highlight the detections of \CIV, while the rest are for nondetections. \textsl{Overall, at EW$\geq100$ m$\AA$ (3$\sigma$), we find low detection rates of \cfCIV\ in \CIV, \cfCII\ in \CII, \cfSiIV\ in \SiIV, and \cfSiIII\ in \SiIII, and \cfSiII\ in \SiII, all of which occur within 0.5$\rvirm$. Meanwhile, \HI\ (via Ly$\alpha$ absorption) is ubiquitously observed throughout the CGM with a detection rate of \cfHI}. See \S\ref{sec:result_covering_fraction} for more detail. }
	\label{fig:logM_rho_xy}
\end{figure*}

To further enlarge our data sample, we adopt additional {\it HST}/COS measurements from the literature, including a total of \npairliterature\ pairs from \citetalias{qu22}, \citetalias{zheng20b}, \citetalias{johnson17}, \citetalias{bordoloi14}+\citetalias{bordoloi18}, and \citetalias{liang14}, collectively. The information regarding each work has been summarized in Table \ref{tb:literature} and Section \ref{sec:intro}, the adopted pairs are shown in Figures \ref{fig:logM_rho_all} and \ref{fig:logM_rho_xy}, and relevant information is tabulated in Table \ref{tb:pair_info}. 

\textsl{Literature Compilation Rules}: to ensure comparable results, we only include CGM measurements whose QSO spectra have S/N$\geqslant8$. Because different works may define a spectrum's S/N differently (e.g., local S/N near a line of interest \textsl{vs.} averaged S/N over a wide wavelength range), we recalculate the S/N for each QSO spectrum as described in \S\ref{sec:data_new_pairs}. 
Then, $\mstar$ from the literature is converted to \citeauthor{kroupa01} IMF from either \citeauthor{salpeter55} (as used in \citetalias{bordoloi14}+\citetalias{bordoloi18}) or \citeauthor{chabrier03} IMF (as used in \citetalias{johnson17} and \citetalias{liang14}) following the analysis in \citetalias{zheng20b}. Only dwarf galaxies with $\mstar\leqslant$10$^{9.5}~\msun$ are selected. Lastly, we adopt QSO sight lines with $\impactpara/\rvir\leqslant1$, with $\rvir$ and $\mhalo$ recalculated from the adopted $\mstar$ using the stellar mass--halo mass relation shown in \S\ref{sec:galaxy_property}. Below we include additional details on the specific treatment to each literature sample beyond what is already shown in Table \ref{tb:literature} and Section \ref{sec:intro}. Note that we do not include the dwarf galaxy sample from \cite{richter17} as most of the dwarfs are spheroidal type and thus do not contain gas. 

For the \npairbordoloi\ dwarf-QSO pairs from the COS-Dwarfs survey (\citetalias{bordoloi14}), we obtain the \CIV\ $\log N$ measurements from \citetalias{bordoloi14} and \HI\ Ly$\alpha$ $\log N$ values from \citetalias{bordoloi18}. For each ion measurement, we convert the original $2\sigma$ upper limits to $3\sigma$ values to be consistent with the rest of our data sample. For \citetalias{liang14}'s sample, as most of their galaxy-QSO pairs are either with high $\mstar$, large $\impactpara$, and/or low S/N (see Figure \ref{fig:logM_rho_all}), we find only five pairs that meet our literature compilation rules. Since only EW values are given in \citetalias{liang14}, we convert EW to $\log N$ assuming the absorbers are optically thin, which is reasonable given that most absorbers are either weak or nondetections. For nondetections, we convert their quoted $2\sigma$ upper limits to $3\sigma$ for consistency. 


We adopt \npairjohnson\ pairs from \citetalias{johnson17}. We convert their $2\sigma$ nondetection EW values to $3\sigma$ and then convert the EW to $\log N$ assuming optically thin. For detections in galaxies D1 and D2 in their sample, we adopt \citetalias{johnson17}'s Voigt profile fit results. In cases where multiple absorbers are reported along a given sight line, we calculate the total $\log N$ of all available components in each ion. 

Within the Local Group, we include \citetalias{zheng20b}'s measurements of three QSOs at 0.05--0.5$\rvir$ from IC 1613, which is on the outskirts of the LG with no known galaxies within 400 kpc. We exclude the rest of the three QSOs in \citetalias{zheng20b}'s sample at $\sim0.6-0.7\rvir$ to avoid potential contamination from the Magellanic Stream (MS) in the foreground that may skew the column density profiles. We also do not include \cite{zheng19b}'s tentative detection in WLM where potential contamination from the MS may affect the result here too. We further include three QSO measurements (out of 6) from \citetalias{qu22} at 0.2--0.7$\rvir$ from Sextans A, Sextans B, and NGC 3109. The values for the nondetections toward these QSOs are recalculated over $\pm50~\kms$ velocity intervals (Qu; 2023, private communication).

To summarize, our literature compilation yields \npairliterature\ high-quality dwarf-QSO pairs with $\mstar\leqslant$10$^{9.5}~\msun$, $\impactpara/\rvir\leqslant1.0$, and S/N$\geq8$. These 34 literature pairs, in combination with the \npairthiswork\ new pairs described in \S\ref{sec:data_new_pairs}, form the Full Sample of \npairtotal\ dwarf-QSO pairs in this work. We note that while dwarf galaxies at $z\sim0$ in the Full Sample are \HI\ rich (see \S\ref{sec:data_new_pairs} and Table \ref{tb:dwarf_info}), those dwarfs at higher redshifts from \citetalias{bordoloi14}, \citetalias{liang14}, and \citetalias{johnson17} are star-forming by design but generally do not have available \HI\ measurements. We argue that the star-forming dwarfs at higher z should also be \HI\ rich because it has been found that galaxies with H$\alpha$ emission (hence star-forming) are usually observable in \HI\ (when available) and vice versa \citep{meurer06, VanSistine16}. Therefore, it is reasonable to combine the $z\sim0$ and higher-z dwarfs to form the Full Sample in this work.

\subsection{Determination of Galaxy Properties}
\label{sec:galaxy_property}


There are \nunidwarf\ unique dwarf galaxies among the \npairtotal\ dwarf-QSO pairs in our Full Sample. The galaxies' properties are tabulated in Tables \ref{tb:pair_info} and \ref{tb:dwarf_info}. Here we summarize the range and median galaxy property values of these \nunidwarf\ dwarf galaxies, which have a stellar mass range of $\mstar=\samplemstarrange~\msun$ with a median value at $\langle\mstar\rangle=\medmstar~\msun$, a halo mass range of $\mhalom=\samplemhalorange~\msun$ with a median value at $\langle\mhalom\rangle=\medmhalo~\msun$, an \HI\ gas mass range of $\mhi=10^{6.1-8.5}~\msun$ with a median value of $\langle\mhi\rangle=\medmHI~\msun$, a SFR range of $\mdotstar=10^{-4.3}$ to $10^{-0.2}~\msunyr$ with a median value of $\langle\mdotstar\rangle=10^{\medlogsfr}~\msunyr$, and a virial radius range of $\rvirm=68-213$ kpc with a median value of $\langle\rvir\rangle=\medrvirval$ kpc. Below we describe the approach to obtain $\mstar$ and $\mhalom$, which are the most important properties in this work, and defer the discussion on other properties to Appendix \ref{sec:app_gal_prop}.

For dwarf galaxies from \citetalias{liang14}, \citetalias{bordoloi14}, or \citetalias{johnson17}, we adopt the quoted $\mstar$ from the corresponding works and convert the values to \citet{kroupa01} IMF (see \S\ref{sec:data_literature_pairs}). For a majority of galaxies within the Local Volume, we calculate their $\mstar$ from {\it Spitzer} 3.6$\micron$ fluxes from the Local Volume Survey \citep{dale09}. Since the 3.6$\micron$ mostly traces infrared light from old stellar population, it is not sensitive to internal extinction of a galaxy's ISM as compared to young star populations. Therefore, the $\mstar$ value derived based on 3.6$\micron$ best represents most of the mass in a galaxy. Instead of directly using $\mstar$ values from the Local Volume Survey catalog \citep{Cook14}, we recalculate $\mstar$ based on the galaxies' 3.6$\micron$ fluxes and the distances we adopt in this work for consistency. We assume a mass to light ratio of $\mstar/L_{3.6\micron} = 0.5 ~\msun/L_\odot$ \citep{Cook14}. When {\it Spitzer} photometry is unavailable, we adopt $\mstar$ from \cite{mcconnachie12}, with values rescaled based on our newly adopted distances. If a galaxy is not included in either \cite{Cook14} or \cite{mcconnachie12}, nor can we find an appropriate value in the literature, we compute $\mstar$ from the galaxy's $K$s band magnitude from the Two Micron All Sky Survey \citep[2MASS;][]{jarrett03, Karachentsev13}. For the $K$s band, we assume a mass to light ratio of 0.6 \citep{McGaugh14}.

To derive a galaxy's halo mass from $\mstar$, one needs to assume an SMHM relation. The SMHM relation at low mass is found to be stochastic \citep{rey19, mcquinn22, sales22}, and the scatter in $\mstar$ increases with decreasing halo mass \citep{garrison-kimmel17,munshi21}. Among the literature works that expand the SMHM relation to as low as $\mstar\sim10^{6-7}~\msun$ where a power-law fit of $\mstar\propto\mhalo^{\alpha}$ is often assumed, it has been shown that the power law becomes steeper toward lower masses with $\alpha$ ranging from 1.4 to 3.5 \citep[e.g., ][]{garrison-kimmel14,garrison-kimmel17,read17,jethwa18,nadler20,munshi21}. In this work we adopt a broken power-law relation from \cite{munshi21}:
\begin{equation}
\log_{10} \mstar =\log_{10} M_{\rm *, 0} + \alpha (\frac{M_{\rm 200c}}{M_{\rm 200c,0}})
\label{eq:smhm}
\end{equation}
where $M_{\rm *,0}=10^{6.9}~\msun$, and $\mhaloc$ is the halo mass enclosed within a virial radius inside which the average halo density is 200 times the critical density of the Universe at $z=0$. The power-law index is $\alpha=2.8$ when $\mhaloc\leq M_{\rm 200c,0}$ and $\alpha=1.9$ when $\mhaloc>M_{\rm 200c,0}$, with $M_{\rm 200c,0}=10^{10}~\msun$. Because in this work we adopt a ``200m" definition for our virial radius (i.e., 200 times the mean matter density instead of critical density), we convert the $\mhaloc$ values derived from Eq. \ref{eq:smhm} to $\mhalom$ using $\log_{10} \mhalom =\log_{10} \mhaloc+0.15~{\rm dex}$. The conversions are estimated by comparing the $\mhaloc$ and $\mhalom$ values of a suite of simulated dwarf galaxies from an EAGLE high-resolution simulation volume \citep{oppenheimer18c} that we elaborate on in \S\ref{sec:sims_eagle}.

Lastly, we comment on the uncertainties involved in the galaxy properties that may impact the evaluation of the galaxies' surrounding environment and isolation criteria (\S\ref{sec:data_new_pairs}). There are two main sources of uncertainties: the intrinsic scatter in the SMHM relation and the uncertainties in the distances. The SMHM relation is typically with an intrinsic scatter of $\sim$0.3 dex \citep{munshi21}, which leads to an uncertainty of a few tens of percent in $\rvirm$. On the other hand, most galaxies at $z\sim0$ are with distances estimated based on the tip of the red-giant branch (TRGB) method except that two galaxies (PGC039646 and UGC07485) only have distances from the Tully--Fisher (TF) relation (Table \ref{tb:dwarf_info}). The TRGB method yields robust distance estimates with very small uncertainties ($\sim$1--7\% except for UGCA292 which is 14\%), while the TF method is typically with an uncertainty of $\sim$20\%. For those galaxies with distance uncertainties more than 10\%, we have checked that there are no neighbors near the galaxies within the virial radii when both the uncertainties in distances and $\rvirm$ are considered.

\section{Result: Ion Distributions in the CGM} 
\label{sec:result}

\subsection{Ion Covering Fractions $\Cf$}
\label{sec:result_covering_fraction}

\begin{figure*}[t]
	\centering
	\includegraphics[width=\linewidth]{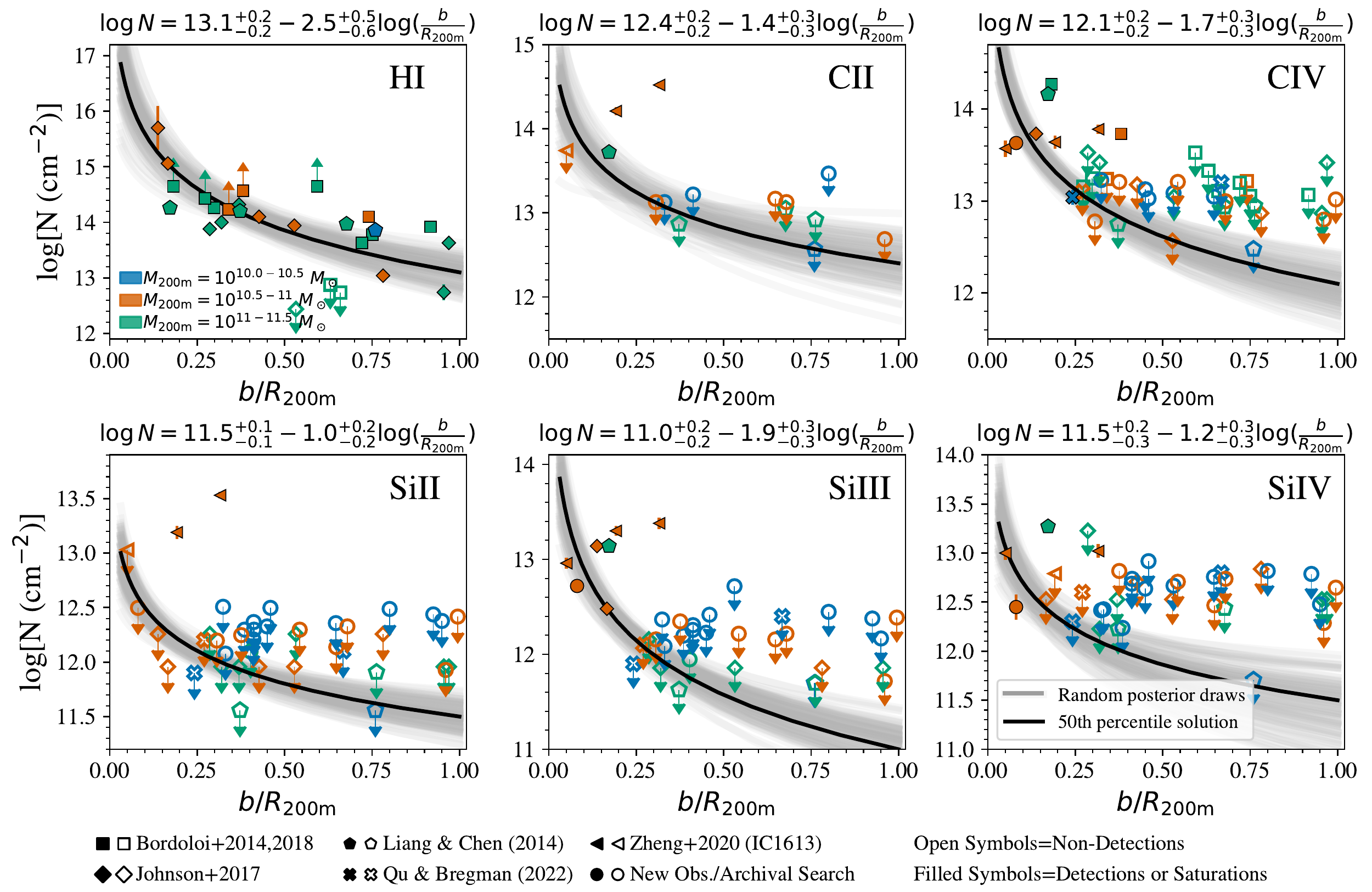}
	\caption{Ion column densities \textsl{vs.} $\impactpara/\rvir$ for \HI, \CII, \CIV, \SiII, \SiIII, and \SiIV. Filled symbols indicate detections, filled symbols with upward arrows for saturations (lower limits), and open symbols with downward arrows for nondetections ($3\sigma$ upper limits). Note that the y-axis scale varies from panel to panel. The data points are color-coded into 3 halo mass bins: $\mhalom=10^{10-10.5}~\msun$ (blue), $\mhalom=10^{10.5-11}~\msun$ (orange), and $\mhalom=10^{11-11.5}~\msun$ (green). We fit the $\log N$--$\impactpara/\rvir$ relation of each ion as a power law based on a Bayesian linear regression algorithm that accounts for the upper/lower limits as censored data (see \S\ref{sec:result_pymc}). The gray curves show 100 random draws from the posterior distributions, and the black solid curve in each panel indicates the 50th percentile solution. The title of each panel shows the 50th percentile solution and the errors in the coefficients indicates the values of the 50th--16th and the 84th--50th differences. \textsl{Overall, we find ion column density decreases with $\impactpara/\rvir$; except for \HI\ that is ubiquitously detected in the CGM, other ions are only detected within $\sim$0.5$\rvir$.}}
	\label{fig:observed_logN_b}
\end{figure*}

In Figure \ref{fig:logM_rho_xy}, we show the distribution of QSO sight lines with respect to host dwarf galaxies in our Full Sample. Those sight lines with detections of \CIV\ are highlighted with magenta edges. Then, in Figure \ref{fig:observed_logN_b}, we show the relation of $\log N$ \textsl{vs.} $\impactpara/\rvir$ for \HI\ (via Ly$\alpha$ absorption), \CII, \CIV, \SiII, \SiIII, and \SiIV. Except for \HI, which is ubiquitously detected in the halos of low-mass dwarf galaxies, the rest of the ions typically show nondetections unless the sight lines are at small impact parameters (i.e., $\impactpara\lesssim$0.5$\rvir$), consistent with findings from \citetalias{bordoloi14}+\citetalias{bordoloi18}, \citetalias{liang14}, and \citetalias{johnson17}. Similar to the COS-Dwarfs survey (\citetalias{bordoloi14}), our typical detection threshold for ion absorbers is EW$\approx$50-100 ${\rm m \AA}$ at $3\sigma$. At EW$\geq$100 ${\rm m \AA}$, we find detection rates (or covering fraction $\Cf$) of $\cfHI$ for \HI, $\cfCII$ for \CII, $\cfCIV$ for \CIV, $\cfSiII$ for \SiII, $\cfSiIII$ for \SiIII, and $\cfSiIV$ for \SiIV\ in our Full Sample. 
Our measurements show that the metals as probed by \SiII, \SiIII, \SiIV, \CII, and \CIV\ in the outer CGM of dwarf galaxies are too diffuse to be detected with {\it HST}/COS at a column density limit of $\log N_{\rm ion}\sim12.5-13.5$ at 3$\sigma$.

When compared to literature values, our \CIV\ detection rate is lower than what is found in the COS-Dwarfs survey (\citetalias{bordoloi14}), which reported $\Cf$(\CIV)$\approx$40\% at EW$\geq$100 $\rm m\AA$. Similarly, \citetalias{liang14} found $\Cf\approx$30--50\% for \CIV\ with EW$\geq$100 ${\rm m \AA}$ at $\impactpara<0.7\rvir$. When taking into account the difference in the data samples (see Figure \ref{fig:logM_rho_all}), we find that the higher \CIV\ detection rates by \citetalias{bordoloi14} and \citetalias{liang14} are likely because of the inclusion of higher-mass galaxies with $\mstar\geq10^{9.5}~\msun$ which contribute to the majority of their detections. Our \CIV\ detection rate is similar to that found by \cite{burchett16}, which shows that low-mass galaxies ($\mstar\leqslant10^{9.5}~\msun$) in their sample show very little detection of metal absorbers, with a covering fraction of $\sim9^{+12}_{-6}\%$ (for 11 galaxies). 

For other absorbers, \citetalias{liang14} estimated covering fractions of nearly 100\% for \HI\ and generally $\sim40-60\%$ for \SiII, \SiIII, \SiIV, and \CII\ at $\impactpara/\rvir\sim$(0.25--0.6), as evaluated at a detection threshold of EW$\geq$50 $\rm m\AA$ at $2\sigma$. However, in the outer halo ($\impactpara/\rvir\gtrsim$0.6--1.1), their ions' covering fractions drop below 14\% except for \HI\ (96\%) and \CIV\ (38\%). Their high detection rates at $\impactpara/\rvir\sim$(0.25--0.6) are likely caused by (1) a less restrictive detection threshold when evaluating the detections (i.e., EW$\geq$50 $\rm m\AA$ at 2$\sigma$), and (2) contributions from absorbers detected at small impact parameters in galaxies with higher masses ($\mstar>10^{9.5}~\msun$).

Lastly, among the 18 star-forming dwarf galaxies in \citetalias{johnson17}'s sample, $\Cf<$17\% is detected with \SiII\ within $\rvir$ at EW$>$100 $\rm m\AA$ ($2\sigma$ threshold), 10\% for \SiIII, $<$19\% for \SiIV, 23\% for \CIV, and 94\% for \HI, respectively. Generally we find good consistency in covering fractions between ours and \citetalias{johnson17}'s given that both samples cover a similar low-mass range  (Figure \ref{fig:logM_rho_all}).

In all, our measurements show that in the CGM of dwarf galaxies with $\mstar=$10$^{6.5-9.5}~\msun$, \HI\ (via Ly$\alpha$ 1215 \AA\ line) is ubiquitously detected (89\%) at a column density level of $\log N_{\rm HI}\approx13-16$. On the other hand, ions such as \SiII, \SiIII, \SiIV, \CII, and \CIV\ are typically found with low detection rates of $\Cf\approx5-22$\%, and the detections of metal absorbers only occur in the inner CGM ($b\lesssim0.5\rvir$). On the outskirts of dwarf galaxies' CGM ($b\gtrsim0.5\rvir$), the column densities of these ions are too low to be detected at {\sl HST}/COS's sensitivity of $\log N(3\sigma)\sim12.5-13.5$ at S/N$\geq8$.

\subsection{Ion Column Density Profiles: $\log N$ \textsl{vs.} $\impactpara/\rvir$}
\label{sec:result_pymc}

To parameterize the relation of $\log N$ versus $\impactpara/\rvir$, which will be used for further understanding the ion distribution in dwarf galaxies' CGM (see \S\ref{sec:theory}), we fit the data points in Figure \ref{fig:observed_logN_b} assuming a power-law relation: 
\begin{equation}
N  = N_0 (\impactpara/\rvir)^{k} ~~~,
\label{eq:power_law}
\end{equation}
where $N$ is the column density profile for the ion of interest, including \HI, \CII, \CIV, \SiII, \SiIII, and \SiIV. $N_0$ is an ion's column density evaluated at $\impactpara=\rvir$, and $k$ is the power-law index. We adopt a power-law form because it has been widely used in previous CGM studies \citep[e.g.][]{thom12, werk13, bordoloi14, keeney17}, and it is consistent with theoretical ion radial profiles predicted from simulations (see Figure \ref{fig:eagle_logN_b} and \S\ref{sec:sims_eagle}). 
Additionally, it provides the simplest form to fit for a dataset in log--log space with minimal number of parameters, providing a clean way to parameterize the dataset where most of the information is hidden in nondetections (see below). 
In a log--log space, Eq. \ref{eq:power_law} turns into a linear form: $\log N=\log N_{0} + kx$, where $x\equiv \log (\impactpara/\rvir)$.

We adopt a censored regression algorithm\footnote{\href{https://docs.pymc.io/en/v3/pymc-examples/examples/generalized_linear_models/GLM-truncated-censored-regression.html}{see a python notebook: Bayesian regression with truncated or censored data}, by Benjamin T. Vincent} implemented with \texttt{PyMC3} \citep{pymc3} that treats the nondetection upper limits and saturation lower limits appropriately. The dataset is split into three groups: detections, nondetections (3$\sigma$ upper limits), and saturations (lower limits). We set up priors for $k$ and $\log N_0$ and likelihood functions for each group following the instructions in the aforementioned \texttt{PyMC3} algorithm. Specifically, the likelihood function for the detection group is calculated as the joint probability of the observed data given the set of parameters ($k$, $\log N_0$), while the likelihood functions for censored data with upper or lower limits are computed by integrating the probability to the measured limits. We note that when constructing the likelihood functions for our dataset, we have included an intrinsic density scatter term to account for likely clumpy gas distribution in dwarf galaxies' CGM (see Eq. \ref{eq:lnL_det}). For clarity and readability, we defer the details of our algorithm setup to Appendix \ref{sec:app_pymc3} and discuss the results of our censored regression below.

In Figure \ref{fig:observed_logN_b}, we show the 50th percentile solution of our \texttt{PyMC3} runs with the solid black curve and also 100 random draws from the posterior distributions. From the \HI\ panel, it is straightforward to see that when most data points are detections, our power-law parameterization and \texttt{PyMC3} solutions well predict not only the $\log N$ versus $\impactpara/\rvir$ trend but also the scatter in the data points. This means that the priors and the likelihood functions we set have been able to sufficiently capture the information in the data. 

In cases where a majority of the data points are nondetections, as is the case for \CII, \CIV, \SiII, \SiIII, and \SiIV, we find that the \texttt{PyMC3} curves often occupy the space below the $3\sigma$ upper limit values, especially at $\impactpara/\rvir>0.5$ where there are no detections. This is reasonable because the $3\sigma$ upper-limit data points indicate the thresholds below which the actual values would occur with 99.7\% probability, if with sufficient sensitivity. In other words, given the assumption of a power-law relation, the \texttt{PyMC3} curves predict the trend of $\log N$ at $\impactpara/\rvir>0.5$ where the metal content of the dwarf galaxies' CGM is mostly unavailable to observers at the current sensitivity of {\it HST}/COS. The upper limit constraints from our data and \texttt{PyMC3} analysis suggest that future UV spectrographs with sensitivity 1 order of magnitude better are most likely to be needed to probe diffuse metals in the CGM of dwarf galaxies, especially at low masses (see \S\ref{sec:theory} and \S\ref{sec:sims_eagle} for theoretical predictions on CGM metal properties in dwarf galaxies).

When compared to literature values, we find that our power-law fit to \HI\ shows a similar slope ($k=-2.5^{+0.5}_{-0.6}$) to those \HI\ distributions in the CGM of MW-mass galaxies ($k=-2.7\pm0.3$; \citealt{keeney17}). For other ions, when available, we find steeper slopes in our fits than typically reported. For example, the COS-Dwarfs survey (\citetalias{bordoloi14}) characterized their EW(\CIV) versus $\impactpara$ distribution with a $k=-1$ power law, while our algorithm finds a steeper slope of $k=-1.5\pm0.3$ when fitting for the EW--$\impactpara/\rvir$ relation (not shown here). The discrepancy is most likely because we take into account the nondetections of \CIV\ as censored data in our fits, while \citetalias{bordoloi14}'s fit may be biased toward detected values in the inner CGM. Another example is the $\log N_{\rm SiIII}$ versus $\impactpara$ fit from the COS-Halos survey \citep{werk13}, where a slope of $k=-1.11\pm0.29$ was found for detected absorbers, while our \SiIII\ fits in Figure \ref{fig:observed_logN_b} show a steeper slope of $k=-1.9\pm0.3$. Similarly, the different treatment of nondetections likely contributes to the discrepancy here, although we note that the COS-Halos survey targets MW-mass galaxies which may harbor a CGM with ion properties different from lower-mass galaxies.

Because of the relatively small sample size, we do not conduct detailed analyses on how each ion's $\log N$ versus $b$ profile may depend on galaxy masses or star formation rates (SFRs). As shown in Figure \ref{fig:observed_logN_b}, when we split the Full Sample into three halo mass bins, there is little difference in the profiles. The lack of difference here is most likely caused by the small number of data points in each bin, especially in the lowest mass bin. As shown in \S\ref{sec:sims_eagle}, when a large sample of simulated dwarf galaxies (207) from the EAGLE simulation are considered, the \HI\ and low-to-intermediate ion column densities are found to increase with galaxy masses. Similarly, for higher ions like \OVI, \citet{tchernyshyov22} showed that the \OVI\ column densities increase as a function of $\mstar$ at a given impact parameter in the inner CGM  when a larger sample of dwarf galaxies ($\sim$100--150) are considered. Therefore, a larger sample of high-quality sight lines in the CGM of dwarf galaxies, especially at small impact parameters, is needed for further investigation on the mass-dependency of the ion column density profiles.

\section{An Empirical Model for Dwarf Galaxies' CGM}
\label{sec:theory}

In \S\ref{sec:result}, we present power-law fits to the observed column density profiles, which parameterize the ions' projected distributions and can be compared to different CGM studies. These fits are performed for each ion individually, and avoid assumptions about the physical conditions in the CGM. Building upon the power-law fits, in this Section, we aim to construct a physically motivated empirical model that relates the observed ion absorption to the underlying spatial distribution and ionization states of the gas in dwarf galaxies' CGM. Given that the galaxies in our Full Sample span a wide mass range from $\mstar=10^{6.5}~\msun$ to $10^{9.5}~\msun$, as a first approximation, we construct an empirical model for a typical dwarf galaxy with $\mhalom=10^{10.9}~\msun$ ($\mstar=10^{8.3}~\msun$, $\rvirm=136$ kpc), which are the median values among the Full Sample. The model can be applied in future work to different halo masses or individual galaxies.

This Section is structured as follows. We first describe the model setup and parameters in \S\ref{sec:empirical_model}, and use the observed \HI\ $\log N$ profile to constrain the parameters in \S\ref{sec:empirical_constraints}. Then, in \S\ref{sec:empirical_columns}, we show how the \HI-constrained model parameters predict metal ion column densities. In \S\ref{sec:empirical_mass_inference}, we discuss the implications of the empirical model for the estimated gas and metal masses in the CGM, and for the baryon and metal budget of dwarf galaxies. 
 
\subsection{Model Setup}
\label{sec:empirical_model}

The observed ion distributions in the dwarf galaxies' CGM are likely due to a delicate balance between the underlying density distribution, volume filling factor, and ionization structures in different gas phases. For example, the ion column densities shown in Figure \ref{fig:observed_logN_b} have different profiles as functions of the impact parameter, as evidenced by the different fitted slopes, suggesting variation in the gas ionization state with radius. In the following, we describe the assumptions and parameters of our empirical model, addressing the gas properties.

\begin{figure*}[t]
\centering
\includegraphics[width=\textwidth]{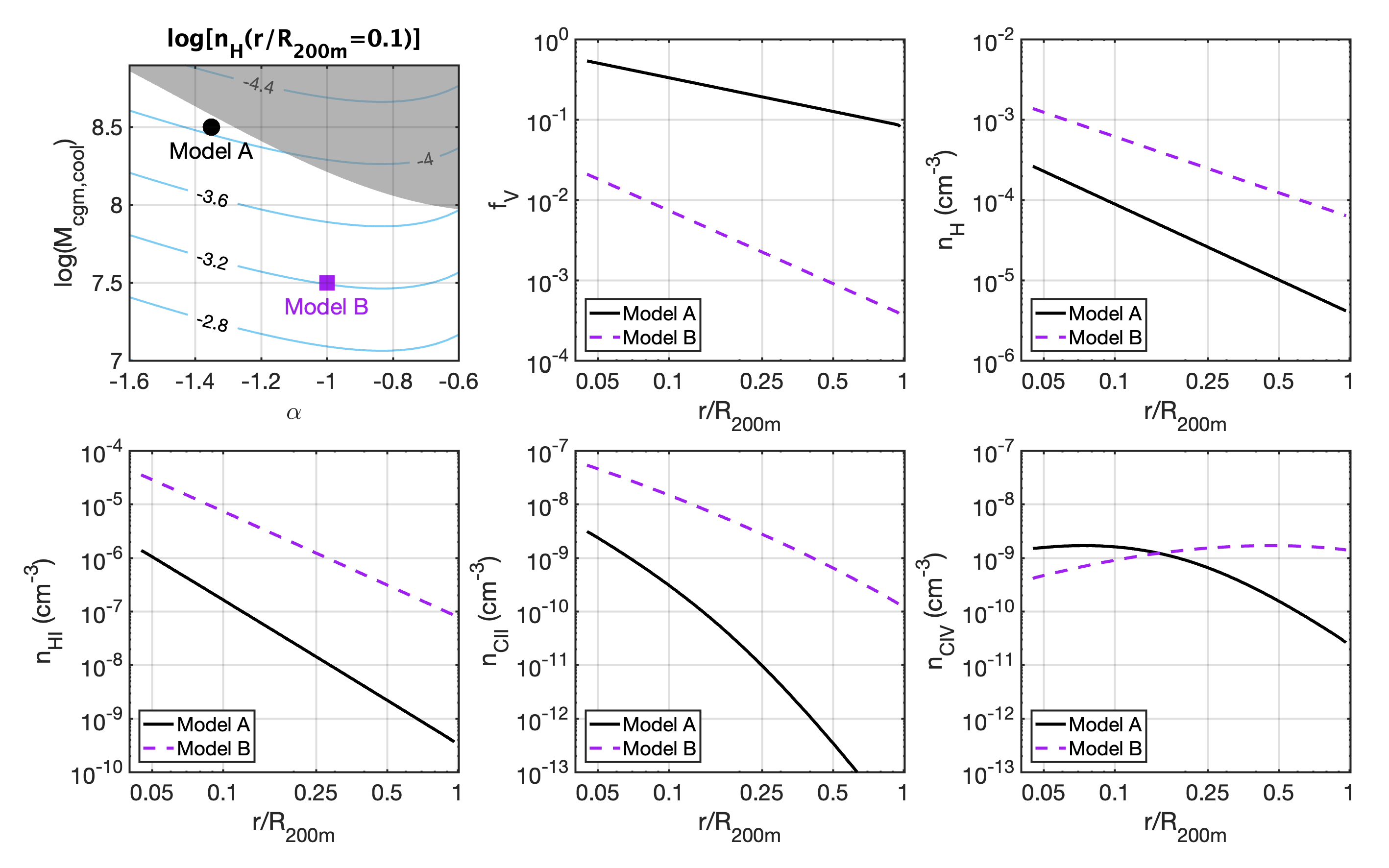}
\caption{Empirical model gas distributions. \textbf{Top left}: $\mcgmcool$ \textsl{vs.} $\alpha$ parameter space, with the contours showing the hydrogen volume densities $n_{\rm H}$ evaluated at $r=0.1\rvir$ (see Eq. \ref{eq:mod_profiles}). Two representative models (A and B) are highlighted as a black circle and a purple square, respectively, both of which reproduce the observed \HI\ column density profile (left panel, Figure \ref{fig:empirical_logN_b}). The shaded gray area indicates $\f0 > 1$ which is prohibited in our model. \textbf{Top middle and right}: radial distributions of gas volume filling factor (\fv, middle) and  total hydrogen volume density (\nh, right) for Model A (solid black lines) and Model B (dashed purple lines).
\textbf{Bottom}: radial distributions of ion volume densities for the two models, for \HI\ (left), \CII\ (middle) and \CIV\ (right) The gas ionization state is set by the local volume density, and Model A, with lower gas densities, has higher ratios of $n_{\rm CIV}/n_{\rm CII}$ than Model B. We note that Model A, with volume filling factor close to unity at small radii, represents a case in which the cool baryonic gas mass in the CGM $\mcgmcool$ is close to the maximum value allowed by the observed \HI\ column densities.
\label{fig:empirical_model_setup}}
\end{figure*}

The measurements reported in this study are of \HI\ and low-to-intermediate metal ions, and we assume they trace the cool, photoionized phase of the CGM at $T \approx 10^4$~K (see also \citealp{fw23} for a model of the cool CGM of $L_*$ galaxies). We set the gas spatial distribution in our model through two functions: (1) the hydrogen number density $\nh$, and (2) the volume filling fraction, defined as the local volume fraction occupied by the cool gas clouds, $\fv \equiv dV_{\rm cool}/dV\leq1$, where $dV_{\rm cool}$ and $dV = 4 \pi r^2dr$ are the cool gas and total volume of a shell at a given radius, respectively. As shown in Figure \ref{fig:empirical_model_setup}, we allow both $\fv=1$ (a volume filling phase) and lower values ($\fv \sim 10^{-1} - 10^{-3}$) to account for potential variations in cool CGM gas properties in dwarf galaxies (see also Figure \ref{fig:eagle-phase}). We assume spherical symmetry and model both $\nh$ and $\fv$ as power-law functions of the radial distance from the halo center: 
\beq \label{eq:mod_profiles}
\nh(r) = \n0 \left(\frac{r}{r_0}\right)^\alpha ~~,~~ \fv(r) = \f0 \left(\frac{r}{r_0}\right)^\beta ~~~,
\eeq
where $\alpha,\beta<0$, and $\n0$ and $\f0$ are the hydrogen number density and volume filling fraction evaluated at $r=r_0$, respectively. Here $r_0$ is a reference radius that is set at $r_0 = \impactpara_{\rm min} \approx 6$~kpc, which is the minimum impact parameter among the dwarf-QSO pairs in the Full Sample. We note that throughout this work the notation $r$ is referred to as a gas cloud's radial distance from its host galaxy in 3D, while $b$ indicates the corresponding 2D projected radial distance (i.e., impact parameter) as viewed along a line of sight.

We assume the gas temperature to be constant with radius, and adopt $T=10^4$~K, which is set by heating and cooling equilibrium with the metagalactic radiation field \citep{haardt12, werk16}. Given the local gas density and temperature, we calculate the gas ionization state using Cloudy 17.00 \citep{Ferland17}. Overall, the model setup in Eq. \ref{eq:mod_profiles} gives a total of four free parameters ($n_{\rm H,0}$, $f_{\rm V,0}$, $\alpha$, $\beta$). By integrating the $n_{\rm H}(r)$ and $f_{\rm V}(r)$ profiles to $\rvirm$, we can write down the total cool CGM mass as
\beq
\mcgmcool \approx \frac{4 \pi \mp}{(3+\alpha+\beta)X_{\rm H}} \n0 \f0 r_0^3 \left(\frac{\rvir}{r_0}\right)^{3+\alpha+\beta} ~,
\eeq
where we used $X_{\rm H}=0.74$ for the hydrogen mass fraction, accounting for the contribution of helium. In practice, we take $\mcgmcool$ instead of $\n0$ as one of the four free parameters, which has broader implications on baryon and metal masses in the CGM (see \S\ref{sec:empirical_mass_inference}). As stated above, our model is anchored at a typical halo mass of $\mhalom=10^{10.9}~\msun$, which corresponds to a cosmological baryonic mass budget of $M_{\rm bar}=\mhalom\Omega_b/\Omega_m \approx 10^{10.1}~\msun$.


\subsection{Constraining the Model with the Observed \HI\ Profile}
\label{sec:empirical_constraints}
 
We first use the observed \HI\ column densities to constrain the empirical model. We focus on \HI\ because it is ubiquitously detected throughout the halos (see Figure \ref{fig:observed_logN_b}), resulting in a well-constrained column density profile. Furthermore, the \HI\ distribution does not depend on assumptions regarding gas metallicity. Based on Eq. \ref{eq:mod_profiles}, the \HI\ column density through the halo projected at an impact parameter $\impactpara$ is calculated as 
\beq
N_{\rm HI}(\impactpara) = 2\int_{r=\impactpara}^{\rvir}f_{V}(r){n_{\rm H}(r) f_{\rm HI}(n_{\rm H}) ds} ~~~,
\label{eq:mod_columns} 
\eeq
where $s = \sqrt{r^2-\impactpara^2}$ is the path length along the line of sight within a galaxy's halo, and $f_{\rm HI}(n_{\rm H})$ is the \HI\ ion fraction which is a function of the gas density $n_{\rm H}(r)$.

\begin{figure*}
\includegraphics[width=\textwidth]
{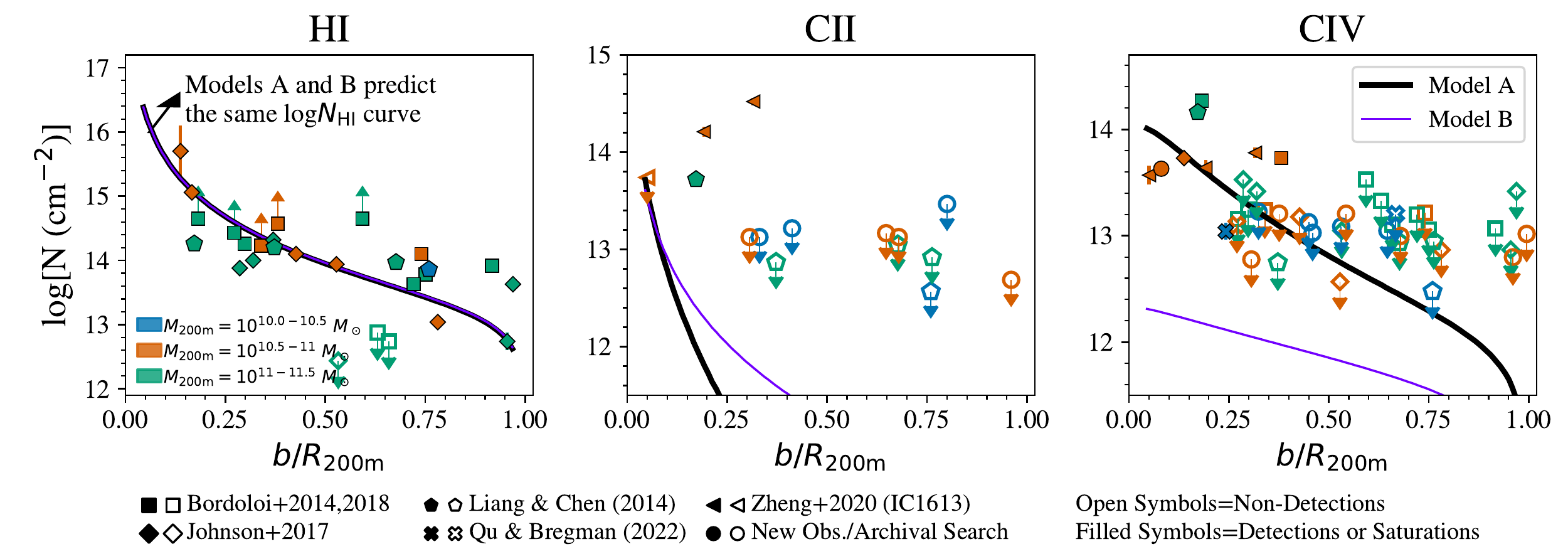}
\caption{Predicted \HI, \CII\ and \CIV\ column densities from Model A (thick black line) and Model B (thin purple line), in comparison with observed values. In our model, the cool CGM is photoionized by the metagalactic radiation \citep{haardt12, werk16} and the gas temperature is set at $T=10^4$ K (\S\ref{sec:empirical_model}). The \HI\ profiles of Models A and B (left panel) are identical by construction to match the observed values (\S\ref{sec:empirical_constraints}). The \CII\ and \CIV\ column densities are estimated at $Z'=0.3{\rm Z_\odot}$; we note that \SiII, \SiIII, and \SiIV\ are not shown here, but display similar radial profiles (\S\ref{sec:empirical_columns}). Model A has low gas density but high volume filling factor, while Model B is constructed with high gas density but low volume filling factor (Figure \ref{fig:empirical_model_setup}). In general, the low gas density ($n_{\rm H}\sim 10^{-4}$ cm$^{-3}$) in Model A leads to more \CIV\ (\CII) being formed (removed) by photoionization, resulting in higher \CIV\ column densities that are more consistent with observed values, especially in the inner CGM.
\label{fig:empirical_logN_b}}
\end{figure*}

In our model, we vary the parameters such that the model $N_{\rm HI}$ profile, given by Eq. \ref{eq:mod_columns}, matches the 50th-percentile power-law fit of \HI\ from \S\ref{sec:result_pymc} and Figure \ref{fig:observed_logN_b}. In the left panel of Figure \ref{fig:empirical_model_setup}, we show how the hydrogen volume density $n_{\rm H}(r)$ at $r=0.1\rvir$ changes due to different combinations of $\mcgmcool$ and $\alpha$. In particular, we highlight two models: one with high $\mcgmcool$ but low $\alpha$ (steep density profile, hereafter Model A), and another with low $\mcgmcool$ and high $\alpha$ (flatter density profile, hereafter Model B). The purpose of these two models is to demonstrate different possible distributions of cool gas in the CGM of low-mass halos, and their distinct observational signatures. Both models reproduce the observed \HI\ column density profile but predict different metal ion columns (Fig. \ref{fig:empirical_logN_b}).

To demonstrate how the model parameters translate to gas distributions, we plot the radial profiles of $\fv(r)$ and $\nh(r)$ for Model A (solid black curves) and Model B (dashed purple) in the top middle and right panels of Figure \ref{fig:empirical_model_setup}. Model A has a steep $\nh$ profile (right panel), leading to low densities at large radii ($\nh\sim 10^{-5}$~cm$^{-3}$). These are compensated by the high volume filling factors (middle panel, $\fv\sim 0.1-1$), giving a large \mcgmcool\ that is close to the maximum value allowed by the observed \HI\ column densities. Model B, in contrast, has higher gas densities but much lower volume filling factors, leading to low \mcgmcool.

The gas densities in each model set the gas ionization state, and in the bottom panels we plot the radial profiles of \HI\ (left panel) \CII\ (middle) and \CIV\ (right) volume densities, assuming a constant metallicity of $Z'=0.3~Z_\odot$ (see details in \S\ref{sec:empirical_columns}). In Model A (solid black), the low $\nh$ densities lead to high-ionization, low \HI\ fractions, and low \HI\ densities.
When multiplied by the high $\fv$ (top middle), these reproduce the observed \HI\ column density profile (Figure \ref{fig:empirical_logN_b}).
In Model B (dashed purple), the high gas densities result in higher \HI\ fractions and densities, which when multiplied by the low $\fv$ produce the same \HI\ columns as Model A (by construction). Similarly, the \CII\ densities (middle panel) in Model B are significantly higher, due to the model's higher gas densities and lower ionization. The \CIV\ densities (right panel) are more similar between the two models. However, as we will see in \S\ref{sec:empirical_columns}, Model A predicts much higher \CIV\ column densities than Model B thanks to its higher volume filling factor, $\fv$.

\subsection{Predicted Metal Column Densities}
\label{sec:empirical_columns}

We now address the predictions of Models A and B for the column densities of low to intermediate ions, and compare them to the observed values. At an impact parameter $b$, the line-of-sight metal column density is
\beq \label{eq:mod_columns_ion} 
N_{\rm X,i}(\impactpara) = 2\int_{r=\impactpara}^{\rvir}{f_{V}(r) n_{\rm H}(r) Z' a_{X} f_{\rm X,i}(n_{\rm H}) ds} ~~~,
\eeq
where $a_X$ is the elemental abundance of element $X$ relative to hydrogen, and $f_{\rm X,i}(n)$ is the ionization fraction of ion ${\rm X}_i$ as a function of the gas density $n_{\rm H}(r)$. In this study we assume a constant metallicity ($Z'$) with radius, and the metal column density scales linearly with $Z'$. As shown by the metal ion volume densities in the bottom panels of Figure~\ref{fig:empirical_model_setup}, even with the same presumed metallicity and predicted \HI\ column densities (left panel of Figure~\ref{fig:empirical_logN_b}), the different gas ionization states in Models A and B yield different metal column density profiles, which we discuss below.

The middle and right panels in Figure \ref{fig:empirical_logN_b} show the predicted \CII\ and \CIV\ column densities assuming $Z'=0.3~{\rm Z_\odot}$, respectively. The choice of $Z'=0.3~{\rm Z_\odot}$ is to anchor the subsequently derived CGM mass at a similar metallicity as commonly assumed or estimated in the CGM literature, although we note that the metallicity of CGM gas can be found over a wide range \citep[e.g.][]{prochaska17, zahedy21}. As we will show in \S\ref{sec:sims_eagle}, the assumed metallicity is also similar to the metallicity derived for $T\sim10^4$ K gas in the CGM of simulated dwarf galaxies from the EAGLE simulation. The middle panel shows that both models predict very low column densities in \CII, especially at large impact parameters, consistent with the nondetections we see in the data. In the right panel, we show that Model A with low CGM densities \nh\ and high volume filling factors \fv\ produces \CIV\ column densities that are consistent with the detected values in the inner CGM, and the model suggests low \CIV\ column densities at $b\gtrsim0.5\rvir$, also consistent with the upper limits that we observe. In contrast, Model B with high \nh\ densities but low $\fv$ values predicts \CIV\ column densities that are significantly below the measured columns at $b\lesssim 0.5 \rvir$, but are consistent with the upper limits at larger projected distances. While not shown here, our models also predict \SiII, \SiIII, and \SiIV\ column density profiles with values lower than the observed upper limits. Note that in our empirical models, ion column densities scale linearly with metallicity, so assuming a higher or lower metallicity than $Z'=0.3{\rm Z_\odot}$ would result in changes in predicted column densities accordingly.

 In all, while the two models produce the same \HI\ column densities, Model A with low densities (\nh) and high volume filling factors (\fv) leads to a cool CGM that is more ionized than that of Model B. These two model setups demonstrate the flexibility of the empirical model in predicting different scenarios for CGM ionization states, which can be applied to galaxies with different halo masses in future work.

\subsection{Inferences on CGM Baryon and Metal Masses}
\label{sec:empirical_mass_inference}

As we have seen in the previous section, Model A (low gas density, high volume filling factor) is more consistent with the measured \CIV\ columns at small impact parameters, whereas Model B underpredicts them. Thus, in this section we focus on Model A and discuss its properties in more detail. We infer the total masses of \HI\ and low-to-intermediate ions in dwarf galaxies' CGM based Model A, and tabulate them in Table \ref{tb:cgm_mass}. Note that the masses estimated here are for the cool phase of the CGM at $T\approx 10^4$ K, and the masses are integrated over a spherical halo volume from $0.1\rvirm$ to $\rvirm$ for a typical dwarf galaxy with $\mhalom=10^{10.9}~\msun$.


The gas density and volume filling profiles in Model A are given by 
\begin{equation}
\begin{split}
    n_{\rm H}(r) & =8.9 \times 10^{-5} \times (r/0.1\rvir)^{-1.35} \\ 
    f_{\rm V}(r) & =0.33 \times (r/0.1\rvir)^{-0.60} ~~~.
\end{split}
\end{equation}

The total hydrogen mass in Model A is $M_{\rm CGM,H}\approx10^{8.3}~\msun$, which corresponds to a total cool CGM mass of $\mcgmcool=M_{\rm CGM, H}/X_{\rm H}\approx10^{8.4}~\msun$. As shown in Figure \ref{fig:empirical_model_setup}, in Model A the volume filling factor is close to unity at small radii, which gives an $\mcgmcool$ (top-left panel) that is close to the maximum value allowed by the observed \HI\ column densities. When compared to the total baryon budget of a typical dwarf galaxy with $\mhalom=10^{10.9}~\msun$, this corresponds to $f_{\rm CGM,cool}=M_{\rm CGM,cool}/(\mhalom \Omega_b/\Omega_m)\sim2$\%\footnote{We note that while the empirical model presented here reproduces an average \HI\ column density profile, the observed \HI\ columns show some scatter of order $\sim 0.5$~dex (Figure \ref{fig:observed_logN_b}). Therefore, the derived cool CGM gas mass and the corresponding 2\%\ cool CGM mass fraction in individual galaxies may differ by a similar amount at a given halo mass.}. The low fraction indicates that the total gas mass in the cool phase CGM only accounts for $\sim$2\% of the total galactic baryonic mass. 
Lastly, as shown in Table \ref{tb:cgm_mass}, we find the total \HI\ mass in the typical dwarf galaxy's CGM to be $M_{\rm HI}=10^{4.7}~\msun$, which means the hydrogen ionization fraction is $f_{\rm HI, Model A}=M_{\rm CGM, HI}/M_{\rm CGM, H}\sim 3\times10^{-4}$.

When considering the metals, the total carbon and silicon masses in the CGM as predicted by Model A are $M_{\rm CGM, C}\approx10^{5.3}~\msun$ and $M_{\rm CGM, Si}\approx10^{4.8}~\msun$. Using global chemical yields inferred from the EAGLE simulation (see \S\ref{sec:sims_eagle}), we estimate that the total amount of carbon (silicon) yielded from star formation (i.e., Type II Supernova, AGB stars) is $y_{\rm C}\sim1.0$\% ($y_{\rm Si}\sim0.32$\%) of the present-day stellar mass\footnote{We do not calculate the stellar yields from dwarf galaxies that have lower metallicities and later star formation histories than the typical galaxy; therefore this represents only a rough calculation.}. Therefore, the total amount of carbon (silicon), including those still locked in stars, in the ISM, CGM and beyond is $M_{\rm tot, C}=\mstar\times y_{\rm C}\approx10^{6.3}~\msun$ ($M_{\rm tot, Si}=\mstar \times y_{\rm Si}\approx10^{5.8}~\msun$). This indicates that only $\sim$10\% ($=M_{\rm CGM, C}/M_{\rm tot, C}$) of the total amount of carbon still resides in the CGM in the cool $T=10^4$ K phase. And a similar CGM mass fraction is found for silicon. Our estimates on the total metal mass fraction in the CGM is consistent with simulations of dwarf galaxies, which generally find that the CGM for dwarfs at $\mstar\sim10^{7-9.5}~\msun$ retain $\sim$10--40\% of the metals generated by stars \citep{muratov17, christensen18}.

\begin{deluxetable}{lccccc}[t]
\tabletypesize{\footnotesize}
\label{tb:cgm_mass}
\tablecaption{CGM Ion Masses within $\langle \rvirm \rangle$}
\tablehead{
\colhead{Ion} & \colhead{Model A} & \colhead{EAGLE} & \colhead{\citetalias{bordoloi14}} & \colhead{\citetalias{johnson17}} & \colhead{\citetalias{tchernyshyov22}} \\
\colhead{} & \colhead{($\log\msun$)} & \colhead{($\log\msun$)} & \colhead{($\log\msun$)} & \colhead{($\log\msun$)} & \colhead{($\log\msun$)} 
}
\startdata 
\hline 
\HI\    & $4.7$ & ($5.1$, $6.3$, $7.7$) & $-$          & $-$           & $-$ \\ 
\SiII\  & $1.3$ & ($1.1$, $3.4$, $5.2$) & $-$          & $\lesssim4$   & $-$ \\ 
\SiIII\ & $2.5$ & ($1.9$, $3.6$, $4.9$) & $-$          & $4.4$         & $-$ \\
\SiIV\  & $3.1$ & ($1.7$, $3.0$, $4.2$) & $-$          & $\lesssim4.4$ & $-$ \\
\CII\   & $2.6$ & ($1.9$, $4.0$, $5.7$) & $-$          & $-$           & $-$ \\
\CIV\   & $4.5$ & ($3.2$, $4.2$, $5.0$) & $5.3$ & $4.8$         & $-$\\
\OVI\   & $5.1$ & ($4.1$, $4.8$, $5.5$) & $-$          & $5.8$         & $5.7$ \\ 
\hline
\enddata
\tablenotetext{}{Note: Total ion masses in logarithmic values. For Model A (\S\ref{sec:empirical_mass_inference}), the ion masses are estimated for the cool phase at $T\approx10^4$ K. For the EAGLE simulation (\S\ref{sec:sims_eagle}), the ion masses are for all phases (cool+warm, Fig. \ref{fig:eagle-phase}), and the three values are for the three halo mass bins, $\mhalom=10^{10.0-10.5}$, $10^{10.5-11.0}$, $10^{11.0-11.5}~\msun$, respectively. We also include mass estimates from \citetalias{bordoloi14} \citep{bordoloi14}, \citetalias{johnson17} \citep{johnson17}, and \citetalias{tchernyshyov22} \citep{tchernyshyov22} when available; their masses have been rescaled to our adopted median $\medrvir$ value, which we elaborate in \S\ref{sec:discuss_mass}. All masses are based on a cylindrical geometry to be consistent with observational values derived from ion column densities projected along given lines of sight and integrated over a galaxy's surface area (i.e, $M\propto N_{\rm ion} R^2$). See Fig. \ref{fig:compare_ion_mass} for a comparison among these values. 
}
\end{deluxetable}


When considering that less than 5--10\% of metals are retained in the ISM and stars of dwarf galaxies \citep{kirby11, kirby13, mcquinn15b, zheng19b}, our CGM mass estimate suggests that dwarf galaxies at present day only retain a total of $\sim$15--20\% of metals in their stars, ISM, and CGM (cool phase). The remaining $\sim$80--85\% either has been transported to the IGM or is in more ionized states that are not available in our study of cool ions (but 
 see \citetalias{johnson17} and \citealt{tchernyshyov22} for \OVI\ observations).


\section{Examining a Volume-Limited Sample of Dwarfs from the EAGLE Simulation}
\label{sec:sims_eagle}

The empirical modeling as described in \S\ref{sec:theory} provides a straightforward way to parameterize the gas density distributions and ionization states in dwarf galaxies' CGM. In this section we use a volume-limited sample of dwarf galaxies from the EAGLE Simulation Project \citep{schaye15} to briefly explore how the inclusion of feedback processes may impact the CGM (and subsequently variation of gas temperature as the galaxy evolves).

\begin{figure*}[t]
\centering
\includegraphics[width=\columnwidth]{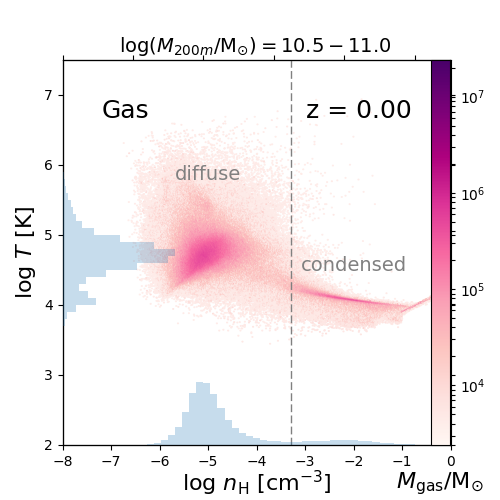}
\includegraphics[width=\columnwidth]{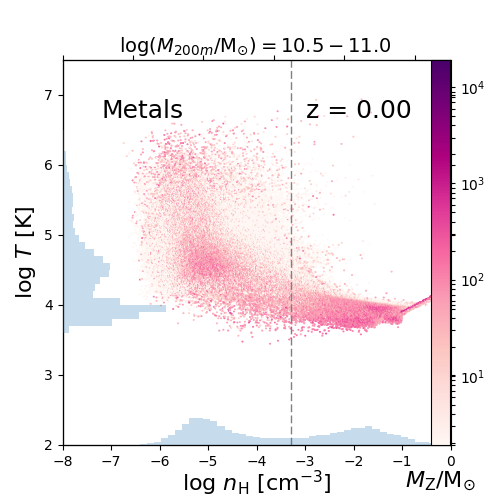}
\caption{Phase diagrams (\nh\ versus $T$) for gas (left panel) and metals (right panel) in the CGM of dwarf galaxies from the EAGLE simulation (see \S\ref{sec:sims_eagle}). The values in the phase diagrams are averaged over the CGM of 53 dwarf galaxies with halo masses of $\mhalom=10^{10.5-11.0}~\msun$, encompassing the medium halo mass of the dwarf galaxies in our observational sample. The color bar in the left panel indicates the masses of gas particles, while the color bar in the right panel shows the corresponding metal masses. The inset blue histograms on the x and y axes show the marginalized mass-weighted density \nh\ and temperature distributions, respectively. Overall, we find bimodal distributions in both the gas and metals in dwarf galaxies' CGM. 
\label{fig:eagle-phase}
}
\end{figure*}

Modern cosmological hydrodynamical simulations tune their feedback prescriptions to reproduce an array of observed properties, mainly regarding galaxies as opposed to cosmic gas reservoirs (i.e., CGM or IGM). The EAGLE Simulation Project tunes their stellar and supermassive black hole feedback prescriptions to reproduce the observed galactic stellar mass function as well as several other galactic properties \citep{crain15}. At the masses of dwarf galaxies, the main tuned constraint is the slope of the galactic stellar mass function, which has an observed slope of $dn/d\mstar \sim \mstar^{-1.4}$ \citep{baldry12} at $\mstar=10^{8-9}~\msun$. Given that the slope of the dark matter halo mass function in cold dark matter ($\Lambda$CDM) cosmology is proportional to $\mstar^{-2}$, this implies a steady decline in the efficiency of gaseous baryons being converted to stars going down the mass scale.  In fact, even before the establishment of CDM as the baseline cosmology, observations of declining dwarf galaxy surface brightnesses and metallicities toward lower masses \citep[e.g.,][]{skillman89, tremonti04, lee06, andrews13, kirby13} motivated theoretical arguments that dwarf galaxies are different from their more-massive counterparts and that gas loss via stellar-driven superwinds may be required \citep[e.g., ][]{dekel86, sales22}.

Our estimates on the baryonic and metal contents of dwarf galaxies' CGM (\S\ref{sec:empirical_mass_inference}) also support the idea that outflows driven by star formation activities are efficient at transporting gas mass and metal mass out of dwarf galaxies and into the CGM and beyond. In fact, our mass estimates find that the cool phase of the CGM ($T\approx10^4$ K) surrounding dwarf galaxies only contains $\sim10$\% of the metals that were ever produced. To aid in the physical interpretation of these CGM reservoirs, we use an EAGLE high-resolution (12.5 Mpc)$^3$ simulation volume that follows nonequilibrium ionization and cooling in diffuse gas. The simulation, introduced in \citet{oppenheimer18c}, follows $376^3$ fluid and dark matter particles with a gas mass resolution of $2.2\times10^5\;{\rm M}_{\odot}$. The nonequilibrium module \citep{richings14a} tracks 136 ionization states across 11 elements, but is found not to deviate significantly from equilibrium assumptions when assuming a constant UV background \citep{oppenheimer18c}. As an example, for silicon, the simulation self-consistently follows all 15 ion states, despite us observing only three states available in UV (\SiII, \SiIII, \& \SiIV).

\begin{figure*}[t]
\includegraphics[width=\textwidth]{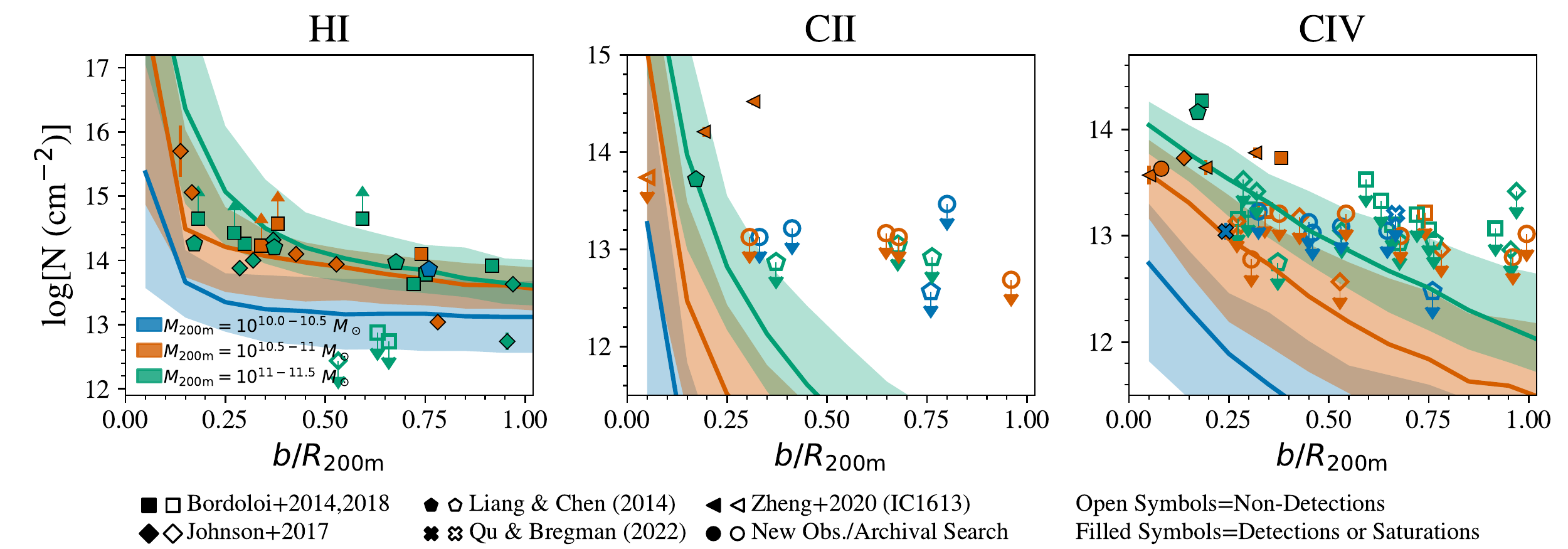}
\caption{Comparison between observed ion column density profiles and predicted curves from the EAGLE simulation. Data symbols are the same as in Figure \ref{fig:observed_logN_b}. The solid lines in each panel indicate the median values, while the patches enclose the 16th-84th percentile distribution. Both observations and simulations are color-coded into three halo mass bins with $\mhalom=10^{10-10.5}$, $10^{10.5-11}$, $10^{11-11.5}~\msun$. Overall, the EAGLE simulation predicts \HI\ column densities consistent with the observed values, while for other low-to-intermediate ions (\CII, \CIV, \SiII, \SiIII, and \SiIV) the simulation predicts almost no detectable metal ion absorbers in the cool CGM of dwarf galaxies except in the innermost impact parameter. See \S\ref{sec:sims_eagle} for more details.
\label{fig:eagle_logN_b}}
\end{figure*}

Since most of our observed dwarf galaxies are in relatively isolated environments without massive nearby halos (see \S\ref{sec:data}), we select from the simulation dwarf galaxies that are centrals of their halos, and that have no contaminating galaxies within impact parameters of 150 kpc with $\mstar$ of greater than 10\% of the targeted galaxy's. The selected galaxy halos are projected along three Cartesian axes ($x$, $y$, $z$), discarding any axis with contaminating galaxies in their projected CGM. This selection is also effective at discarding dwarf galaxies in denser environments outside the virial radii of more-massive galaxies. A total of 207 simulated galaxies and 445 projections are included in the following analysis. Dividing the sample into the same three halo mass bins as shown in Figure \ref{fig:observed_logN_b}, we examine 126 halos with $\mhalom=10^{10.0-10.5}~\msun$, 53 halos with $10^{10.5-11.0}~\msun$, and 28 halos with $10^{11.0-11.5}~\msun$.

We first examine the phase diagrams ($n_{\rm H}$ versus $T$) of gas particles and metals in the CGM of the selected EAGLE dwarf galaxies. While the $n_{\rm H}$ versus $T$ distributions of individual halos vary from low to high halo masses, collectively we find that over the mass range of dwarf galaxies that we investigate, the CGM gas and metals predominately reside at a temperature of $T\sim10^{4-5}$ K, close to the virial temperature at the corresponding halo mass. In Figure \ref{fig:eagle-phase}, we show the averaged phase diagrams of gas (left) and metals (right) for a sample of 53 EAGLE dwarf galaxies with $\mhalom=10^{10.5-11.0}~\msun$, encompassing the median halo mass ($\langle\mhalom\rangle=10^{10.9}~\msun$) of our observational sample (i.e., the Full Sample). Overall, both the gas and metals in the CGM show bimodal distributions with a warm ``diffuse'' phase at $T\sim 10^{4.5-4.8}$ K and gas densities peaking at $n_{\rm H}\sim 10^{-5}~\rm cm^{-3}$, and a cool ``condensed'' phase with $T\sim 10^{4.0}$ K and $n_{\rm H}\sim 10^{-3}-10^{-1}~\rm cm^{-3}$. The CGM mass in the warm diffuse phase is more than seven times higher than that in the cool condensed phase; however, the metal mass is more equitably distributed between the two phases. When compared to the empirical model (\S\ref{sec:theory}), the cool condensed phase in EAGLE is similar in density to Model B, while the warm diffuse phase is more similar to that of Model A with $n_{\rm H}\sim10^{-5}$ cm$^{-3}$. However, when we consider the gas metallicity, the cool condensed phase is found with $Z\sim0.2~Z_\odot$, while the warm diffuse phase is more metal-poor with $Z\sim0.04~Z_\odot$. This difference highlights the fact that while the simulation produces a multiphase CGM, the empirical model presented here is constructed for a cool phase with $T\approx10^4$ K. In a follow-up study, we will extend the empirical model to describe the multiphase CGM of dwarf galaxies.

We then examine the projected column densities of \HI\ and low-to-intermediate ions as a function of impact parameter for the selected dwarf galaxies over the same three halo mass bins in Figure \ref{fig:eagle_logN_b}. For each halo mass bin, we show in solid lines the 50th percentile $\log N_{\rm ion}$ value at a given $\impactpara$ from all available dwarf halos, while the shaded regions encompass the 16th-84th percentile ranges. Here we focus on the column density profiles of \HI, \CII, and \CIV, but note that \SiII, \SiIII, and \SiIV\ exhibit similar radial profiles. We also note that while we are selecting based on halo masses, the mean stellar masses in the three halo mass bins are $10^{6.8}$, $10^{8.0}$, and $10^{8.9}~\msun$, which is in agreement with the SMHM relation from \cite{munshi21} that we adopted in \S\ref{sec:galaxy_property}.

Overall, the left panel in Figure \ref{fig:eagle_logN_b} shows that the EAGLE simulation reproduces well both the profile shape and the magnitudes of the \HI\ column densities. Although in observations we do not find obvious $\log N_{\rm HI}$ variations among the three halo mass bins at a given $\impactpara$, the simulations show that lower-mass halos tend to have lower $\log N_{\rm HI}$, especially at small impact parameters. This is not surprising given the smaller baryonic mass reservoirs available to lower-mass halos. We also note that even though the simulation matches the overall distribution of the observed \HI, it appears unable to fully reproduce the scatter in $\log N_{\rm HI}$. Specifically, when considering the lower and upper limits, the $\log N_{\rm HI}$ dispersion in the high (green) mass bin is about a factor of 2 higher than the simulated values at $\gtrsim0.5b/\rvir$. The lack of mass dependence in the observed data points and the larger scatters are likely due to the small sample size, while in the simulation we combine column density profiles over a large sample of dwarf galaxy halos.

In the middle and right panels of Figure \ref{fig:eagle_logN_b}, we further compare the simulated \CII\ and \CIV\ column density profiles to their observed counterparts, respectively. For both \CII\ and \CIV, the simulation slightly underestimates the ion column densities inside $\sim0.5\rvir$. At $b\gtrsim0.5\rvir$, the simulation predicts column density values either consistent with or lower than the upper limits. Similar trends can be seen when examining \SiII, \SiIII, and \SiIV. Overall, the simulation agrees with the observations (as well as the empirical model in \S\ref{sec:theory}) that except in the innermost impact parameter ($\lesssim0.5\rvir$) or in the halos of higher-mass dwarf galaxies (e.g., $\mhalom=10^{11-11.5}~\msun$), there exist almost no detectable metal absorbers in the cool CGM of dwarf galaxies over the mass range we probe.

\begin{figure}
\includegraphics[width=\columnwidth]{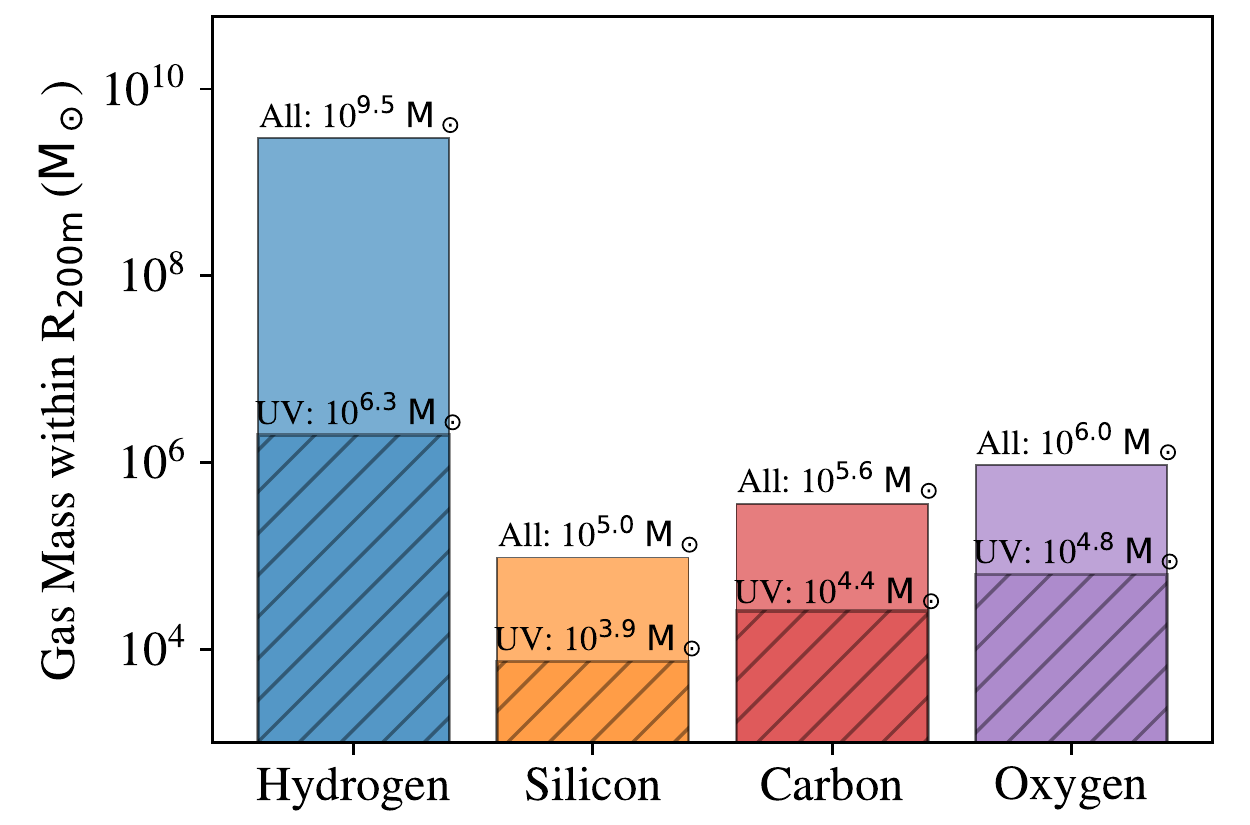}
\caption{Element mass observable in UV at $z\sim0-0.3$ (hatched region) compared to the element total mass integrated from 10 kpc to $\rvir$, estimated for EAGLE dwarf galaxies with $\mhalom=10^{10.5-11.0}~\msun$. For hydrogen, ``observable in UV" is defined as gas in the form of \HI\ (Ly$\alpha$). For carbon, it means \CII\ and \CIV; for silicon, it means \SiII, \SiIII, and \SiIV; and for oxygen, it means \OVI. We find that at the median halo mass of our sample, the ions that are observable in UV only trace a small fraction of the corresponding metal mass in the dwarf galaxies' CGM. 
\label{fig:eagle_uv_mass}}
\end{figure}

Another insight from the simulation is the total gas mass that we can directly probe via UV absorption lines at $z\sim0-0.3$, which we show in Figure \ref{fig:eagle_uv_mass}. For each element, we calculate the total ion mass in the CGM available through a set of common UV transition lines, and compare that with the total gas mass of the element. For example, for hydrogen, the observable gas mass in UV ($M_{\rm H, UV}\approx10^{6.3}~\msun$) is mainly detected through \HI\ Ly$\alpha$ absorption when the line is redshifted into the far UV range at appropriate redshifts. This indicates an ionization correction of $f_{\rm HI}\sim M_{\rm H,UV}/M_{\rm H}\sim6\times10^{-4}$, similar to the hydrogen ionization fraction we derive from Model A ($\sim3\times10^{-4}$; \S\ref{sec:empirical_mass_inference}). 


Furthermore, Figure \ref{fig:eagle_uv_mass} implies the existence of a significant metal reservoir that is mostly missed by the ions we survey in UV. For example, for the EAGLE dwarf galaxies with $\mhalom=10^{10.5-11.0}~\msun$, the total mass probed by UV-observable silicon ions (i.e., \SiII, \SiIII, and \SiIV\ combined) accounts for $\sim$8\%\ of the total silicon mass, with the remaining mass in higher-ionization states. This indicates that the EAGLE galaxies' CGM has a warmer/more diffuse component containing higher-ionization silicon species not observed in the UV. For more-massive, $\sim 10^{12}~\msun$ halos, \citet[][their fig. 8]{oppenheimer18c} found that while the low silicon ions dominate the inner CGM silicon content, high ions that are not detected in the UV overwhelmingly dominate the volume beyond 50 kpc. Similarly, carbon has fewer ions, allowing \CII~and \CIV~to trace $\sim$6\% of this element, but a significant gap is missing with \CIII~being unobservable in our survey. The single ion \OVI~available in the far UV traces oxygen with $\sim$6\% of the total mass arising from the primarily photoionized \OVI~in the dwarf regime as discussed by \citet{oppenheimer16}. From the simulation perspective, the higher detection rates of \OVI\ absorbers as observed by \citetalias{johnson17} and \cite{tchernyshyov22} are due to a smoother distribution of this ion arising from a more diffuse phase combined with the higher abundance of oxygen relative to other metals. 

In contrast to the empirical Models A and B, ions in the CGM of the EAGLE dwarf galaxies arise from multiple phases (see Figure \ref{fig:eagle-phase}). \CIV\ originates from the warm diffuse gas with densities primarily below $\nh=10^{-4}~\cmc$, while \CII\ comes from the cool condensed gas with densities above $\nh=10^{-3}~\cmc$, as does \SiII\ and \HI. The simulated \CIV\ appears to reflect the lower-density characteristics of Model A, but \HI, \CII, and \SiII\ would arise from the separate condensed version of Model B that may have higher density in the inner CGM with a steeper radial decline for the volume filling factor. We will explore further the comparisons between empirical models and simulations in future work.

\section{Comparison of Dwarf CGM Mass among Various Sources}
\label{sec:discuss_mass}

\begin{figure}
\includegraphics[width=\columnwidth]{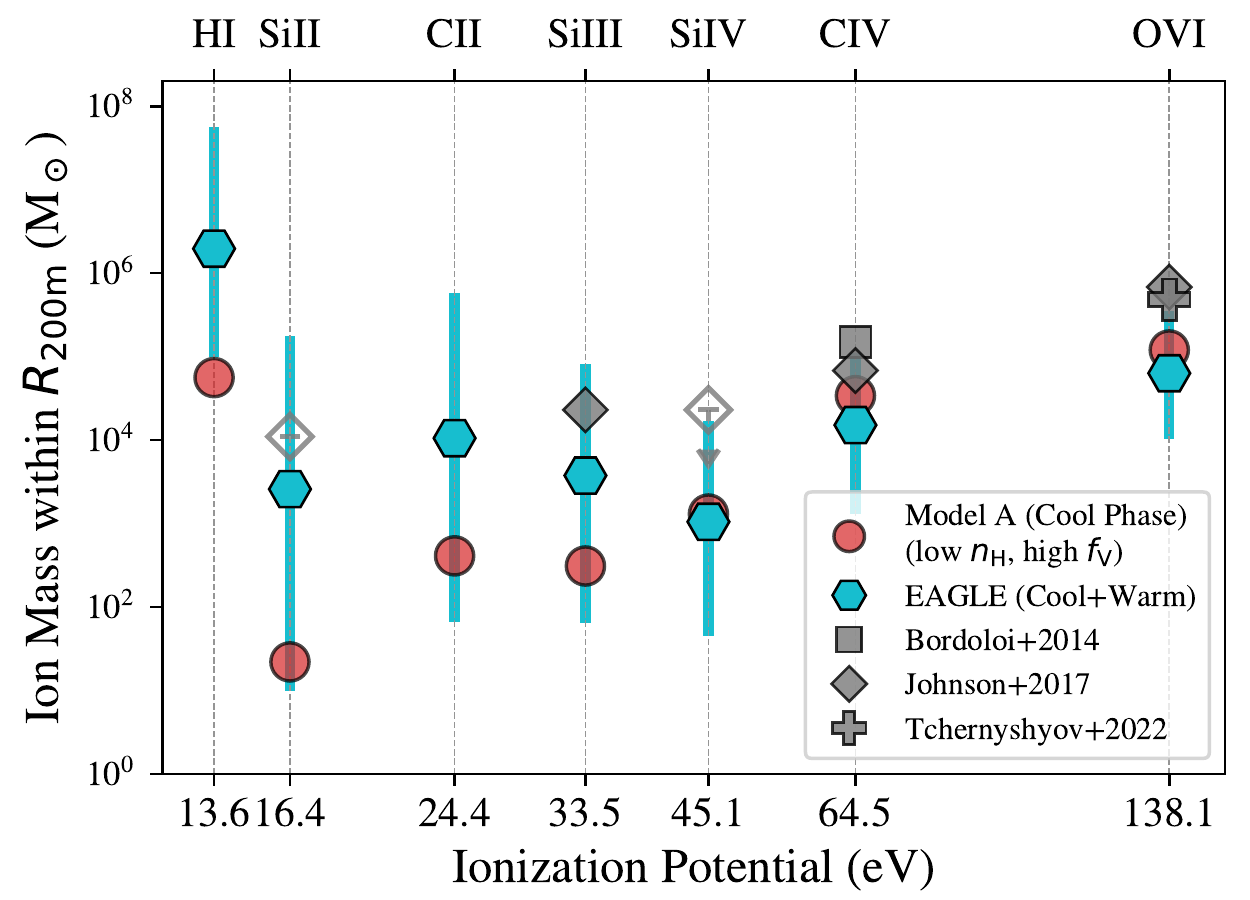}
\caption{CGM ion mass vs.\ ionization potential (with logarithmic scaling) from Model A (red circle; \S\ref{sec:empirical_mass_inference}), the EAGLE simulation (teal hexagon; \S\ref{sec:sims_eagle}), \citetalias{bordoloi14}'s COS-Dwarfs (gray square), \citetalias{johnson17} (gray diamond), and \citet[][gray cross]{tchernyshyov22}. The corresponding mass values can be found in Table \ref{tb:cgm_mass}. For the EAGLE simulation values, the lower and upper bounds show the ion masses for the low ($\mhalom=10^{10.0-10.5}~\msun$) and high ($\mhalom=10^{11.0-11.5}~\msun$) halo mass bins, respectively; while the hexagon symbols indicate the values for the middle halo mass bin ($\mhalom=10^{10.5-11.0}~\msun$). 
The range of ion masses from EAGLE suggest that the CGM gas and metal content in dwarf galaxies may vary with halo masses, which cannot be inferred directly from observations or empirical models given the limitation in sample sizes and absorber detection rates. See more details in \S\ref{sec:discuss_mass}. The ion masses are tabulated in Table \ref{tb:cgm_mass}. 
\label{fig:compare_ion_mass}}
\end{figure}

In the previous Sections, we constructed an empirical model (\S\ref{sec:theory}) and examined a suite of simulated halos from the EAGLE simulation (\S\ref{sec:sims_eagle}) to understand the physical properties of gas and metals in the CGM of dwarf galaxies. Both sections provide CGM gas and metal mass estimates. In the following, we compare these mass estimates with observational constraints from the literature to provide a comprehensive picture on the baryon and metal budgets of dwarf galaxies. 
In Figure \ref{fig:compare_ion_mass}, we compare the total ion masses within $\medrvir$ between this work and previous observational estimates by \citetalias{bordoloi14}, \citetalias{johnson17}, and \cite{tchernyshyov22}. While in this work we only study \HI\ and low-to-intermediate ions (i.e., \CII, \CIV, \SiII, \SiIII, and \SiIV), for completeness we also include the \OVI\ mass estimates for dwarf galaxies of similar masses from \citetalias{johnson17} and \cite{tchernyshyov22}. The ion masses quoted here are also tabulated in Table \ref{tb:cgm_mass}. 

For their low-mass sample ($\mstar=10^{8-9.5}~\msun$), \citetalias{bordoloi14} found a total carbon mass of $\gtrsim0.4\times10^{6}~\msun$ within an impact parameter of 110 kpc, assuming a \CIV\ ionization fraction of $f_{\rm CIV}=0.3$ (see their Table 2). To make a consistent comparison, we convert their carbon mass back to \CIV\ mass as $M_{\rm CIV}=M_{\rm C}\times f_{\rm CIV}(\medrvir/110~{\rm kpc})^2\sim 10^{5.3}~\msun$, where $(\medrvir/110~{\rm kpc})^2$ is used to scale the mass value to our median virial radius. When compared to the estimated \CIV\ mass of $\approx10^{4.5}~\msun$ from Model A (or $10^{4.2}$ from EAGLE's middle halo mass bin), we find that \citetalias{bordoloi14}'s value is roughly a factor of 6 (12 for EAGLE) higher. 

\citetalias{johnson17}'s ion mass values are based on their absorber detections from their galaxies D1 and D2, which indicate a total of $\approx$10$^4~\msun$ in \SiIII, 
3$\times$10$^4~\msun$ for \CIV, 
3$\times$10$^5~\msun$ for \OVI, 
$<$10$^4~\msun$ for \SiIV, 
and $<$5$\times$10$^3~\msun$ for \SiII\ within a virial radius of 90 kpc. 
After scaling their ion mass values from the assumed virial radius of 90 kpc to our median $\medrvir$, we find that their estimated ion masses are generally higher than those from Model A and EAGLE within a factor of a few. 

For completeness, here we also include \OVI\ mass measurements from \cite{tchernyshyov22} that examine the relation between CGM \OVI\ column density and host galaxies' stellar mass, SFR, as well as impact parameters of the absorbers from the galaxies. We adopt \OVI\ halo mass from their lowest mass bin ($\mstar=10^{7.8-8.5}~\msun$) that covers our median stellar mass range. After scaling the mass from $\rvirc$ to $\rvir$, we find a total \OVI\ mass of $\sim10^{5.7}~\msun$. 

Figure \ref{fig:compare_ion_mass} shows that the mass estimates of high ions (i.e., \CIV, \OVI) from previous observations, the empirical Model A, and the EAGLE simulation are generally in agreement with each other within a factor of a few. For weakly ionized species such as \CII, \SiII, \SiIII, and \SiIV, Model A and the EAGLE simulation's low and median halo mass bins both predict much lower values with ion column densities too low to be detected observationally (Figures \ref{fig:empirical_logN_b} and \ref{fig:eagle_logN_b}). It is worth noting that the EAGLE simulation predicts a range of CGM ion masses when different halo mass bins are considered. This indicates that the ion masses (and the corresponding ionization states) in the CGM of dwarf galaxies may vary with their halo masses. However, we note that the comparison in Figure \ref{fig:compare_ion_mass} should be interpreted with caution for the following reasons. 

We reflect on the assumptions made on the CGM of dwarf galaxies when constructing the empirical model in \S\ref{sec:theory}. We note that the model focuses on the cool CGM, and does not address gas in warmer phases. Meanwhile, the EAGLE simulation shows that the warm diffuse gas phase dominates the CGM baryonic mass (Figure \ref{fig:eagle-phase}), and has a significant contribution to the column densities of ions that are highly ionized (e.g., \OVI). However, the relatively low resolution of the EAGLE simulation limits its interpretive power because it has been realized in recent years that a higher resolution in the CGM leads to the production of more gas clouds in cool phases and small sizes \citep{vandevoort19, Peeples19, hummels19, suresh19}. In future work, we will further develop the empirical model to consider dwarf galaxies' CGM with temperature profiles regulated by the extragalactic UV background as well as feedback from star-formation activities in the galaxies. 


\section{Conclusion}
\label{sec:summary}

We investigate the baryonic and metal content in the CGM of a sample of \nunidwarf\ low-mass, isolated, and gas-rich dwarf galaxies within 8 Mpc of the Sun and at $z<0.3$ using {\it HST}/COS. Our sample includes \npairtotal\ dwarf-QSO pairs that cover a unique parameter space of $\impactpara/\rvir=0.05-1.0$ and $\mstar=10^{6.5-9.5}~\msun$ that has rarely been explored in previous work (Figure \ref{fig:logM_rho_all}, Table \ref{tb:literature}, and \S\ref{sec:data}). The median properties of the dwarf galaxies in our Full Sample are $\langle\mstar\rangle=10^{8.3}~\msun$, $\langle\mhalom\rangle=10^{10.9}~\msun$, $\medrvir=\medrvirval$ kpc, $\langle\mhi\rangle=10^{7.4}~\msun$, and $\langle{\rm SFR}\rangle=10^{-1.73}~\msunyr$. The main findings of this work are summarized as follows.

At a sensitivity of EW$\geq$100 m\AA\ at 3$\sigma$, we find ubiquitous detections of \HI\ (via Ly$\alpha$ 1215\AA\ line) in the CGM of dwarf galaxies with a detection rate of \cfHI\ (see \S\ref{sec:result_covering_fraction}). The \HI\ gas is typically detected at a column density of $\log N_{\rm HI}=13-16$. On the other hand, metal ions are generally detected at much lower rates, with \cfCII\ in \CII, \cfCIV\ in \CIV, \cfSiII\ in \SiII,  \cfSiIII\ in \SiIII, and \cfSiIV\ in \SiIV. All of the metal detections occur within $\sim0.5\rvir$, largely consistent with existing literature values. We note that the low ion detection rates occur despite the high-quality QSO sight lines used in this work, some of which reaching S/N$\sim20-100$ (see Table \ref{tb:pair_info}). This suggests that for dwarf galaxies at the low-mass range probed in this study ($\mstar=10^{6.5-9.5}~\msun$), the metals in the galaxies' CGM, especially those at $\impactpara/\rvirm\gtrsim0.5$, may be too diffuse to be detected by {\it HST}/COS. 

We construct an empirical model for the cool phase of the dwarf galaxies' CGM (T$\approx$10$^4$ K; \S\ref{sec:theory}), parameterizing the gas density and volume filling fraction as power-law functions of the radius, and assuming photoionization equilibrium. For the median halo mass $\langle\mhalom\rangle=10^{10.9}~\msun$ in our sample, we present two parameter combinations (Models A and B) that match the observed \HI\ column density profile with different CGM masses and gas densities (Figures~\ref{fig:empirical_model_setup} and \ref{fig:empirical_logN_b}). Assuming a metallicity of $Z'=0.3~Z_\odot$, Model A is more consistent with the measured metal columns, and has a cool gas mass of $\mcgmcool\sim10^{8.4}~\msun$, which accounts for $\sim2$\% of the baryon budget of the median halo mass. When considering metals in the cool CGM, we find a total of $\sim10^{5.3}~\msun$ in carbon and $\sim10^{4.8}~\msun$ in silicon. This corresponds to $\sim$10\% of the metals that have ever been produced throughout the dwarf galaxy's star formation history.

We further examine a volume-limited sample of dwarf galaxies in the EAGLE simulation to understand how dwarf galaxies' CGM may be impacted when considering feedback processes (\S\ref{sec:sims_eagle}). In general, we find that the selected EAGLE dwarf galaxies are able to reproduce the observed \HI\ and ion column density profiles. When considering the mass distribution in different phases, we find the EAGLE dwarf galaxies' CGM is dominated by a more diffuse and warmer ($T\sim10^{4.5-4.8}$ K; Figure \ref{fig:eagle-phase}) phase, and that only $\sim$6\%--8\% of the element masses (e.g., silicon, carbon, and oxygen) can be observed in UV at $z\sim0-0.3$. The remaining element masses are mainly in high-ionization states that are often not available in UV (Figure \ref{fig:eagle_uv_mass}). However, this conclusion is tempered by the low resolution of the EAGLE simulation, which might artificially suppress the production of cool gas.

Lastly, we compare the ion mass estimates from the empirical Model A and the EAGLE simulation to observational values from the literature in Figure \ref{fig:compare_ion_mass}. In general, we find good agreement in masses for high ions \CIV\ and \OVI\ (within a factor of a few). On the other hand, Model A and the EAGLE simulation predict much lower mass values for \SiII, \SiIII, \SiIV, and \CII, which are too diffuse to be easily detected with {\it HST}/COS.

Much work remains to be done to fully understand the CGM of dwarf galaxies from both observational and theoretical perspectives. Overall, our analyses presented in this paper suggest that: (1) dwarf galaxies' CGM only harbors $\sim10$\% of the metals in the cool $T\approx10^4$ K phase, with the rest either in higher-ionization states in the CGM or have been lost to the IGM; (2) at the current sensitivity of {\it HST}/COS which is the prime instrument for UV absorption-line studies at $z\sim0$, only the inner CGM of dwarf galaxies can be well probed; (3) a larger dwarf galaxy sample size, especially at $\mstar<10^8~\msun$, is needed to better illustrate how CGM properties scale with host galaxy properties; and (4) more sophisticated empirical models as well as dwarf galaxy simulations with higher resolution are necessary to better understand the physical processes governing the CGM of dwarf galaxies.

\acknowledgments
Y.Z. thanks D. Weisz for his advice and mentorship, and thanks her office mate, A. Savino, and many staff at UC Berkeley and Miller Institute for their support during the completion of this manuscript. Y.Z. thanks Z.J. Qu for providing updated measurements on their QSO sight lines adopted in this work, thanks R. Bordoloi for sharing \HI\ Ly$\alpha$ measurements, and thanks K. Tchernyshyov and N. Sandford for discussions on treatment of censored data using \texttt{PyMC3}. This research has made use of NASA's Astrophysics Data System. This work is based on observations made
with the NASA/ESA Hubble Space Telescope (program
ID: No. 16301, No. 15156, and No. 15227). Support for HST-GO-16301, HST-GO-15156, and HST-GO-15227 was provided
by NASA through a grant from the Space Telescope Science Institute (STScI). STScI is operated by the Association of Universities for Research in Astronomy, Inc.,
under NASA contract NAS5-26555. E.N.K. acknowledges support from the National Science Foundation under grant No.\ AST-2233781. Y.F. acknowledges support from NASA award 19-ATP19-0023 and NSF award AST-2007012. J.K.W. acknowledges support from NSF-CAREER 2044303. The revision of this manuscript was performed in part at Aspen Center for Physics, which is supported by National Science Foundation grant PHY-2210452. This research has made use of the HSLA database, developed and maintained at STScI, Baltimore, USA.

\facilities{Hubble Space Telescope/Cosmic Origins Spectrograph; Mikulski Archive for Space Telescopes; The Parkes telescope is part of the Australia Telescope which is funded by the Commonwealth of Australia for operation as a National Facility managed by CSIRO.}

\software{Astropy \citep{astropy2}, Numpy \citep{numpy}, Matplotlib \citep{matplotlib}, CLOUDY \citep{Ferland17}, IDL, PyMC3 \citep{pymc3}.}

Data Availability: {\it HST}/COS spectra can be found on \href{https://archive.stsci.edu/missions-and-data/hsla/hsla-target-tables}{HSLA} and in MAST: \dataset[10.17909/ve0k-ps78]{http://dx.doi.org/10.17909/ve0k-ps78}. Measurements and galaxy properties used in this work (as well as the broader literature search) and relevant codes can be found at: \href{https://github.com/yzhenggit/zheng_dwarfcgm_survey.git}{yzhenggit/zheng\_dwarfcgm\_survey}. 


\bibliographystyle{aasjournal}
\bibliography{main}

\begin{thebibliography}{}
\expandafter\ifx\csname natexlab\endcsname\relax\def\natexlab#1{#1}\fi
\providecommand{\url}[1]{\href{#1}{#1}}
\providecommand{\dodoi}[1]{doi:~\href{http://doi.org/#1}{\nolinkurl{#1}}}
\providecommand{\doeprint}[1]{\href{http://ascl.net/#1}{\nolinkurl{http://ascl.net/#1}}}
\providecommand{\doarXiv}[1]{\href{https://arxiv.org/abs/#1}{\nolinkurl{https://arxiv.org/abs/#1}}}

\bibitem[{{Agertz} {et~al.}(2020){Agertz}, {Pontzen}, {Read}, {Rey}, {Orkney},
  {Rosdahl}, {Teyssier}, {Verbeke}, {Kretschmer}, \& {Nickerson}}]{agertz20}
{Agertz}, O., {Pontzen}, A., {Read}, J.~I., {et~al.} 2020, \mnras, 491, 1656,
  \dodoi{10.1093/mnras/stz3053}

\bibitem[{{Andersson} {et~al.}(2023){Andersson}, {Agertz}, {Renaud}, \&
  {Teyssier}}]{andersson22}
{Andersson}, E.~P., {Agertz}, O., {Renaud}, F., \& {Teyssier}, R. 2023, \mnras,
  521, 2196, \dodoi{10.1093/mnras/stad692}

\bibitem[{{Andrews} \& {Martini}(2013)}]{andrews13}
{Andrews}, B.~H., \& {Martini}, P. 2013, \apj, 765, 140,
  \dodoi{10.1088/0004-637X/765/2/140}

\bibitem[{{Baldry} {et~al.}(2012){Baldry}, {Driver}, {Loveday}, {Taylor},
  {Kelvin}, {Liske}, {Norberg}, {Robotham}, {Brough}, {Hopkins}, {Bamford},
  {Peacock}, {Bland-Hawthorn}, {Conselice}, {Croom}, {Jones}, {Parkinson},
  {Popescu}, {Prescott}, {Sharp}, \& {Tuffs}}]{baldry12}
{Baldry}, I.~K., {Driver}, S.~P., {Loveday}, J., {et~al.} 2012, \mnras, 421,
  621, \dodoi{10.1111/j.1365-2966.2012.20340.x}

\bibitem[{{Barnes} {et~al.}(2001){Barnes}, {Staveley-Smith}, {de Blok},
  {Oosterloo}, {Stewart}, {Wright}, {Banks}, {Bhathal}, {Boyce}, {Calabretta},
  {Disney}, {Drinkwater}, {Ekers}, {Freeman}, {Gibson}, {Green}, {Haynes}, {te
  Lintel Hekkert}, {Henning}, {Jerjen}, {Juraszek}, {Kesteven}, {Kilborn},
  {Knezek}, {Koribalski}, {Kraan-Korteweg}, {Malin}, {Marquarding}, {Minchin},
  {Mould}, {Price}, {Putman}, {Ryder}, {Sadler}, {Schr{\"o}der}, {Stootman},
  {Webster}, {Wilson}, \& {Ye}}]{hipass_data}
{Barnes}, D.~G., {Staveley-Smith}, L., {de Blok}, W.~J.~G., {et~al.} 2001,
  \mnras, 322, 486, \dodoi{10.1046/j.1365-8711.2001.04102.x}

\bibitem[{{Begum} {et~al.}(2008){Begum}, {Chengalur}, {Karachentsev},
  {Sharina}, \& {Kaisin}}]{begum08}
{Begum}, A., {Chengalur}, J.~N., {Karachentsev}, I.~D., {Sharina}, M.~E., \&
  {Kaisin}, S.~S. 2008, \mnras, 386, 1667,
  \dodoi{10.1111/j.1365-2966.2008.13150.x}

\bibitem[{{Bordoloi} {et~al.}(2018){Bordoloi}, {Prochaska}, {Tumlinson},
  {Werk}, {Tripp}, \& {Burchett}}]{bordoloi18}
{Bordoloi}, R., {Prochaska}, J.~X., {Tumlinson}, J., {et~al.} 2018, \apj, 864,
  132, \dodoi{10.3847/1538-4357/aad8ac}

\bibitem[{{Bordoloi} {et~al.}(2014){Bordoloi}, {Tumlinson}, {Werk},
  {Oppenheimer}, {Peeples}, {Prochaska}, {Tripp}, {Katz}, {Dav{\'e}}, {Fox},
  {Thom}, {Ford}, {Weinberg}, {Burchett}, \& {Kollmeier}}]{bordoloi14}
{Bordoloi}, R., {Tumlinson}, J., {Werk}, J.~K., {et~al.} 2014, \apj, 796, 136,
  \dodoi{10.1088/0004-637X/796/2/136}

\bibitem[{{Burchett} {et~al.}(2016){Burchett}, {Tripp}, {Bordoloi}, {Werk},
  {Prochaska}, {Tumlinson}, {Willmer}, {O'Meara}, \& {Katz}}]{burchett16}
{Burchett}, J.~N., {Tripp}, T.~M., {Bordoloi}, R., {et~al.} 2016, \apj, 832,
  124, \dodoi{10.3847/0004-637X/832/2/124}

\bibitem[{{Chabrier}(2003)}]{chabrier03}
{Chabrier}, G. 2003, \pasp, 115, 763, \dodoi{10.1086/376392}

\bibitem[{{Christensen} {et~al.}(2018){Christensen}, {Dav{\'e}}, {Brooks},
  {Quinn}, \& {Shen}}]{christensen18}
{Christensen}, C.~R., {Dav{\'e}}, R., {Brooks}, A., {Quinn}, T., \& {Shen}, S.
  2018, \apj, 867, 142, \dodoi{10.3847/1538-4357/aae374}

\bibitem[{{Christensen} {et~al.}(2016){Christensen}, {Dav{\'e}}, {Governato},
  {Pontzen}, {Brooks}, {Munshi}, {Quinn}, \& {Wadsley}}]{christensen16}
{Christensen}, C.~R., {Dav{\'e}}, R., {Governato}, F., {et~al.} 2016, \apj,
  824, 57, \dodoi{10.3847/0004-637X/824/1/57}

\bibitem[{{Cook} {et~al.}(2014){Cook}, {Dale}, {Johnson}, {Van Zee}, {Lee},
  {Kennicutt}, {Calzetti}, {Staudaher}, \& {Engelbracht}}]{Cook14}
{Cook}, D.~O., {Dale}, D.~A., {Johnson}, B.~D., {et~al.} 2014, \mnras, 445,
  899, \dodoi{10.1093/mnras/stu1787}

\bibitem[{{Corbelli}(2003)}]{corbelli03}
{Corbelli}, E. 2003, \mnras, 342, 199, \dodoi{10.1046/j.1365-8711.2003.06531.x}

\bibitem[{{Crain} {et~al.}(2015){Crain}, {Schaye}, {Bower}, {Furlong},
  {Schaller}, {Theuns}, {Dalla Vecchia}, {Frenk}, {McCarthy}, {Helly},
  {Jenkins}, {Rosas-Guevara}, {White}, \& {Trayford}}]{crain15}
{Crain}, R.~A., {Schaye}, J., {Bower}, R.~G., {et~al.} 2015, \mnras, 450, 1937,
  \dodoi{10.1093/mnras/stv725}

\bibitem[{{Dalcanton} {et~al.}(2009){Dalcanton}, {Williams}, {Seth}, {Dolphin},
  {Holtzman}, {Rosema}, {Skillman}, {Cole}, {Girardi}, {Gogarten},
  {Karachentsev}, {Olsen}, {Weisz}, {Christensen}, {Freeman}, {Gilbert},
  {Gallart}, {Harris}, {Hodge}, {de Jong}, {Karachentseva}, {Mateo}, {Stetson},
  {Tavarez}, {Zaritsky}, {Governato}, \& {Quinn}}]{dalcanton09}
{Dalcanton}, J.~J., {Williams}, B.~F., {Seth}, A.~C., {et~al.} 2009, \apjs,
  183, 67, \dodoi{10.1088/0067-0049/183/1/67}

\bibitem[{{Dale} {et~al.}(2009){Dale}, {Cohen}, {Johnson}, {Schuster},
  {Calzetti}, {Engelbracht}, {Gil de Paz}, {Kennicutt}, {Lee}, {Begum},
  {Block}, {Dalcanton}, {Funes}, {Gordon}, {Johnson}, {Marble}, {Sakai},
  {Skillman}, {van Zee}, {Walter}, {Weisz}, {Williams}, {Wu}, \& {Wu}}]{dale09}
{Dale}, D.~A., {Cohen}, S.~A., {Johnson}, L.~C., {et~al.} 2009, \apj, 703, 517,
  \dodoi{10.1088/0004-637X/703/1/517}

\bibitem[{{Danforth} {et~al.}(2010){Danforth}, {Keeney}, {Stocke}, {Shull}, \&
  {Yao}}]{danforth10}
{Danforth}, C.~W., {Keeney}, B.~A., {Stocke}, J.~T., {Shull}, J.~M., \& {Yao},
  Y. 2010, \apj, 720, 976, \dodoi{10.1088/0004-637X/720/1/976}

\bibitem[{{Dekel} \& {Silk}(1986)}]{dekel86}
{Dekel}, A., \& {Silk}, J. 1986, \apj, 303, 39, \dodoi{10.1086/164050}

\bibitem[{{Faerman} \& {Werk}(2023)}]{fw23}
{Faerman}, Y., \& {Werk}, J.~K. 2023, \apj, 956, 92,
  \dodoi{10.3847/1538-4357/acf217}

\bibitem[{{Ferland} {et~al.}(2017){Ferland}, {Chatzikos}, {Guzm{\'a}n},
  {Lykins}, {van Hoof}, {Williams}, {Abel}, {Badnell}, {Keenan}, {Porter}, \&
  {Stancil}}]{Ferland17}
{Ferland}, G.~J., {Chatzikos}, M., {Guzm{\'a}n}, F., {et~al.} 2017, \rmxaa, 53,
  385.
\newblock \doarXiv{1705.10877}

\bibitem[{{Gallazzi} {et~al.}(2005){Gallazzi}, {Charlot}, {Brinchmann},
  {White}, \& {Tremonti}}]{gallazzi05}
{Gallazzi}, A., {Charlot}, S., {Brinchmann}, J., {White}, S. D.~M., \&
  {Tremonti}, C.~A. 2005, \mnras, 362, 41,
  \dodoi{10.1111/j.1365-2966.2005.09321.x}

\bibitem[{{Garrison-Kimmel} {et~al.}(2014){Garrison-Kimmel}, {Boylan-Kolchin},
  {Bullock}, \& {Lee}}]{garrison-kimmel14}
{Garrison-Kimmel}, S., {Boylan-Kolchin}, M., {Bullock}, J.~S., \& {Lee}, K.
  2014, \mnras, 438, 2578, \dodoi{10.1093/mnras/stt2377}

\bibitem[{{Garrison-Kimmel} {et~al.}(2017){Garrison-Kimmel}, {Bullock},
  {Boylan-Kolchin}, \& {Bardwell}}]{garrison-kimmel17}
{Garrison-Kimmel}, S., {Bullock}, J.~S., {Boylan-Kolchin}, M., \& {Bardwell},
  E. 2017, \mnras, 464, 3108, \dodoi{10.1093/mnras/stw2564}

\bibitem[{{Haardt} \& {Madau}(2012)}]{haardt12}
{Haardt}, F., \& {Madau}, P. 2012, \apj, 746, 125,
  \dodoi{10.1088/0004-637X/746/2/125}

\bibitem[{Harris {et~al.}(2020)Harris, Millman, van~der Walt, Gommers,
  Virtanen, Cournapeau, Wieser, Taylor, Berg, Smith, Kern, Picus, Hoyer, van
  Kerkwijk, Brett, Haldane, del R{'{\i}}o, Wiebe, Peterson,
  G{'{e}}rard-Marchant, Sheppard, Reddy, Weckesser, Abbasi, Gohlke, \&
  Oliphant}]{numpy}
Harris, C.~R., Millman, K.~J., van~der Walt, S.~J., {et~al.} 2020, Nature, 585,
  357, \dodoi{10.1038/s41586-020-2649-2}

\bibitem[{{Haynes} {et~al.}(2018){Haynes}, {Giovanelli}, {Kent}, {Adams},
  {Balonek}, {Craig}, {Fertig}, {Finn}, {Giovanardi}, {Hallenbeck}, {Hess},
  {Hoffman}, {Huang}, {Jones}, {Koopmann}, {Kornreich}, {Leisman}, {Miller},
  {Moorman}, {O'Connor}, {O'Donoghue}, {Papastergis}, {Troischt}, {Stark}, \&
  {Xiao}}]{Haynes18}
{Haynes}, M.~P., {Giovanelli}, R., {Kent}, B.~R., {et~al.} 2018, \apj, 861, 49,
  \dodoi{10.3847/1538-4357/aac956}

\bibitem[{{HI4PI Collaboration} {et~al.}(2016){HI4PI Collaboration}, {Ben
  Bekhti}, {Fl{\"o}er}, {Keller}, {Kerp}, {Lenz}, {Winkel}, {Bailin},
  {Calabretta}, {Dedes}, {Ford}, {Gibson}, {Haud}, {Janowiecki}, {Kalberla},
  {Lockman}, {McClure-Griffiths}, {Murphy}, {Nakanishi}, {Pisano}, \&
  {Staveley-Smith}}]{hi4pi16}
{HI4PI Collaboration}, {Ben Bekhti}, N., {Fl{\"o}er}, L., {et~al.} 2016, \aap,
  594, A116, \dodoi{10.1051/0004-6361/201629178}

\bibitem[{{Hummels} {et~al.}(2019){Hummels}, {Smith}, {Hopkins}, {O'Shea},
  {Silvia}, {Werk}, {Lehner}, {Wise}, {Collins}, \& {Butsky}}]{hummels19}
{Hummels}, C.~B., {Smith}, B.~D., {Hopkins}, P.~F., {et~al.} 2019, \apj, 882,
  156, \dodoi{10.3847/1538-4357/ab378f}

\bibitem[{{Hunter} {et~al.}(2012){Hunter}, {Ficut-Vicas}, {Ashley}, {Brinks},
  {Cigan}, {Elmegreen}, {Heesen}, {Herrmann}, {Johnson}, {Oh}, {Rupen},
  {Schruba}, {Simpson}, {Walter}, {Westpfahl}, {Young}, \& {Zhang}}]{hunter12}
{Hunter}, D.~A., {Ficut-Vicas}, D., {Ashley}, T., {et~al.} 2012, \aj, 144, 134,
  \dodoi{10.1088/0004-6256/144/5/134}

\bibitem[{Hunter(2007)}]{matplotlib}
Hunter, J.~D. 2007, Computing in Science \& Engineering, 9, 90,
  \dodoi{10.1109/MCSE.2007.55}

\bibitem[{{Jacobs} {et~al.}(2009){Jacobs}, {Rizzi}, {Tully}, {Shaya},
  {Makarov}, \& {Makarova}}]{jacobs09}
{Jacobs}, B.~A., {Rizzi}, L., {Tully}, R.~B., {et~al.} 2009, \aj, 138, 332,
  \dodoi{10.1088/0004-6256/138/2/332}

\bibitem[{{Jarrett} {et~al.}(2003){Jarrett}, {Chester}, {Cutri}, {Schneider},
  \& {Huchra}}]{jarrett03}
{Jarrett}, T.~H., {Chester}, T., {Cutri}, R., {Schneider}, S.~E., \& {Huchra},
  J.~P. 2003, \aj, 125, 525, \dodoi{10.1086/345794}

\bibitem[{{Jethwa} {et~al.}(2018){Jethwa}, {Erkal}, \& {Belokurov}}]{jethwa18}
{Jethwa}, P., {Erkal}, D., \& {Belokurov}, V. 2018, \mnras, 473, 2060,
  \dodoi{10.1093/mnras/stx2330}

\bibitem[{{Johnson} {et~al.}(2017){Johnson}, {Chen}, {Mulchaey}, {Schaye}, \&
  {Straka}}]{johnson17}
{Johnson}, S.~D., {Chen}, H.-W., {Mulchaey}, J.~S., {Schaye}, J., \& {Straka},
  L.~A. 2017, \apjl, 850, L10, \dodoi{10.3847/2041-8213/aa9370}

\bibitem[{{Karachentsev} {et~al.}(2013){Karachentsev}, {Makarov}, \&
  {Kaisina}}]{Karachentsev13}
{Karachentsev}, I.~D., {Makarov}, D.~I., \& {Kaisina}, E.~I. 2013, \aj, 145,
  101, \dodoi{10.1088/0004-6256/145/4/101}

\bibitem[{{Keeney} {et~al.}(2017){Keeney}, {Stocke}, {Danforth}, {Shull},
  {Pratt}, {Froning}, {Green}, {Penton}, \& {Savage}}]{keeney17}
{Keeney}, B.~A., {Stocke}, J.~T., {Danforth}, C.~W., {et~al.} 2017, \apjs, 230,
  6, \dodoi{10.3847/1538-4365/aa6b59}

\bibitem[{{Kennicutt}(1998)}]{kennicutt98}
{Kennicutt}, Robert~C., J. 1998, \araa, 36, 189,
  \dodoi{10.1146/annurev.astro.36.1.189}

\bibitem[{{Kennicutt} {et~al.}(2008){Kennicutt}, {Lee}, {Funes}, {J.}, {Sakai},
  \& {Akiyama}}]{kennicutt08}
{Kennicutt}, Robert~C., J., {Lee}, J.~C., {Funes}, J.~G., {et~al.} 2008, \apjs,
  178, 247, \dodoi{10.1086/590058}

\bibitem[{{Kennicutt} \& {Evans}(2012)}]{kennicutt12}
{Kennicutt}, R.~C., \& {Evans}, N.~J. 2012, \araa, 50, 531,
  \dodoi{10.1146/annurev-astro-081811-125610}

\bibitem[{{Kirby} {et~al.}(2013){Kirby}, {Cohen}, {Guhathakurta}, {Cheng},
  {Bullock}, \& {Gallazzi}}]{kirby13}
{Kirby}, E.~N., {Cohen}, J.~G., {Guhathakurta}, P., {et~al.} 2013, \apj, 779,
  102, \dodoi{10.1088/0004-637X/779/2/102}

\bibitem[{{Kirby} {et~al.}(2011){Kirby}, {Martin}, \& {Finlator}}]{kirby11}
{Kirby}, E.~N., {Martin}, C.~L., \& {Finlator}, K. 2011, \apjl, 742, L25,
  \dodoi{10.1088/2041-8205/742/2/L25}

\bibitem[{{Kroupa}(2001)}]{kroupa01}
{Kroupa}, P. 2001, \mnras, 322, 231, \dodoi{10.1046/j.1365-8711.2001.04022.x}

\bibitem[{{Lee} {et~al.}(2006){Lee}, {Skillman}, {Cannon}, {Jackson}, {Gehrz},
  {Polomski}, \& {Woodward}}]{lee06}
{Lee}, H., {Skillman}, E.~D., {Cannon}, J.~M., {et~al.} 2006, \apj, 647, 970,
  \dodoi{10.1086/505573}

\bibitem[{{Lee} {et~al.}(2009){Lee}, {Gil de Paz}, {Tremonti}, {Kennicutt},
  {Salim}, {Bothwell}, {Calzetti}, {Dalcanton}, {Dale}, {Engelbracht}, {Funes},
  {Johnson}, {Sakai}, {Skillman}, {van Zee}, {Walter}, \& {Weisz}}]{lee09}
{Lee}, J.~C., {Gil de Paz}, A., {Tremonti}, C., {et~al.} 2009, \apj, 706, 599,
  \dodoi{10.1088/0004-637X/706/1/599}

\bibitem[{{Lee} {et~al.}(2011){Lee}, {Gil de Paz}, {Kennicutt}, {Bothwell},
  {Dalcanton}, {Jos{\'e} G. Funes S.}, {Johnson}, {Sakai}, {Skillman},
  {Tremonti}, \& {van Zee}}]{lee11}
{Lee}, J.~C., {Gil de Paz}, A., {Kennicutt}, Robert~C., J., {et~al.} 2011,
  \apjs, 192, 6, \dodoi{10.1088/0067-0049/192/1/6}

\bibitem[{{Liang} \& {Chen}(2014)}]{liang14}
{Liang}, C.~J., \& {Chen}, H.-W. 2014, \mnras, 445, 2061,
  \dodoi{10.1093/mnras/stu1901}

\bibitem[{{McConnachie}(2012)}]{mcconnachie12}
{McConnachie}, A.~W. 2012, \aj, 144, 4, \dodoi{10.1088/0004-6256/144/1/4}

\bibitem[{{McGaugh} \& {Schombert}(2014)}]{McGaugh14}
{McGaugh}, S.~S., \& {Schombert}, J.~M. 2014, \aj, 148, 77,
  \dodoi{10.1088/0004-6256/148/5/77}

\bibitem[{{McQuinn} {et~al.}(2015){McQuinn}, {Skillman}, {Dolphin}, {Cannon},
  {Salzer}, {Rhode}, {Adams}, {Berg}, {Giovanelli}, \& {Haynes}}]{mcquinn15b}
{McQuinn}, K. B.~W., {Skillman}, E.~D., {Dolphin}, A., {et~al.} 2015, \apjl,
  815, L17, \dodoi{10.1088/2041-8205/815/2/L17}

\bibitem[{{McQuinn} {et~al.}(2022){McQuinn}, {Adams}, {Cannon}, {Fuson},
  {Skillman}, {Brooks}, {Rhode}, {Haynes}, {Inoue}, {Marine}, {Salzer}, \&
  {Talluri}}]{mcquinn22}
{McQuinn}, K. B.~W., {Adams}, E. A.~K., {Cannon}, J.~M., {et~al.} 2022, \apj,
  940, 8, \dodoi{10.3847/1538-4357/ac9285}

\bibitem[{{Meurer} {et~al.}(2006){Meurer}, {Hanish}, {Ferguson}, {Knezek},
  {Kilborn}, {Putman}, {Smith}, {Koribalski}, {Meyer}, {Oey}, {Ryan-Weber},
  {Zwaan}, {Heckman}, {Kennicutt}, {Lee}, {Webster}, {Bland-Hawthorn},
  {Dopita}, {Freeman}, {Doyle}, {Drinkwater}, {Staveley-Smith}, \&
  {Werk}}]{meurer06}
{Meurer}, G.~R., {Hanish}, D.~J., {Ferguson}, H.~C., {et~al.} 2006, \apjs, 165,
  307, \dodoi{10.1086/504685}

\bibitem[{{Mina} {et~al.}(2021){Mina}, {Shen}, {Keller}, {Mayer}, {Madau}, \&
  {Wadsley}}]{mina21}
{Mina}, M., {Shen}, S., {Keller}, B.~W., {et~al.} 2021, \aap, 655, A22,
  \dodoi{10.1051/0004-6361/202039420}

\bibitem[{{Munshi} {et~al.}(2021){Munshi}, {Brooks}, {Applebaum},
  {Christensen}, {Quinn}, \& {Sligh}}]{munshi21}
{Munshi}, F., {Brooks}, A.~M., {Applebaum}, E., {et~al.} 2021, \apj, 923, 35,
  \dodoi{10.3847/1538-4357/ac0db6}

\bibitem[{{Muratov} {et~al.}(2017){Muratov}, {Kere{\v{s}}},
  {Faucher-Gigu{\`e}re}, {Hopkins}, {Ma}, {Angl{\'e}s-Alc{\'a}zar}, {Chan},
  {Torrey}, {Hafen}, {Quataert}, \& {Murray}}]{muratov17}
{Muratov}, A.~L., {Kere{\v{s}}}, D., {Faucher-Gigu{\`e}re}, C.-A., {et~al.}
  2017, \mnras, 468, 4170, \dodoi{10.1093/mnras/stx667}

\bibitem[{{Nadler} {et~al.}(2020){Nadler}, {Wechsler}, {Bechtol}, {Mao},
  {Green}, {Drlica-Wagner}, {McNanna}, {Mau}, {Pace}, {Simon}, {Kravtsov},
  {Dodelson}, {Li}, {Riley}, {Wang}, {Abbott}, {Aguena}, {Allam}, {Annis},
  {Avila}, {Bernstein}, {Bertin}, {Brooks}, {Burke}, {Rosell}, {Kind},
  {Carretero}, {Costanzi}, {da Costa}, {De Vicente}, {Desai}, {Evrard},
  {Flaugher}, {Fosalba}, {Frieman}, {Garc{\'\i}a-Bellido}, {Gaztanaga},
  {Gerdes}, {Gruen}, {Gschwend}, {Gutierrez}, {Hartley}, {Hinton}, {Honscheid},
  {Krause}, {Kuehn}, {Kuropatkin}, {Lahav}, {Maia}, {Marshall}, {Menanteau},
  {Miquel}, {Palmese}, {Paz-Chinch{\'o}n}, {Plazas}, {Romer}, {Sanchez},
  {Santiago}, {Scarpine}, {Serrano}, {Smith}, {Soares-Santos}, {Suchyta},
  {Tarle}, {Thomas}, {Varga}, {Walker}, \& {DES Collaboration}}]{nadler20}
{Nadler}, E.~O., {Wechsler}, R.~H., {Bechtol}, K., {et~al.} 2020, \apj, 893,
  48, \dodoi{10.3847/1538-4357/ab846a}

\bibitem[{{Oppenheimer} {et~al.}(2018){Oppenheimer}, {Schaye}, {Crain}, {Werk},
  \& {Richings}}]{oppenheimer18c}
{Oppenheimer}, B.~D., {Schaye}, J., {Crain}, R.~A., {Werk}, J.~K., \&
  {Richings}, A.~J. 2018, \mnras, 481, 835, \dodoi{10.1093/mnras/sty2281}

\bibitem[{{Oppenheimer} {et~al.}(2016){Oppenheimer}, {Crain}, {Schaye},
  {Rahmati}, {Richings}, {Trayford}, {Tumlinson}, {Bower}, {Schaller}, \&
  {Theuns}}]{oppenheimer16}
{Oppenheimer}, B.~D., {Crain}, R.~A., {Schaye}, J., {et~al.} 2016, \mnras, 460,
  2157, \dodoi{10.1093/mnras/stw1066}

\bibitem[{{Peeples} {et~al.}(2017){Peeples}, {Tumlinson}, {Fox}, {Aloisi},
  {Fleming}, {Jedrzejewski}, {Oliveira}, {Ayres}, {Danforth}, {Keeney}, \&
  {Jenkins}}]{Peeples17}
{Peeples}, M., {Tumlinson}, J., {Fox}, A., {et~al.} 2017, {The Hubble
  Spectroscopic Legacy Archive}, Instrument Science Report COS 2017-4, 8 pages

\bibitem[{{Peeples} {et~al.}(2019){Peeples}, {Corlies}, {Tumlinson}, {O'Shea},
  {Lehner}, {O'Meara}, {Howk}, {Earl}, {Smith}, {Wise}, \&
  {Hummels}}]{Peeples19}
{Peeples}, M.~S., {Corlies}, L., {Tumlinson}, J., {et~al.} 2019, \apj, 873,
  129, \dodoi{10.3847/1538-4357/ab0654}

\bibitem[{{P{\'e}roux} \& {Howk}(2020)}]{peroux20}
{P{\'e}roux}, C., \& {Howk}, J.~C. 2020, \araa, 58, 363,
  \dodoi{10.1146/annurev-astro-021820-120014}

\bibitem[{{Planck Collaboration} {et~al.}(2016){Planck Collaboration}, {Ade},
  {Aghanim}, {Arnaud}, {Ashdown}, {Aumont}, {Baccigalupi}, {Banday},
  {Barreiro}, {Bartlett}, {Bartolo}, {Battaner}, {Battye}, {Benabed},
  {Beno{\^\i}t}, {Benoit-L{\'e}vy}, {Bernard}, {Bersanelli}, {Bielewicz},
  {Bock}, {Bonaldi}, {Bonavera}, {Bond}, {Borrill}, {Bouchet}, {Boulanger},
  {Bucher}, {Burigana}, {Butler}, {Calabrese}, {Cardoso}, {Catalano},
  {Challinor}, {Chamballu}, {Chary}, {Chiang}, {Chluba}, {Christensen},
  {Church}, {Clements}, {Colombi}, {Colombo}, {Combet}, {Coulais}, {Crill},
  {Curto}, {Cuttaia}, {Danese}, {Davies}, {Davis}, {de Bernardis}, {de Rosa},
  {de Zotti}, {Delabrouille}, {D{\'e}sert}, {Di Valentino}, {Dickinson},
  {Diego}, {Dolag}, {Dole}, {Donzelli}, {Dor{\'e}}, {Douspis}, {Ducout},
  {Dunkley}, {Dupac}, {Efstathiou}, {Elsner}, {En{\ss}lin}, {Eriksen},
  {Farhang}, {Fergusson}, {Finelli}, {Forni}, {Frailis}, {Fraisse},
  {Franceschi}, {Frejsel}, {Galeotta}, {Galli}, {Ganga}, {Gauthier}, {Gerbino},
  {Ghosh}, {Giard}, {Giraud-H{\'e}raud}, {Giusarma}, {Gjerl{\o}w},
  {Gonz{\'a}lez-Nuevo}, {G{\'o}rski}, {Gratton}, {Gregorio}, {Gruppuso},
  {Gudmundsson}, {Hamann}, {Hansen}, {Hanson}, {Harrison}, {Helou},
  {Henrot-Versill{\'e}}, {Hern{\'a}ndez-Monteagudo}, {Herranz}, {Hildebrandt},
  {Hivon}, {Hobson}, {Holmes}, {Hornstrup}, {Hovest}, {Huang}, {Huffenberger},
  {Hurier}, {Jaffe}, {Jaffe}, {Jones}, {Juvela}, {Keih{\"a}nen}, {Keskitalo},
  {Kisner}, {Kneissl}, {Knoche}, {Knox}, {Kunz}, {Kurki-Suonio}, {Lagache},
  {L{\"a}hteenm{\"a}ki}, {Lamarre}, {Lasenby}, {Lattanzi}, {Lawrence}, {Leahy},
  {Leonardi}, {Lesgourgues}, {Levrier}, {Lewis}, {Liguori}, {Lilje},
  {Linden-V{\o}rnle}, {L{\'o}pez-Caniego}, {Lubin}, {Mac{\'\i}as-P{\'e}rez},
  {Maggio}, {Maino}, {Mandolesi}, {Mangilli}, {Marchini}, {Maris}, {Martin},
  {Martinelli}, {Mart{\'\i}nez-Gonz{\'a}lez}, {Masi}, {Matarrese}, {McGehee},
  {Meinhold}, {Melchiorri}, {Melin}, {Mendes}, {Mennella}, {Migliaccio},
  {Millea}, {Mitra}, {Miville-Desch{\^e}nes}, {Moneti}, {Montier}, {Morgante},
  {Mortlock}, {Moss}, {Munshi}, {Murphy}, {Naselsky}, {Nati}, {Natoli},
  {Netterfield}, {N{\o}rgaard-Nielsen}, {Noviello}, {Novikov}, {Novikov},
  {Oxborrow}, {Paci}, {Pagano}, {Pajot}, {Paladini}, {Paoletti}, {Partridge},
  {Pasian}, {Patanchon}, {Pearson}, {Perdereau}, {Perotto}, {Perrotta},
  {Pettorino}, {Piacentini}, {Piat}, {Pierpaoli}, {Pietrobon}, {Plaszczynski},
  {Pointecouteau}, {Polenta}, {Popa}, {Pratt}, {Pr{\'e}zeau}, {Prunet},
  {Puget}, {Rachen}, {Reach}, {Rebolo}, {Reinecke}, {Remazeilles}, {Renault},
  {Renzi}, {Ristorcelli}, {Rocha}, {Rosset}, {Rossetti}, {Roudier},
  {Rouill{\'e} d'Orfeuil}, {Rowan-Robinson}, {Rubi{\~n}o-Mart{\'\i}n},
  {Rusholme}, {Said}, {Salvatelli}, {Salvati}, {Sandri}, {Santos},
  {Savelainen}, {Savini}, {Scott}, {Seiffert}, {Serra}, {Shellard}, {Spencer},
  {Spinelli}, {Stolyarov}, {Stompor}, {Sudiwala}, {Sunyaev}, {Sutton},
  {Suur-Uski}, {Sygnet}, {Tauber}, {Terenzi}, {Toffolatti}, {Tomasi},
  {Tristram}, {Trombetti}, {Tucci}, {Tuovinen}, {T{\"u}rler}, {Umana},
  {Valenziano}, {Valiviita}, {Van Tent}, {Vielva}, {Villa}, {Wade}, {Wandelt},
  {Wehus}, {White}, {White}, {Wilkinson}, {Yvon}, {Zacchei}, \&
  {Zonca}}]{planck15}
{Planck Collaboration}, {Ade}, P.~A.~R., {Aghanim}, N., {et~al.} 2016, \aap,
  594, A13, \dodoi{10.1051/0004-6361/201525830}

\bibitem[{{Prochaska} {et~al.}(2016){Prochaska}, {Tejos}, {Crighton},
  {jnburchett}, {Tuo-Ji}, {tiffanyhsyu}, {ktirimba}, {jhennawi}, {O'Meara}, \&
  {Werk}}]{prochaska16}
{Prochaska}, J.~X., {Tejos}, N., {Crighton}, N., {et~al.} 2016,
  {Linetools/Linetools: Second Major Release}, v0.2, Zenodo,  Zenodo,
  \dodoi{10.5281/zenodo.168270}

\bibitem[{{Prochaska} {et~al.}(2017){Prochaska}, {Werk}, {Worseck}, {Tripp},
  {Tumlinson}, {Burchett}, {Fox}, {Fumagalli}, {Lehner}, {Peeples}, \&
  {Tejos}}]{prochaska17}
{Prochaska}, J.~X., {Werk}, J.~K., {Worseck}, G., {et~al.} 2017, \apj, 837,
  169, \dodoi{10.3847/1538-4357/aa6007}

\bibitem[{{Putman} {et~al.}(2012){Putman}, {Peek}, \& {Joung}}]{putman12}
{Putman}, M.~E., {Peek}, J.~E.~G., \& {Joung}, M.~R. 2012, \araa, 50, 491,
  \dodoi{10.1146/annurev-astro-081811-125612}

\bibitem[{{Qu} \& {Bregman}(2018{\natexlab{a}})}]{Qu18a}
{Qu}, Z., \& {Bregman}, J.~N. 2018{\natexlab{a}}, \apj, 856, 5,
  \dodoi{10.3847/1538-4357/aaafd4}

\bibitem[{{Qu} \& {Bregman}(2018{\natexlab{b}})}]{Qu18b}
---. 2018{\natexlab{b}}, \apj, 862, 23, \dodoi{10.3847/1538-4357/aaccec}

\bibitem[{{Qu} \& {Bregman}(2022)}]{qu22}
---. 2022, \apj, 927, 228, \dodoi{10.3847/1538-4357/ac51df}

\bibitem[{{Read} {et~al.}(2017){Read}, {Iorio}, {Agertz}, \&
  {Fraternali}}]{read17}
{Read}, J.~I., {Iorio}, G., {Agertz}, O., \& {Fraternali}, F. 2017, \mnras,
  467, 2019, \dodoi{10.1093/mnras/stx147}

\bibitem[{{Rey} {et~al.}(2020){Rey}, {Pontzen}, {Agertz}, {Orkney}, {Read}, \&
  {Rosdahl}}]{rey20}
{Rey}, M.~P., {Pontzen}, A., {Agertz}, O., {et~al.} 2020, \mnras, 497, 1508,
  \dodoi{10.1093/mnras/staa1640}

\bibitem[{{Rey} {et~al.}(2019){Rey}, {Pontzen}, {Agertz}, {Orkney}, {Read},
  {Saintonge}, \& {Pedersen}}]{rey19}
---. 2019, \apjl, 886, L3, \dodoi{10.3847/2041-8213/ab53dd}

\bibitem[{{Richings} {et~al.}(2014){Richings}, {Schaye}, \&
  {Oppenheimer}}]{richings14a}
{Richings}, A.~J., {Schaye}, J., \& {Oppenheimer}, B.~D. 2014, \mnras, 440,
  3349, \dodoi{10.1093/mnras/stu525}

\bibitem[{{Richter} {et~al.}(2017){Richter}, {Nuza}, {Fox}, {Wakker}, {Lehner},
  {Ben Bekhti}, {Fechner}, {Wendt}, {Howk}, {Muzahid}, {Ganguly}, \&
  {Charlton}}]{richter17}
{Richter}, P., {Nuza}, S.~E., {Fox}, A.~J., {et~al.} 2017, \aap, 607, A48,
  \dodoi{10.1051/0004-6361/201630081}

\bibitem[{{Sales} {et~al.}(2022){Sales}, {Wetzel}, \& {Fattahi}}]{sales22}
{Sales}, L.~V., {Wetzel}, A., \& {Fattahi}, A. 2022, Nature Astronomy, 6, 897,
  \dodoi{10.1038/s41550-022-01689-w}

\bibitem[{{Salpeter}(1955)}]{salpeter55}
{Salpeter}, E.~E. 1955, \apj, 121, 161, \dodoi{10.1086/145971}

\bibitem[{{Salvatier} {et~al.}(2016){Salvatier}, {Wiecki{\^a}}, \&
  {Fonnesbeck}}]{pymc3}
{Salvatier}, J., {Wiecki{\^a}}, T.~V., \& {Fonnesbeck}, C. 2016, {PyMC3:
  Python probabilistic programming framework}, Astrophysics Source Code
  Library, record ascl:1610.016.
\newblock \doeprint{1610.016}

\bibitem[{{Savage} \& {Sembach}(1996)}]{savage96}
{Savage}, B.~D., \& {Sembach}, K.~R. 1996, Annual Review of Astronomy and
  Astrophysics, 34, 279, \dodoi{10.1146/annurev.astro.34.1.279}

\bibitem[{{Schaye} {et~al.}(2015){Schaye}, {Crain}, {Bower}, {Furlong},
  {Schaller}, {Theuns}, {Dalla Vecchia}, {Frenk}, {McCarthy}, {Helly},
  {Jenkins}, {Rosas-Guevara}, {White}, {Baes}, {Booth}, {Camps}, {Navarro},
  {Qu}, {Rahmati}, {Sawala}, {Thomas}, \& {Trayford}}]{schaye15}
{Schaye}, J., {Crain}, R.~A., {Bower}, R.~G., {et~al.} 2015, \mnras, 446, 521,
  \dodoi{10.1093/mnras/stu2058}

\bibitem[{{Shen} {et~al.}(2014){Shen}, {Madau}, {Conroy}, {Governato}, \&
  {Mayer}}]{shen14}
{Shen}, S., {Madau}, P., {Conroy}, C., {Governato}, F., \& {Mayer}, L. 2014,
  \apj, 792, 99, \dodoi{10.1088/0004-637X/792/2/99}

\bibitem[{{Skillman} {et~al.}(1989){Skillman}, {Kennicutt}, \&
  {Hodge}}]{skillman89}
{Skillman}, E.~D., {Kennicutt}, R.~C., \& {Hodge}, P.~W. 1989, \apj, 347, 875,
  \dodoi{10.1086/168178}

\bibitem[{{Soderblom}(2021)}]{cos_data}
{Soderblom}, D.~R. 2021, in COS Data Handbook v. 5.0, Vol.~5, 5

\bibitem[{{Stil} \& {Israel}(2002)}]{SI02}
{Stil}, J.~M., \& {Israel}, F.~P. 2002, \aap, 389, 29,
  \dodoi{10.1051/0004-6361:20020352}

\bibitem[{{Suresh} {et~al.}(2019){Suresh}, {Nelson}, {Genel}, {Rubin}, \&
  {Hernquist}}]{suresh19}
{Suresh}, J., {Nelson}, D., {Genel}, S., {Rubin}, K. H.~R., \& {Hernquist}, L.
  2019, \mnras, 483, 4040, \dodoi{10.1093/mnras/sty3402}

\bibitem[{{Tchernyshyov} {et~al.}(2022){Tchernyshyov}, {Werk}, {Wilde},
  {Prochaska}, {Tripp}, {Burchett}, {Bordoloi}, {Howk}, {Lehner}, {O'Meara},
  {Tejos}, \& {Tumlinson}}]{tchernyshyov22}
{Tchernyshyov}, K., {Werk}, J.~K., {Wilde}, M.~C., {et~al.} 2022, \apj, 927,
  147, \dodoi{10.3847/1538-4357/ac450c}

\bibitem[{{The Astropy Collaboration} {et~al.}(2018){The Astropy
  Collaboration}, {Price-Whelan}, {Sip{\H o}cz}, {G{\"u}nther}, {Lim},
  {Crawford}, {Conseil}, {Shupe}, {Craig}, {Dencheva}, {Ginsburg},
  {VanderPlas}, {Bradley}, {P{\'e}rez-Su{\'a}rez}, {de Val-Borro}, {Aldcroft},
  {Cruz}, {Robitaille}, {Tollerud}, {Ardelean}, {Babej}, {Bachetti}, {Bakanov},
  {Bamford}, {Barentsen}, {Barmby}, {Baumbach}, {Berry}, {Biscani}, {Boquien},
  {Bostroem}, {Bouma}, {Brammer}, {Bray}, {Breytenbach}, {Buddelmeijer},
  {Burke}, {Calderone}, {Cano Rodr{\'{\i}}guez}, {Cara}, {Cardoso},
  {Cheedella}, {Copin}, {Crichton}, {D{\'A}vella}, {Deil}, {Depagne},
  {Dietrich}, {Donath}, {Droettboom}, {Earl}, {Erben}, {Fabbro}, {Ferreira},
  {Finethy}, {Fox}, {Garrison}, {Gibbons}, {Goldstein}, {Gommers}, {Greco},
  {Greenfield}, {Groener}, {Grollier}, {Hagen}, {Hirst}, {Homeier}, {Horton},
  {Hosseinzadeh}, {Hu}, {Hunkeler}, {Ivezi{\'c}}, {Jain}, {Jenness}, {Kanarek},
  {Kendrew}, {Kern}, {Kerzendorf}, {Khvalko}, {King}, {Kirkby}, {Kulkarni},
  {Kumar}, {Lee}, {Lenz}, {Littlefair}, {Ma}, {Macleod}, {Mastropietro},
  {McCully}, {Montagnac}, {Morris}, {Mueller}, {Mumford}, {Muna}, {Murphy},
  {Nelson}, {Nguyen}, {Ninan}, {N{\"o}the}, {Ogaz}, {Oh}, {Parejko}, {Parley},
  {Pascual}, {Patil}, {Patil}, {Plunkett}, {Prochaska}, {Rastogi}, {Reddy
  Janga}, {Sabater}, {Sakurikar}, {Seifert}, {Sherbert}, {Sherwood-Taylor},
  {Shih}, {Sick}, {Silbiger}, {Singanamalla}, {Singer}, {Sladen}, {Sooley},
  {Sornarajah}, {Streicher}, {Teuben}, {Thomas}, {Tremblay}, {Turner},
  {Terr{\'o}n}, {van Kerkwijk}, {de la Vega}, {Watkins}, {Weaver}, {Whitmore},
  {Woillez}, \& {Zabalza}}]{astropy2}
{The Astropy Collaboration}, {Price-Whelan}, A.~M., {Sip{\H o}cz}, B.~M.,
  {et~al.} 2018, ArXiv e-prints.
\newblock \doarXiv{1801.02634}

\bibitem[{{Thom} {et~al.}(2012){Thom}, {Tumlinson}, {Werk}, {Prochaska},
  {Oppenheimer}, {Peeples}, {Tripp}, {Katz}, {O'Meara}, {Ford}, {Dav{\'e}},
  {Sembach}, \& {Weinberg}}]{thom12}
{Thom}, C., {Tumlinson}, J., {Werk}, J.~K., {et~al.} 2012, \apjl, 758, L41,
  \dodoi{10.1088/2041-8205/758/2/L41}

\bibitem[{{Tremonti} {et~al.}(2004){Tremonti}, {Heckman}, {Kauffmann},
  {Brinchmann}, {Charlot}, {White}, {Seibert}, {Peng}, {Schlegel}, {Uomoto},
  {Fukugita}, \& {Brinkmann}}]{tremonti04}
{Tremonti}, C.~A., {Heckman}, T.~M., {Kauffmann}, G., {et~al.} 2004, \apj, 613,
  898, \dodoi{10.1086/423264}

\bibitem[{{Tully} {et~al.}(2016){Tully}, {Courtois}, \& {Sorce}}]{tully16}
{Tully}, R.~B., {Courtois}, H.~M., \& {Sorce}, J.~G. 2016, \aj, 152, 50,
  \dodoi{10.3847/0004-6256/152/2/50}

\bibitem[{{Tumlinson} {et~al.}(2017){Tumlinson}, {Peeples}, \&
  {Werk}}]{tumlinson17}
{Tumlinson}, J., {Peeples}, M.~S., \& {Werk}, J.~K. 2017, \araa, 55, 389,
  \dodoi{10.1146/annurev-astro-091916-055240}

\bibitem[{{van de Voort} {et~al.}(2019){van de Voort}, {Springel}, {Mandelker},
  {van den Bosch}, \& {Pakmor}}]{vandevoort19}
{van de Voort}, F., {Springel}, V., {Mandelker}, N., {van den Bosch}, F.~C., \&
  {Pakmor}, R. 2019, \mnras, 482, L85, \dodoi{10.1093/mnrasl/sly190}

\bibitem[{{Van Sistine} {et~al.}(2016){Van Sistine}, {Salzer}, {Sugden},
  {Giovanelli}, {Haynes}, {Janowiecki}, {Jaskot}, \& {Wilcots}}]{VanSistine16}
{Van Sistine}, A., {Salzer}, J.~J., {Sugden}, A., {et~al.} 2016, \apj, 824, 25,
  \dodoi{10.3847/0004-637X/824/1/25}

\bibitem[{{Werk} {et~al.}(2013){Werk}, {Prochaska}, {Thom}, {Tumlinson},
  {Tripp}, {O'Meara}, \& {Peeples}}]{werk13}
{Werk}, J.~K., {Prochaska}, J.~X., {Thom}, C., {et~al.} 2013, \apjs, 204, 17,
  \dodoi{10.1088/0067-0049/204/2/17}

\bibitem[{{Werk} {et~al.}(2016){Werk}, {Prochaska}, {Cantalupo}, {Fox},
  {Oppenheimer}, {Tumlinson}, {Tripp}, {Lehner}, \& {McQuinn}}]{werk16}
{Werk}, J.~K., {Prochaska}, J.~X., {Cantalupo}, S., {et~al.} 2016, \apj, 833,
  54, \dodoi{10.3847/1538-4357/833/1/54}

\bibitem[{{Wheeler} {et~al.}(2019){Wheeler}, {Hopkins}, {Pace},
  {Garrison-Kimmel}, {Boylan-Kolchin}, {Wetzel}, {Bullock}, {Kere{\v{s}}},
  {Faucher-Gigu{\`e}re}, \& {Quataert}}]{wheeler19}
{Wheeler}, C., {Hopkins}, P.~F., {Pace}, A.~B., {et~al.} 2019, \mnras, 490,
  4447, \dodoi{10.1093/mnras/stz2887}

\bibitem[{{Wilde} {et~al.}(2021){Wilde}, {Werk}, {Burchett}, {Prochaska},
  {Tchernyshyov}, {Tripp}, {Tejos}, {Lehner}, {Bordoloi}, {O'Meara}, \&
  {Tumlinson}}]{wilde21}
{Wilde}, M.~C., {Werk}, J.~K., {Burchett}, J.~N., {et~al.} 2021, \apj, 912, 9,
  \dodoi{10.3847/1538-4357/abea14}

\bibitem[{{Zahedy} {et~al.}(2021){Zahedy}, {Chen}, {Cooper}, {Boettcher},
  {Johnson}, {Rudie}, {Chen}, {Cantalupo}, {Cooksey}, {Faucher-Gigu{\`e}re},
  {Greene}, {Lopez}, {Mulchaey}, {Penton}, {Petitjean}, {Putman}, {Rafelski},
  {Rauch}, {Schaye}, {Simcoe}, \& {Walth}}]{zahedy21}
{Zahedy}, F.~S., {Chen}, H.-W., {Cooper}, T.~M., {et~al.} 2021, \mnras, 506,
  877, \dodoi{10.1093/mnras/stab1661}

\bibitem[{{Zheng} {et~al.}(2020){Zheng}, {Emerick}, {Putman}, {Werk}, {Kirby},
  \& {Peek}}]{zheng20b}
{Zheng}, Y., {Emerick}, A., {Putman}, M.~E., {et~al.} 2020, \apj, 905, 133,
  \dodoi{10.3847/1538-4357/abc875}

\bibitem[{{Zheng} {et~al.}(2019){Zheng}, {Putman}, {Emerick}, {McQuinn},
  {Werk}, {Lockman}, {Oppenheimer}, {Fox}, {Kirby}, \& {Burchett}}]{zheng19b}
{Zheng}, Y., {Putman}, M.~E., {Emerick}, A., {et~al.} 2019, \mnras, 490, 467,
  \dodoi{10.1093/mnras/stz2563}

\end{thebibliography}

\appendix
\restartappendixnumbering
\section{Absorbers Not Included in This Work}
\label{sec:app_absorber}

In addition to the \npairthiswork\ new pairs as shown in \S\ref{sec:data_new_pairs}, we find a few other absorber detections but decide not to include them in this work for the following reasons. 

We do not include QSOs PMNJ2345-1555 and UVQSJ000009.65-163441.4 near WLM because these two sight lines go through the Magellanic Stream and thus are contaminated. 

We detect absorbers toward QSOs PG0052+251 and RXJ00537+2232, which are at impact parameters of 47 and 28 kpc from galaxy LGS3. Even though the absorbers are $\leqslant50~\kms$ from the systemic velocity of LGS3, it is unclear whether they are associated with the galaxy given that there are some \HI\ clouds closer to these QSO sight lines that may cause absorption.   
 
We do not include a \CIV\ detection toward QSO MS1217.0+0700 near galaxy PGC039646. This is because of the absorber's large offset velocity from the galaxy ($\delta v=73~\kms$) and that only a weak absorber is seen in \CIV\ 1548. 

Toward QSO UVQSJ130808.98-455417.9, we find two absorbers at $\impactpara$=57 kpc from galaxy ESO269-037. We inspect the nearby environment and conclude that the absorbers are most likely associated with Cen A's CGM. 

Toward QSO MARK509, there may exist a weak \CIV\ absorber with $\log N$=12.6 at the systemic velocity of galaxy DDO210. 
Given the low mass of the galaxy ($\mstar=10^6~\msun$), the large impact parameter of this weak absorber ($\impactpara\sim0.7\rvir$), and that the systemic velocity of DDO210 ($v_{\rm helio}=-131.5~\kms$) is well within the MW's high velocity cloud regime, it is unclear whether this absorber belongs to the CGM of the galaxy.  

Lastly, we do not include QSO UVQSJ101124.51-044215.5, which goes through the ISM of Sextans A (J. N. Burchett et al. 2023, in preparation). 

\section{Determination of Galaxy Properties}
\label{sec:app_gal_prop}
Below we describe the methods and references that we use to determine the galaxy properties of the \nunidwarf\ dwarf galaxies in our Full Sample, the values of which are tabulated in Tables \ref{tb:pair_info} and \ref{tb:dwarf_info}. When not specified, the corresponding galaxy properties are from \cite{Karachentsev13}'s Updated Nearby Galaxy Catalog, such as galaxies' coordinates and velocities. 

\textsl{Distances ($d$):} For dwarfs within the Local Volume, a majority of the distances are based on distance modules from the cosmicflow3 database \citep{tully16}. We also supplement the table with better distance values from various sources as listed in Table \ref{tb:dwarf_info}. 
Overall, the dwarf galaxies in our sample are at distances of 0.8--8 Mpc (for Local Volume dwarfs) or at redshifts of 0.003--0.3 (for dwarfs from \citetalias{liang14}, \citetalias{bordoloi14}, or \citetalias{johnson17}). 

\textsl{\HI\ mass ($\mhi$)}: When available, we adopt \HI\ fluxes ($S_{\rm HI}$) from the most recent ALFALFA extragalactic \HI\ source catalog \citep{Haynes18}, and calculate $\mhi$ from $S_{\rm HI}$ with our adopted distances: $\frac{M_{\rm HI}}{\msun}\approx0.236(\frac{S_{\rm HI}}{\rm Jy~\kms})(\frac{d}{\rm kpc})^2$. If the galaxy is not available in ALFALFA, we adopt $\mhi$ from the LITTLE THINGS survey \citep{hunter12}, with the mass value rescaled to our adopted distance for consistency. If a target cannot be found in either of the aforementioned \HI\ references, we adopt $S_{\rm HI}$ from \cite{Karachentsev13} and derive the corresponding $\mhi$ accordingly. Among the dwarf galaxies with available $\mhi$ in our Full Sample, the \HI\ gas mass ranges from $\mhi=10^{6.1}$ to $10^{8.6}~\msun$ with a median value of $\medmHI~\msun$.

\textsl{Star Formation Rate (SFR or $\mdotstar$)}: For galaxies adopted from \citetalias{bordoloi14} and \citetalias{liang14}, we adopt their provided $\mdotstar$ values without any correction. We note that there are not $\mdotstar$ values included by \citetalias{johnson17}. For galaxies within the Local Volume, we calculate each galaxy's $\mdotstar$ based on their GALEX FUV ($\mfuv$) magnitudes from the Local Volume Survey \citep{lee11} as shown in Equation \ref{eq:mfuv_mdotstar} below. If a galaxy is not included in \cite{lee11}, we obtain its $\mfuv$ from \cite{Karachentsev13}. All the $\mfuv$ values have been corrected for extinction from the MW. We choose to estimate $\mdotstar$ from $\mfuv$ because for low-mass galaxies the FUV fluxes provide more accurate measurements of the SFR than H$\alpha$ tracers \citep{lee09}. 
\begin{equation}
\begin{split}
    \mdotstar_{\rm, K98} &= 10^{2.78-0.4\mfuv+2\log_{10}(d/{\rm Mpc})} \\ 
    \mdotstar_{\rm, K12} &= 0.63 \mdotstar_{\rm, K98} ~~~,
    \label{eq:mfuv_mdotstar}
\end{split}
\end{equation}
where $\mdotstar_{\rm, K98}$ is the SFR derived based on \cite{kennicutt98} with \cite{salpeter55} IMF. And in this work, we adopt the new SFR conversion $\mdotstar_{\rm, K12}$ from \cite{kennicutt12}, which is based on a \cite{kroupa01} IMF to be consistent with other galaxy properties used here. Among all the galaxies within the Local Volume, only dwarf galaxy NGC5408 does not have a measured $\mfuv$ in either \cite{lee11} or \cite{Karachentsev13}, in which case we estimate its $\mdotstar$ based on the galaxy's H$\alpha$ luminosity as measured in \cite{kennicutt08} and convert the value to a \cite{kroupa01} IMF accordingly. Among the dwarf galaxies with available $\mdotstar$ in our sample, the SFR ranges from $\mdotstar=10^{-4.3}$ to $10^{-0.2}~\msunyr$ with a median value of $10^{\medlogsfr}~\msunyr$.

\section{PyMC3 Analysis on Ion Column Density Profiles: $\log N$ vs. $\impactpara/\rvir$}
\label{sec:app_pymc3}

As introduced in \S\ref{sec:result_pymc}, we adopt a censored regression algorithm based on \texttt{PyMC3} to fit for the $\log N$ \textsl{vs.} $b/\rvirm$ relation and take into account the nondetection upper limits and saturation lower limits in our data. We first run a maximum-likelihood (ML) fit assuming that all the data points (including upper/lower limits) are detected values, which results in a set of best-fit parameters of ($k^{\rm ML}$, $\log N_0^{\rm ML}$). Note that we only use ($k^{\rm ML}$, $\log N_0^{\rm ML}$) as an intermediate step to guide the set-up of the following MCMC priors, but do not use them anywhere else since this ML fit most likely overpredicts the $\log N_0$ value by treating upper limits as detections.

To set up an MCMC sampling, we first assume priors of truncated normal distribution for the slope $k$ and the intercept $\log N_0$ that we want to solve for. The truncated normal distribution is a probability distribution that derived from a normal distribution $\mathcal{N}(\mu, \sigma^2)$, but with lower and/or upper bounds for the parameter of interest. For the slope $k$, we learn from the data in Figure \ref{fig:observed_logN_b} and from the CGM literature that ion column densities seem to drop as a function of impact parameter, which means the slope $k$ should be less than or equal to 0. On the other hand, in theory the lower bound for $k$ could go to $-\infty$, but in practice we set it at $-10$, which is sufficient here given that none of the solutions we find in Figure \ref{fig:observed_logN_b} are generally less than $-3$. In all, the prior we implement for the unknown $k$ parameter is a truncated normal distribution with bounds of [$-10$, 0]. Since we also do not know the mean $\mu_k$ or the variance $\sigma_k^2$ for the probability distribution of $k$, we set a large variance of $\sigma_k^2$=10 and adopt the ML fit result above as the mean $\mu_k=k^{\rm ML}$. This approach ensures that we implement a prior probability distribution function for $k$ with the limited knowledge available (i.e., $k\leqslant0$ and $k^{\rm ML}$ is a probable solution) without introducing additional bias. 

Similarly, the prior probability distribution for $\log N_0$ is a truncated normal distribution with an upper bound at 100, which is much larger than any possible intercept solution as informed by the data points in Figure \ref{fig:observed_logN_b}. The lower bound for $\log N_0$ is set as 1 for $\log N$, which is much smaller than any possible values as informed from observations (Figure \ref{fig:observed_logN_b}) and simulations (Figure \ref{fig:eagle_logN_b}, \S\ref{sec:sims_eagle}). The variance $\sigma^2_{\log N_0}$ is assumed to be 10, while the mean is taken to be the ML result $\mu_b=b^{\rm ML}$ for the same reason given in the previous paragraph. We have checked that the specific values of these priors do not change the results significantly, as long as they cover a sufficiently large parameter space for the MCMC sampling.

With the priors set up, we then proceed to compute the likelihood function. We divide the dataset into three groups, where the first group consists of detections and the second/third groups with nondetection upper limits/saturation lower limits, respectively. The likelihood function for the first group with detections is as follows: 
\begin{equation}
\begin{split}
    {\rm ln}\mathcal{L_{\rm det}} & =\sum {\rm ln}[p(\log N_i, \sigma_{\log N_i}|k, \log N_0)] \\
    & \propto -\frac{1}{2} \sum \frac{(\log N_i - kx_i - \log N_0)^2}{\sigma_{\log N_i^2}} \\ 
    \sigma_{\log N_i}^2 & = <e_{\log N_i}>^2 + \sigma_{\log N, all-det}^2 ~~~,
\end{split}
\label{eq:lnL_det}
\end{equation}
where $x_i$ is the observed impact parameter $\log (\impactpara/\rvir)_i$, $\log N_i$ the detected values, and $\sigma_{\log N_i}$ the quadrature error of the mean measurement uncertainties ($e_{\log N_i}$; Table \ref{tb:logN}) and the intrinsic scatter ($\sigma_{\log N, all-det}$) calculated as the standard deviation of all available $\log N_i$ values. While $e_{\log N_i}$ is caused by systemic uncertainty related to observations and instruments, $\sigma_{\log N, all-det}$ represents the physical scatter intrinsic to the CGM with clumpy gas distributions. 

On the other hand, the likelihood function for the nondetection data points cannot be evaluated as a normal distribution because we only have upper limit values that indicate the real values occur below such limits at certain probabilities. The likelihood function for the upper limits is set to equal to the integrated probability from $-\infty$ to the upper limit values: 
\begin{equation}
    {\rm ln}\mathcal{L_{\rm non-det}} 
    =\sum {\rm ln}[\int_{-\infty}^{\log N_j} p(\log N_j, \sigma_{\log N_j}|k, \log N_0)] ~~~,
\label{eq:lnL_3sig}    
\end{equation}
where $\log N_j$ are the $3\sigma$ upper limits for the nondetections, suggesting that the real column densities have 99.7\% of probability to be lower than the adopted upper limits. To model the distribution of these nondetection upper limits, we assume that they have similar measurement errors and intrinsic scatters as their detected counterparts and adopt $\sigma_{\log N_j}=\sigma_{\log N_i}$. Similarly, for \HI\ data points that are saturated, the likelihood function is set to equal to the integrated probability from the lower limit values to $+\infty$. The likelihood of the whole dataset, including detections and upper/lower limits are the sum of ${\rm ln}\mathcal{L_{\rm det}}$ and ${\rm ln}\mathcal{L_{\rm non-det}}$ (and ${\rm ln}\mathcal{L_{\rm saturation}}$ for \HI). 

We then run the \texttt{PyMC3} sampler to explore the parameter space of $k$ and $\log N_0$ and calculate the corresponding likelihood with 20k steps. The gray curves in Figure \ref{fig:observed_logN_b} show 100 results that are randomly drawn from the posterior distributions of $k$ and $\log N_0$, while the black solid curves indicate the 50th-percentile solutions. On the title of each panel, we show the 50th-percentile solution with uncertainties derived from the (84th--50th) and (50th--16th) values in the corresponding posterior distributions.

\end{CJK*}
\end{document}